\documentclass{emulateapj}
\usepackage{epstopdf}
\usepackage{color}
\usepackage{CJK}
\newcommand{\beq}{\begin{equation}}
\newcommand{\eeq}{\end{equation}}
\newcommand{\bea}{\begin{eqnarray}}
\newcommand{\eea}{\end{eqnarray}}
\def\bulletskip{0.15cm}

\begin{document}
\begin{CJK}[HL]{KS}{}
\title{The Hydrodynamic Feedback of Cosmic Reionization on Small-Scale Structures and Its Impact on Photon Consumption during the Epoch of Reionization}

%\date{\today}

\author{Hyunbae Park (¹ÚÇö¹è)\altaffilmark{1,2}}
%\author{Hyunbae Park\altaffilmark{1,2}}
\author{Paul R. Shapiro\altaffilmark{1}}
\author{Jun-hwan Choi\altaffilmark{1}}
\author{Naoki Yoshida\altaffilmark{3,4}}
\author{Shingo Hirano\altaffilmark{3}}
\author{Kyungjin Ahn\altaffilmark{5}}

\affil{$^1$Texas Cosmology Center and the Department of Astronomy, The University of Texas at Austin, 1 University Station, C1400, Austin, TX 78712, USA}
\affil{$^2$Korea Astronomy and Space Science Institute, Daejeon  34055, Korea}
\affil{$^3$Department of Physics, University of Tokyo, Bunkyo, Tokyo 113-0033, Japan}
\affil{$^4$Kavli Institute for the Physics and Mathematics of the Universe (WPI), Institutes for Advanced Study, University of Tokyo, Kashiwa, Chiba 277-8583, Japan}
\affil{$^5$Department of Earth Sciences, Chosun University, Gwangju 61452, Korea}

\begin{abstract} 
Density inhomogeneity in the intergalactic medium (IGM) can boost the recombination rate of ionized gas substantially, affecting the growth of HII regions during reionization. Previous attempts to quantify this effect typically failed to resolve down to the Jeans scale in the pre-ionization IGM, which is important in establishing this effect, along with the hydrodynamical back-reaction of reionization on it. Towards that end, we perform a set of fully-coupled, radiation-hydrodynamics simulations from cosmological initial conditions, extending the mass resolution of previous work to the scale of minihalos. Pre-reionization structure is evolved until a redshift $z_i$ at which the ionizing radiation from external sources arrives to sweep an R-type ionization front supersonically across the volume in a few Myr, until it is trapped on the surfaces of minihalos and converted to D-type, after which the minihalo gas is removed by photoevaporative winds. Small-scale density structures during this time lead to a high ($>$10) clumping factor for ionized gas, which hugely boosts the recombination rate until the structures are disrupted by the hydrodynamic feedback after $\sim 10-100~\rm{Myr}$. For incoming stellar radiation with intensity $J_{21}$ in a $200~h^{-1}~\rm{kpc}$ box with the mean density contrast $\bar\delta$, the number of extra recombinations per H atom, on top of what is expected from homogeneously distributed gas, is given by $0.32[J_{21}]^{0.12}[(1+z_i)/11]^{-1.7}[1+\bar\delta]^{2.5}$. In models in which most of the volume is ionized toward the end of reionization, this can add more than one recombination per H atom to the ionizing photon budget to achieve reionization.
\end{abstract}

\section{Introduction} 

With growing computational power, simulations of structure formation and radiative transfer are becoming more and more sophisticated in modeling the details of the epoch of reionization (EoR) when the early galaxies led to the ionization of hydrogen in the intergalactic space during the first billion years after the Big Bang \citep[for reviews, see][]{2006AJ....132..117F,2010Natur.468...49R}. One of the ultimate goals of such simulations is to provide model predictions for observables like 21-cm brightness fluctuations \citep{2013MNRAS.433..639P,2015ApJ...809...61A,2015MNRAS.451.3709A}, secondary CMB anisotropies \citep{2015ApJ...799..177G}, and the luminosity function of Lyman-$\alpha$ emitters at high redshifts \citep{2012ApJ...745..122K} that will help to constrain models of EOR via comparison with observational constraints.

 A distinctive feature observed on large scales during the EoR is the giant H II regions of ionization growing up to tens of Mpc until they overlap to finish reionization \citep{2004ApJ...609..474B,2004ApJ...613....1F,2014MNRAS.439..725I}. The 21cm signal that directly maps the ionization feature was shown to converge in volumes greater than $\sim200~h^{-1}$ Mpc in a side \citep{2014MNRAS.439..725I}. When such a large simulation volume is used, it is usually not computationally feasible to resolve all the baryonic processes related to reionization. Therefore, one has to, for example, rely on sub-grid prescriptions calibrated from small-volume high-resolution simulations accounting for relevant physics. Such an attempt was realized in a large-box ($\sim150$ Mpc) reionization simulation where minihalo sources were implemented by sub-grid physics and was shown to be able to generate a significant number of ionizing photons \citep{2012ApJ...756L..16A}. 

While much attention has been paid to implementing the {\it sources} in simulations, quantitative accounting for the {\it sinks} still requires more study. When a free electron recombines with an ion not directly to the ground state, but cascading through multiple energy levels, it can end up with multiple photons, none of which are able to ionize another atoms. This Case B recombination rate depends on the clumpiness of the intergalactic medium (IGM). Due to the two-body nature of the reaction, the rate in fully ionized gas goes as the square of density with a temperature dependent coefficient in fully ionized gas. Numerical simulations would underestimate the rate if there exists unresolved density structures within resolution elements \citep{2001ApJ...551..599H,2004MNRAS.348..753S,2005ApJ...624..491I}. To factorize this unknown boost, the clumping factor is often defined as $C\equiv \left<n^2\right>/\left<n\right>^2$, where the bracket denotes the volume average and $n$ is the density of ionized gas\footnote{Strictly, the temperature dependence of the recombination coefficient should also be accounted in the clumping factor to accurately estimate the recombination rate although this is often regarded as a minor effect and ignored in studies. We shall present a definition of $C$ that takes into account the temperature dependence in Section~\ref{sec:C} and the quantitative difference made by it in Section~\ref{sec:understanding_clumping_factor}.}. If the volume average is over all of the space, this yields the global clumping factor $C_{\rm global}$, which can also be written as $1+\sigma_i^2$, when $\sigma_i$ is the root-mean-square (RMS) density fluctuations of the ionized gas. Since the gas density and ionization state fluctuate substantially on large scales as the universe undergoes ``patchy" reionization, it is useful to define a spatially-varying local clumping factor, $C_{\rm local}(\vec{x})$. In this factor, the ionized density and its square are averaged over a finite volume $V$ centered on some point $\vec{x}$ in space. We can write this as $1+\sigma^2_{<r}(\vec{x})$, where $\sigma_{<r}(\vec{x})$ is the RMS of the ionized gas density contrast in the volume $V$ of radius $r$, for the density contrast relative to the average ionized density inside $V$. 
%$C$ in a resolution element ($C_{\rm local}$ hereafter) can also be written as $1+\sigma^2_{\rm <r}$ where $\sigma_{\rm <r}$ is the rms of the unresolved ionized gas density contrast within the resolution element. 

Note that the clumping factor for the entire universe ($C_{\rm global}$, hereafter) is often quoted to estimate the number of ionizing photons needed to keep the universe ionized \citep{1999ApJ...514..648M}. In simulations, $C_{\rm global}$ can be expressed as $1+\sigma^2_{\rm <r}+\sigma^2_{\rm >r}$ where $\sigma_{\rm >r}$ is the rms of the ionized gas density of all the resolution elements in the entire universe. While simulations spanning hundreds of Mpc would capture most of the large-scale variation that goes into $\sigma_{\rm >r}^2$, $\sigma_{\rm <r}^2$ could be so significant that simply assuming $C_{\rm global}=1+\sigma^2_{\rm >r}$ would severely underestimate the clumping factor in those simulations. This can be evidenced by the situation that $C_{\rm global}$ fails to converge as the simulation resolution increases \citep[e.g., See Figure~15 of][]{2015MNRAS.453.3593B}. An error in $C_{\rm global}$ is not easily distinguishable in EOR simulations as its effect is largely degenerate with changing the mean efficiency of ionizing sources (i.e., underestimating $C_{\rm global}$ and the source efficiency give similar effects.). But, spatial variation of $C_{\rm local}$ across resolution elements may leave an observable impact by affecting the growth of H II regions. 

Gaseous density structures on sub-Mpc scales are expected to be subject to various baryonic physics, that requires coupled radiative-transfer and hydrodynamics. A number of numerical works dedicated to this problem \citep{1997ApJ...486..581G,2007ApJ...671....1T,2009MNRAS.394.1812P,2011MNRAS.412L..16R,2012MNRAS.427.2464F, 2012ApJ...747..100S,2015ApJ...810..154K,2014ApJ...789..149S} adopted $\sim 10^6M_\odot $ for the mass of the dark matter particle aiming to resolve halos down to $\sim 10^8M_\odot $ corresponding to the mass of $\sim 10^4$ K gas. This however neglects structures formed during the pre-ionization phase in the unheated IGM including {\it minihalos}. Although it is expected that the hydrodynamical feedback from ionization would disrupt such structures formed in low temperature, one needs to quantify the net recombination during the disruption. In particular, minihalos above $\sim 10^6~M_\odot $ can host dense gas that is capable of being self-shielded from ionizing radiation for a significant amount of time ($\gtrsim 10^8~{\rm yr}$) while recombining up to $\sim 10$ per H atom \citep{2004MNRAS.348..753S,2005ApJ...624..491I}. \cite{2004MNRAS.348..753S} and \cite{2005MNRAS.361..405I} were the first to address this problem by performing fully-coupled radiation-hydrodynamics simulations of individual minihalo photo-evaporation during the EoR.

Such dense neutral clumps of gas often last until the post-reionization era and are found as Lyman-limit systems \citep[e.g.,][]{1994ApJ...427L..13S,2010ApJ...718..392P,2015ApJS..221....2P}. Recombination within Lyman-limit systems can be interpreted as finite limit in the mean free path of H-ionizing radiation \citep{2003ApJ...597...66M,2010ApJ...721.1448S}, which in turn impedes the growth of H II regions beyond a certain size \citep{2006ApJ...648....1G,2009MNRAS.394..960C,2012ApJ...747..126A}. Implementing the effect of finite mean free path of ionizing photons have been found to change predictions for EoR observables from EoR models substantially \citep{2011MNRAS.411..289C,2014MNRAS.439..725I,2016MNRAS.tmp...43S}.

Toward this end, \citet[][hereafter ETA13]{2013ApJ...763..146E} posed a question of how finely one has to resolve small-scale structures to obtain convergence of the clumping factor and mean free path of ionizing photons. As the preferable resolution, they reported dark matter particle mass of $50~M_\odot$ that would well resolve structures down to $10^4$ solar masses. With that resolution, ETA13 found a substantially higher clumping factor ($C_{\rm local} \gtrsim 10$) than in other recent works \citep{2011MNRAS.412L..16R,2012MNRAS.427.2464F, 2012ApJ...747..100S,2015ApJ...810..154K,2014ApJ...789..149S} that have reported values around 3. Their simulation however was based on post-processed radiative transfer that should be valid only before the hydrodynamic feedback on the structures following the photoheating of gas comes into effect. Their reported value is likely to decrease when the Jeans mass increase after reionization.

 The goal of this paper is to model $C_{\rm local}$ through simulations that keep track of the hydrodynamic evolution of the gas fully coupled with radiation and that adopt the resolution and methodology similar to those suggested by ETA13. Throughout this paper, the background cosmology is based on the Planck cosmology \citep[$\Omega_M = 0.3175, \Omega_\Lambda = 0.6825, \Omega_b = 0.0490, h = 0.6711, n_s = 0.9624, \sigma_8 = 0.8344$;][]{2014A&A...571A..16P}.

The remainder of this paper is as follows. In Section~\ref{sec:methodology}, we introduce our methodology for simulating the hydrodynamical back-reaction of reionization. In Section~\ref{sec:C}, we give our formal definition of the clumping factor and related expression that we will use throughout the paper. 
In Section~\ref{sec:results}, we present our results. 
In Section~\ref{sec:convergence}, we discuss the effect of finite box size in our results.
In Section~\ref{sec:summary}, we summarize our results and discuss their implications.

\begin{table*} 
\caption{Simulation Parameter \& Results}
\begin{center} 
\begin{tabular}{@{}llllllllllll}  
\hline
label &  Box size & \# of ptls & $z_i$  &  $J_{21}~(\Gamma_{-12})$ &$\bar\delta$& Shielding&Dynamics &$C^{\rm peak}_{\rm r}$\footnote{Because $C_{\rm r}$ in M\_I0\_z10\_NS has a monotonic behavior, the peak value cannot be defined for this case. M\_I0\_z10\_ND also has this problem, but we list its converging value instead.} & $N^{\rm add}_{\rm rec,150}$\footnote{M\_I0\_z10\_ND and L\_I0\_z10 are not run down to $\Delta t=150$ Myr.}&$N^{\rm bg}_{\rm rec,150}$\\
 &  (kpc/$h$) &  & &  & &&$(\rm{cm}^{-3})$&&&\\
\hline
S\_I0\_z10           & \bf{100}\footnote{All numbers in boldface denote a deviation from the parameter choice of the standard run, M\_I0\_z10.} & $\bf{2\times128^3}$ & 10 & 1~(9.2) & 0 & on&on &16.7&0.23&0.12\\
M\_I0\_z10\_NS                 & 200        & $2\times256^3$ & 10      & 1~(9.2)            &   0   & \bf{off}&on& - &0.59 &0.12\\
M\_I0\_z10\_ND                 & 200        & $2\times256^3$ & 10      & 1~(9.2)            &   0   & on&\bf{off}& 26.0 &- &-\\
M\_I0\_z10                         & 200        & $2\times256^3$ & 10      & 1~(9.2)            &   0   & on&on       &  21.0&0.32&0.12\\
M\_I-0.5\_z10                     & 200        & $2\times256^3$ & 10      & \bf{0.3~(2.8)}   &   0   & on&on      &12.7 &0.28&0.13\\
M\_I-1\_z10                        & 200        & $2\times256^3$ & 10      & \bf{0.1~(0.92)} &   0   & on&on      & 7.5 &0.24&0.13\\
M\_I0\_z9                           & 200        & $2\times256^3$ & \bf{9}  & 1~(9.2)            &   0   & on&on      & 28.1 &0.37&0.09\\
M\_I0\_z8                           & 200        & $2\times256^3$ & \bf{8}  & 1~(9.2)            &   0   & on&on      & 37.5 &0.45&0.07\\
M\_I0\_z10\_VL$\delta$      & 200        & $2\times256^3$ & 10       & 1~(9.2)           &\bf{-0.52}& on&on     &  8.8  & 0.05 &0.06\\
M\_I0\_z10\_L$\delta$       & 200        & $2\times256^3$ & 10       & 1~(9.2)            &\bf{-0.26}& on&on      & 15.4 & 0.13 &0.09\\
M\_I0\_z10\_H$\delta$      & 200        & $2\times256^3$ & 10       & 1~(9.2)            &\bf{0.24} & on &on      & 22.8 & 0.44 &0.15\\
M\_I0\_z10\_VH$\delta$     & 200        & $2\times256^3$ & 10       & 1~(9.2)           &\bf{0.59} & on &on      & 21.4 & 1.00 &0.19\\
L\_I0\_z10                          & \bf{400} & $\bf{2\times512^3}$ & 10 & 1~(9.2)            &   0   & on &on      & 21.4 &-     &-      \\
\hline
\end{tabular}  
\end{center}
\label{simulation parameters}
\end{table*}

\section{Methodology} \label{sec:methodology}

\subsection{Gravity, Hydrodynamics, \& Chemistry}  \label{2.1}

%\hline
%label &  Box size & \# of ptls & $z_i$  &  $J_{21}$ & Shielding & $v_{\rm drift}$& $T_{\rm bb}$\footnote{Temperature for the blackbody spectrum of external background radiation}  \\
% &  (kpc/$h$) &  & &  & &(km/s)& (K) \\
%\hline
%S\_I0\_z10           & \bf{100} & $\bf{2\times128^3}$ & 10 & 1 & yes& - & $10^5$ \\
%M\_I0\_z10          & 200 & $2\times256^3$ & 10 & 1 & yes & - & $10^5$ \\
%L\_I0\_z10           & \bf{400} & $\bf{2\times512^3}$ & 10 & 1 & yes & - & $10^5$ \\
%M\_I0\_z9            & 200 & $2\times256^3$ & \bf{9}  & 1 & yes & - & $10^5$ \\
%M\_I0\_z8            & 200 & $2\times256^3$ & \bf{8}  & 1 & yes & - & $10^5$ \\
%M\_I0\_z7            & 200 & $2\times256^3$ & \bf{7}  & 1 & yes & - & $10^5$ \\
%M\_I-0.5\_z10          & 200 & $2\times256^3$ & 10 & \bf{0.3} & yes & - & $10^5$ \\
%M\_I-1\_z10          & 200 & $2\times256^3$ & 10 & \bf{0.1} & yes & - & $10^5$ \\
%M\_I0\_z10\_NS  & 200 & $2\times256^3$ & 10 & 1 & \bf{no} & -  & $10^5$ \\
%M\_I0\_z10\_v30 & 200 & $2\times256^3$ & 10 & 1 & yes & \bf{30} & $10^5$ \\
%M\_I0\_z10\_T5e4   & 200 & $2\times256^3$ & 10 & 1 & yes & - & $\bf{5\times10^4}$ \\
%\hline

%To investigate the problem with a  fixed mass-resolution, 

For the hydrodynamics, we adopt the smoothed particle hydrodynamics (SPH) code GADGET-3 \citep{2001MNRAS.328..726S,2005MNRAS.364.1105S} with non-equilibrium chemistry of 14 primordial species ($\rm e^{-}$, $\rm H$, $\rm H^{+}$, $\rm H^{-}$, $\rm He$, $\rm He^{+}$, $\rm He^{++}$, $\rm H_2$, $\rm H_2^+$, $\rm D$, $\rm D^{+}$, $\rm HD$, $\rm HD^{+}$, $\rm HD^{-}$) as described by \cite{2006ApJ...652....6Y,2007ApJ...663..687Y} with updated cooling rates for $\rm H_2$ and $\rm HD$ from \citet{2013ARA&A..51..163G}. An SPH code like this is suitable for our target problem because it fixes the mass resolution, allowing us to resolve dense structures with a large number of resolution elements. Throughout this paper, the mass resolution is $9.3~M_\odot $ for baryonic particles and $51~M_{\odot }$ for dark matter particles. This resolution was reported to yield converging result for the clumping factor in ETA13. This resolution corresponds to having $256^3$ particles for each of dark matter and baryon in a cubic volume of $(200~h^{-1}~\rm{kpc})^3$. 

We create the initial conditions for $100~h^{-1}~\rm{kpc}$, $200~h^{-1}~\rm{kpc}$, $400~h^{-1}~\rm{kpc}$ and $800~h^{-1}~\rm{kpc}$ boxes for $z=99$ using MUSIC \citep{2011MNRAS.415.2101H}. We first evolve the initial conditions down to $z=19$ without any background radiation. After $z=19$, we suppress formation of molecular hydrogen by turning on a uniform Lyman-Werner (LW) background. The spectrum of the LW background is set to be a blackbody of a temperature $T_{\rm bb}=$100,000 K, and is truncated above $\nu=13.6$ eV/$h_p$, where $h_p$ is the Planck constant. The normalization is set by $J_{\rm 21}=100$ where  $J_{\rm 21}$ is the intensity at $\nu=13.6$ eV/$h_p$ in the unit of  $10^{-21}~\rm{erg}~cm^{-2}~s^{-1}~Hz^{-1}~sr^{-1}$.
As reported in \cite{2015MNRAS.448..568H}, this strongly prohibits dense gas in minihalos from forming ${\rm H}_2$ molecules that would allow the gas to radiatively cool and collapse. This represents our target problem, that of a minihalo which has been deactivated in star formation (SF) throughout its history. 

With star-formation suppressed, the sample cubic volume with $200~h^{-1}~\rm{kpc}$ in a side is evolved down to $z = 8$ and the snapshots are saved at $z=10,9,$ \& $8$. These snapshots are used as the initial conditions for the runs, in which the external ionizing background radiation (EIBR) is turned on at those redshifts. Another sample cubic volumes with $100~h^{-1}~\rm{kpc}$, $400~h^{-1}~\rm{kpc}$ and $800~h^{-1}~\rm{kpc}$ in a side are evolved down to $z=10$ in the same way. Here the mass of all the halos is well below $10^8~M_\odot$, which roughly corresponds to the Jeans mass for 10,000 K. Therefore, we regard all the structures in our simulation as the small-scale structures from the preionization phase.

\begin{figure*}  
  \begin{center} 
    \includegraphics[height=0.15\textheight]{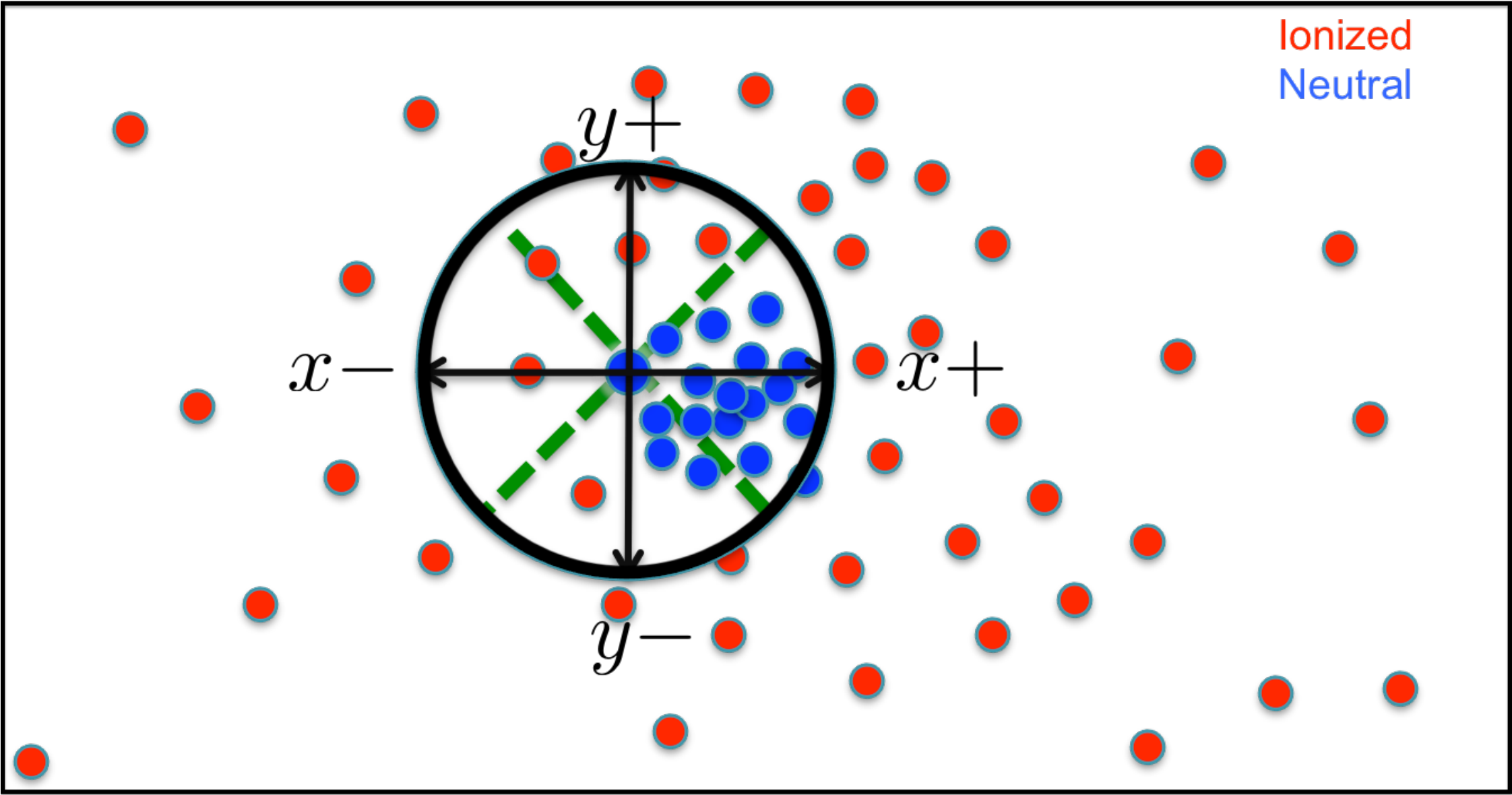}
    \includegraphics[height=0.15\textheight]{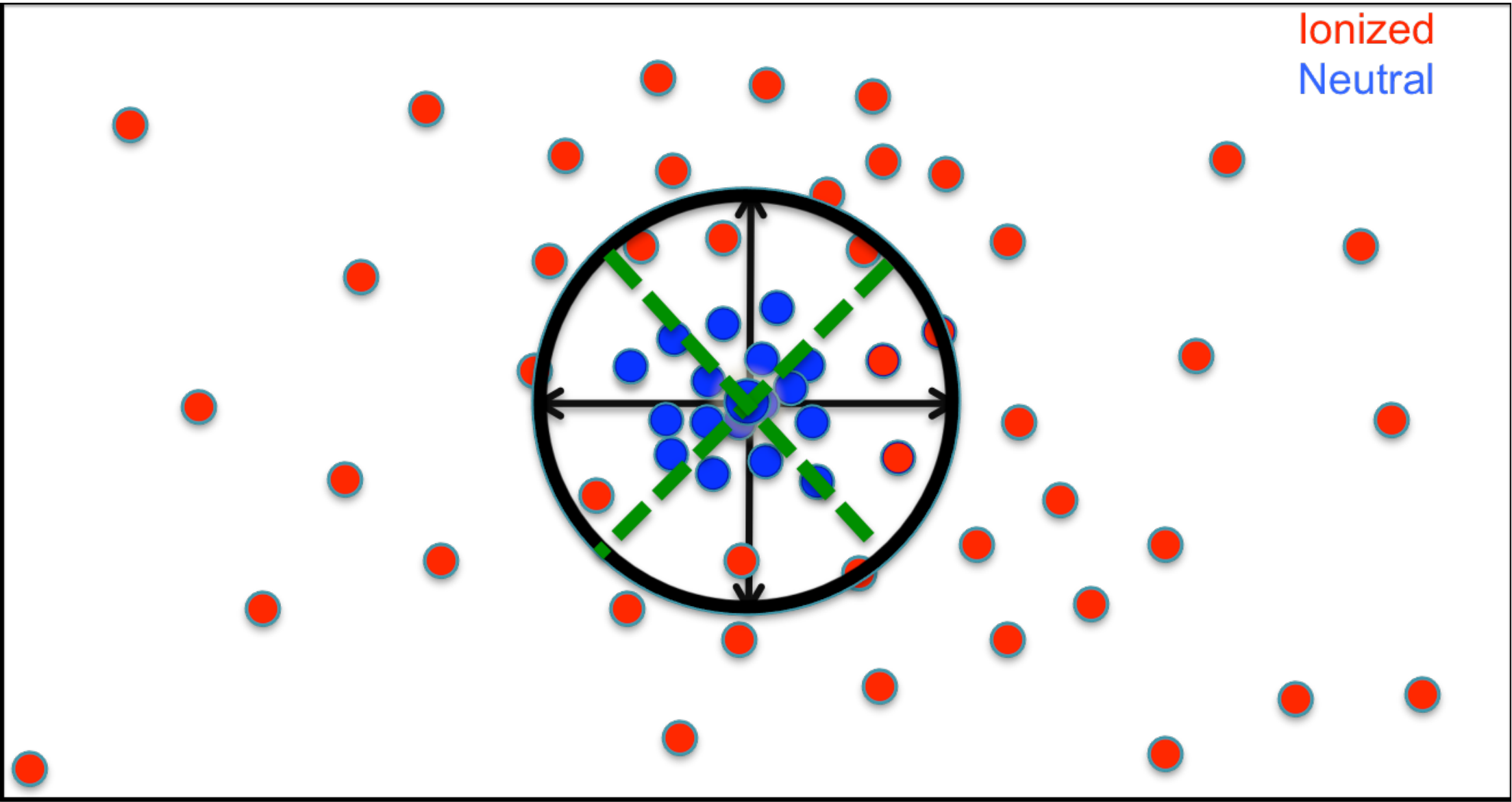}
      \caption{(left) A schematic description for the shielding algorithm used in this work. Blue circles denote neutral SPH particles self-shielded from EIBR whereas red circles denote ionized SPH particles. The blue circle at the center of the black circle represents the target particle that we shall calculate the optical depth to EIBR. In this panel a particle at the outer edge of the clump is chosen as the target. The black circle represents the range within which neighboring particles are allowed to shield target particle. Arrows denote $\pm x,y$ directions on the $xy$-plane that we calculate the optical depth separately. The green dashed lines represent the boundary for each direction. (right) Same as the left panel, but the target particle is located in the center of the clump.}
         \label{fig:shielding}
  \end{center}
\end{figure*}   

\subsection{Algorithm for External Ionizing Background Radiation}

We adopt an uniform and isotropic background for the ionizing radiation. For each particle, the background radiation is shielded by the neighboring particles within a certain distance, $l_s$. Each of the neighboring particles is assigned to the closest one of $\pm x , \pm y~\rm{and}~\pm z$ directions from the target particle to be shielded to calculate the average column densities of neutral hydrogen for these six directions. The column densities are converted to the attenuation fraction for those directions. Figure~\ref{fig:shielding} is a schematic description of how the neighboring particles are assigned to each direction from the target particle. In the left panel, the target particle is located at the left end of the neutral clump and will not be shielded for the radiation coming from the $-x$ direction. On the contrary, the target particle in the right panel will be shielded in all of $\pm x,y$ directions ($\pm z$ directions are omitted in this description) and it will thus remain completely shielded from the radiation until ionization of outer particles eventually expose it to the radiation.

For each neighboring particle shielding the target particle, we add $f_{\rm H I}(m_{\rm gas}/m_p)/(4\pi d_{\rm sh}^2 / 6)$ to H I column density for the direction that the particle is assigned to. Here $m_{\rm gas}$, $m_p$, $f_{\rm H I}$, and $d_{\rm sh}$ are the mass of gas particle, proton mass, the number fraction of hydrogen atom to the number of nucleons and the distance from the shielding particle to the target (shielded) particle, respectively. We assume the neutral fraction of helium follows that of hydrogen and it is only singly ionized when hydrogen is ionized. This is a reasonable assumption for the soft UV spectrum responsible for reionization \citep{2012MNRAS.423..558C}. This algorithm is implemented into the GADGET-3 code to be directly coupled with the gravity, hydrodynamics, and chemistry solvers. We shall call this code GADGET-RT in this work. Our shielding algorithm is similar to the TreeCol algorithm introduced by \cite{2012MNRAS.420..745C}. There they segmented the sky seen by the target particle using the HEALPix algorithm \citep{2005ApJ...622..759G}. We test how accurately this code can keep track of photo-evaporation of a spherical minihalo in Appendix~\ref{sec:Test_Problem}.

\subsection{Simulations} \label{sec:simulation}

We simulate the external ionizing background radiation (EIBR, hereafter) using the snapshot outputs described in Section~\ref{2.1} as the initial conditions. In the left seven columns of Table~\ref{simulation parameters}, we list the name and the parameters of each run. In all the nine simulations, the spectrum of the EIBR is given by the blackbody temperature of $T_{\rm{bb}} $ = 100,000 K with the intensity set by $J_{21}$ = 1, 0.3, or 0.1. Note that this is similar to how we set the LW background in Section~\ref{2.1} except that we do {\it not} truncate the spectrum above $\nu=$13.6 eV/$h_p$. We adopt M\_I0\_z10, which we use $J_{21}=1$ and $z_i = 10$ as the standard run and create other cases by changing one of the parameters to explore the dependency of the results on each parameter. 

For S\_I0\_z10 and L\_I0\_z10, we use $100~h^{-1}~\rm{kpc}$ and $400~h^{-1}~\rm{kpc}$ boxes, respectively, to check the convergence of our results for the box size (See Sec.~\ref{sec:convergence}). Their initial conditions are from different initializations then those used for M\_I0\_z10. For M\_I0\_z8 and M\_I0\_z9, we set $z_i=$ 8 and 9, respectively, to study the dependence of the results on the timing of reionization. We study the dependence of the results on $J_{21}$ by changing it to 0.3 (M\_I-0.5\_z10) and 0.1 (M\_I-1\_z10). We turn off the shielding algorithm for M\_I0\_z10\_NS and disable the dynamics of particles (i.e. freeze particle positions as in post-processed radiative-transfer simulations) for M\_I0\_z10\_ND. 

We also run four simulations (M\_I0\_z10\_VL$\delta$, M\_I0\_z10\_L$\delta$, M\_I0\_z10\_H$\delta$, \& M\_I0\_z10\_VH$\delta$) with their mean densities different from the cosmic mean with the contrast given by $\bar\delta$ = -0.52, -0.26, 0.24, \& 0.59, respectively, in $200~h^{-1}~\rm{kpc}$ boxes and EIBR with $J_{21}$ = 1 and $z_i$ = 10. These simulations share the same box size and EIBR properties with M\_I0\_z10, but differ in the initial conditions. Their initial conditions come from sub-regions of the $800~h^{-1}~\rm{kpc}$ box. We divide the $800~h^{-1}~\rm{kpc}$ box into 64 sub-cubes that are $200~h^{-1}~\rm{kpc}$ in a side and sample four of them to cover a certain range of $\bar\delta$.

\section{Clumping Factor : Definition and How to Calculate} \label{sec:C}

The difference in the ionization rate and recombination rate of hydrogen leads to a change in the number density of ionized hydrogen:
\bea
\frac{dn_{\rm H II}}{dt} = \mathcal{I}-\mathcal{R}.
\eea
The ionization rate can be written as
\bea 
I\equiv n_{\rm H I} \int d\Omega \int d\nu \sigma(\nu) \frac{J_{\gamma}(\hat{\Omega},\nu)}{h\nu},
\eea
where $J_{\gamma}$ is the intensity of the ionizing radiation. 
And, the recombination rate can be written as
\bea \label{eq:rec}
\mathcal{R}\equiv \alpha_{\rm B} (T)n_e n_{\rm H II},
\eea
where $\alpha_{\rm B}=2.6\times 10^{-13} (T/10^4K)^{-0.7} {\rm s}^{-1}{\rm cm}^{3}$ is the case B recombination coefficient, $T$ is the gas temperature, and $n_{\rm X}$ denotes the number density of a species $X$.

For a resolution element like a pixel in numerical simulations, one would usually assume the number density of each species and the temperature is uniform within each resolution element when estimating the recombination rate within the resolution element. In that case, the recombination rate can be expressed in terms of the average values of the physical quantities:
\bea \label{eq:barR}
\bar {\mathcal{R}} =  \alpha_{\rm B} (\bar T) \bar n_e \bar n_{\rm H II}.
\eea
Here $\bar {\mathcal{R}}$,  $\bar n_e$, and $\bar n_{\rm H II}$ are given by the volume weighted average, $\left< \right>_V$. 
And, the average temperature is given by 
\bea \label{eq:T mean}
\bar T = (m_p/k_B) \left< (\gamma -1 )u \right>_M\left<\mu^{-1} \right>_M^{-1}, 
\eea
where $u$ is the specific internal energy, $\mu$ is the mean molecular weight, and $\left< \right>_M$ denotes the mass weighted average.

Equation~(\ref{eq:barR}) however is not accurate when there are unresolved density/temperature fluctuations within the resolution element. So the clumping factor ($C$) is multiplied to the right-hand-side of Equation~(\ref{eq:barR}) to correct for the error. For computational convenience, some works (ETA13, for example) set $n_e = 1.08n_{\rm H II}$ assuming that helium is singly ionized when hydrogen is ionized, and the gas temperature to be constant at $20,000~\rm K$ or similar. Then, the clumping factor is 
\bea 
C_{\rm i}&\equiv&\frac{\left<{n_{\rm H II}}^2\right>_V}{\left<n_{\rm H II}\right>_V^2}.
\eea
And, the recombination rate is 
\bea
\mathcal{R} = C_{\rm i} \alpha_{\rm B} (\bar T) \bar n_e \bar n_{\rm H II}=C_{\rm i} ~\alpha_{\rm B}(T) (1+Y) \bar\chi^2 \bar n_{\rm H}^2,
\eea
where $\chi\equiv n_{\rm H II}/n_{\rm H}$ is the ionized fraction of hydrogen.

In this work, $T$ and $n_e$ are explicitly computed in the simulations. We can therefore define $C_{\rm r}$ in the following way to describe the recombination rate accurately:
\bea \label{eq:C_r_basic}
C_{\rm r}&\equiv&\frac{\left< \alpha_{\rm B} (T)n_e n_{\rm H II}\right>_V}{\left<n_{\rm H II}\right>_V\left<n_{\rm e}\right>_V\alpha_{\rm B} (\bar T)}.
\eea
Here the numerator is the actual recombination rate and the denominator is the hypothetical rate when density, ionization, and temperature are perfectly homogenous without any spatial fluctuation. We refer to the latter as the ``background rate" in this work.

\begin{figure}
  \begin{center}
    \includegraphics[scale=0.52]{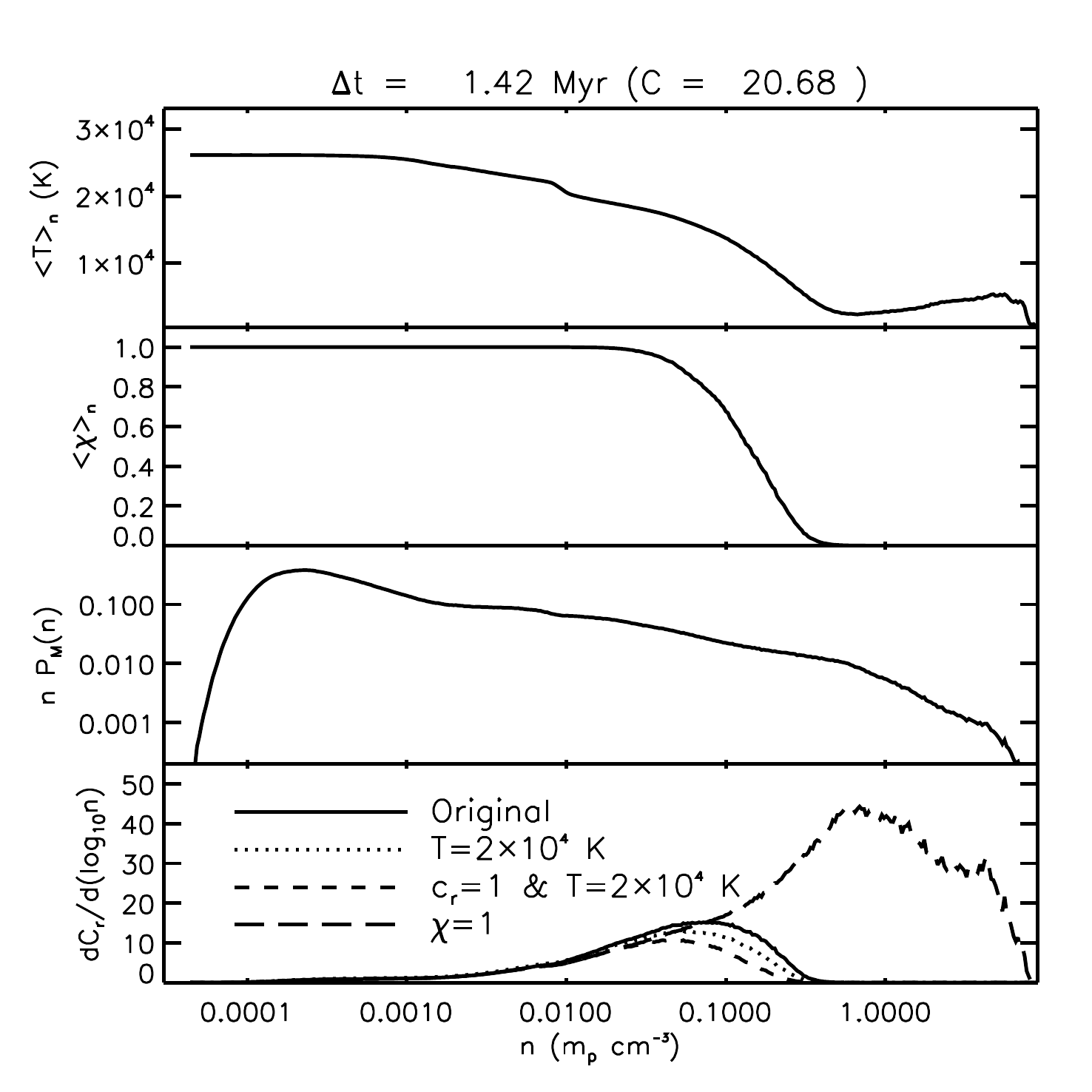}
    \caption{ The mean gas temperature ($\left<T \right>_n$, top panel), mean ionized fraction ($\left<\chi \right>_n$, upper-middle panel), probability density function ($P_M$, lower-casemiddle panel), and clumping factor contribution ($dC_{\rm r}/d\log_{10}n$, bottom panel) at given $n$'s at $\Delta t = 1.42~\rm{Myr}$. In the bottom panel, we display a case that we assume a constant temperature $T=$ 20,000 K (dotted line), a case that we assume $c_r = 1$ as well as $T=$ 20,000 K (dashed line), and a case that we assume complete ionization ($\chi = 1$, long dashed line). The areas under the curves in the bottom panel are proportional to the clumping factor expected for the corresponding cases.}
    \label{fig:X_vs_n}
  \end{center} 
\end{figure} 

To define the SPH smoothed field of a physical quantity $X(\bold{r})$ from particle values of $X$, we adopt a standard method in SPH: 
\bea
X(\bold{r}) = \sum_{i} \frac{X_i}{n_i}W(\bold{r}-\bold{r}_i;h_i),
\eea
where the subscript $i$ denotes the $i$th SPH particle in the simulation,
$n\equiv \rho/m_p$ is the density in the unit of the proton mass $m_p$, 
$\bold{r}$ is the location, 
$W$ is the kernel, and $h_i$ is the adaptive kernel size given by the distance to the 32nd nearest neighbor from the particle. Then, the volume weighted average of this quantity over the simulation volume $V_{\rm sim}$ is given by
\bea
\left< X \right>_V = \frac{1}{V_{\rm sim}}  \sum_{i}\frac{X_i}{n_i} \int_V W(\bold{r}-\bold{r}_i;h_i) d^3r.
\eea
By definition, the volume integral of the kernel in the above should give unity, giving
\bea
\left< X \right>_V = \frac{1}{V_{\rm sim}}  \sum_{i} X_i n_i^{-1}.
\eea
This allows us to calculate $C_{\rm i}$ and $C_{\rm r}$ the following summations.
\bea \label{eq:C_i}
C_{\rm i} &=& \bar{n}^{-1} N_{\rm ptl} \left[ \frac{ \Sigma_{i} n_i \chi_i^2}{(\Sigma_i \chi_i)^2} \right],
\eea
\bea \label{eq:C_r}
C_{\rm r} &=& \bar{n}^{-1} N_{\rm ptl} \left[ \frac{\Sigma_i f_{e,i} f_{\rm{H II},i} \alpha_{\rm B}(T_{i})n_{i}}{ (\Sigma_{i} f_{e,i}) (\Sigma_{i} f_{\rm{H II},i}) \alpha_{\rm B}(\bar T) } \right].
\eea
Here $N_{\rm ptl}$ is the number of SPH particles, $f_X\equiv n_X/n$ is the number density of a species $X$ divided by $n$, and $\bar T$ is given by averaging over the particle values: ${N_{\rm ptl}}^{-1}\Sigma_i T_i$. 

Both $C_{\rm i}$ and $C_{\rm r}$ are calculable from our simulations, but using $C_{\rm r}$ should give the accurate recombination rate. Thus, we by default refer to $C_{\rm r}$ when we mention the clumping factor in the rest of this paper. And, we shall give the value of $C_{\rm i}$ where we look into the the difference between $C_{\rm r}$ and $C_{\rm i}$ such as in Section~\ref{sec:understanding_clumping_factor}.

We also express Equations~(\ref{eq:C_i}) and (\ref{eq:C_r}) as the integrals over $n$ to describe the clumping factor contribution from gas with a certain density. This is done by using a combination of several relevant physical quantities ($\chi$, $f_{e}$ \& $T$) averaged at a given $n$ and the mass-weighted probability density function (PDF) for the SPH densities of SPH particles\footnote{The volumed weighted gas density PDF of \citet{2000ApJ...530....1M} is related to our mass-weighted density PDF by $\Delta P_V(\Delta) d\Delta = P_M(n) dn$, where $\Delta=n/\bar{n}$ is the normalized density. }, $P_M(n^\prime)=dN_{{\rm ptl}}(n<n^\prime)/dn^\prime$:
\bea \label{eq:C_i_int}
C_{\rm i} &=& \frac{\int dn^\prime P_M(n^\prime) \left<\chi \right>_{n=n^\prime}^2   c_{\rm i}(n^\prime) n^\prime }{\bar{n} \left<\chi\right>_M^2}
\eea
\bea \label{eq:C integral}
C_{\rm r} &=& \frac{\int dn^\prime P_M(n^\prime)\left<\chi \right>_{n=n^\prime} \left<f_e \right>_{n=n^\prime} \alpha_{\rm B}(\left<T \right>_{n=n^\prime}) c_{\rm r}(n^\prime) n^\prime}{\bar{n} \left<\chi\right>_M \left<f_e \right>_M \alpha_{\rm B}(\bar{T})}.~~~~~~
\eea
Here the mass weighted average of a quantity X is written as $\left<X\right>_M$ and given by $N_{\rm ptl}^{-1}\sum_i X_i$. Similarly, the average of a quantity at a given density is $\left<X \right>_{n=n^\prime}$. We calculate it by placing all the SPH particles onto 400 logarithmically uniform bins between the maximum and minimum densities, and taking the average within the bins for the quantity of interest.
In addition, we define the following quantities at density $n$:
\bea 
c_{\rm i}(n^\prime) &\equiv& \frac{\left< \chi^2 \right>_{n=n^\prime}}{ \left<\chi \right>_{n=n^\prime}^2}  \\
c_{\rm r}(n^\prime) &\equiv& \frac{\left<\chi f_e \alpha_{\rm B}(T)\right>_{n=n^\prime}}{ \left<\chi \right>_{n=n^\prime} \left<f_e \right>_{n=n^\prime} \alpha_{\rm B}(\left<T \right>_{n=n^\prime}) } 
\eea
These factors arise due to variations in the physical quantities at a given density. They should be included in the integral expressions (Eqs.~\ref{eq:C_i_int} \& \ref{eq:C integral})  to precisely recover $C_{\rm i}$ and $C_{\rm r}$ calculated from Equations~(\ref{eq:C_i}) and (\ref{eq:C_r}), respectively. We shall describe how the clumping factor depends of each physical quantity in more detail in Section~\ref{sec:understanding_clumping_factor}.

\section{Results} \label{sec:results}
%The main results of this paper are presented in this section. 

\begin{figure} 
  \begin{center}
    \includegraphics[scale=0.5]{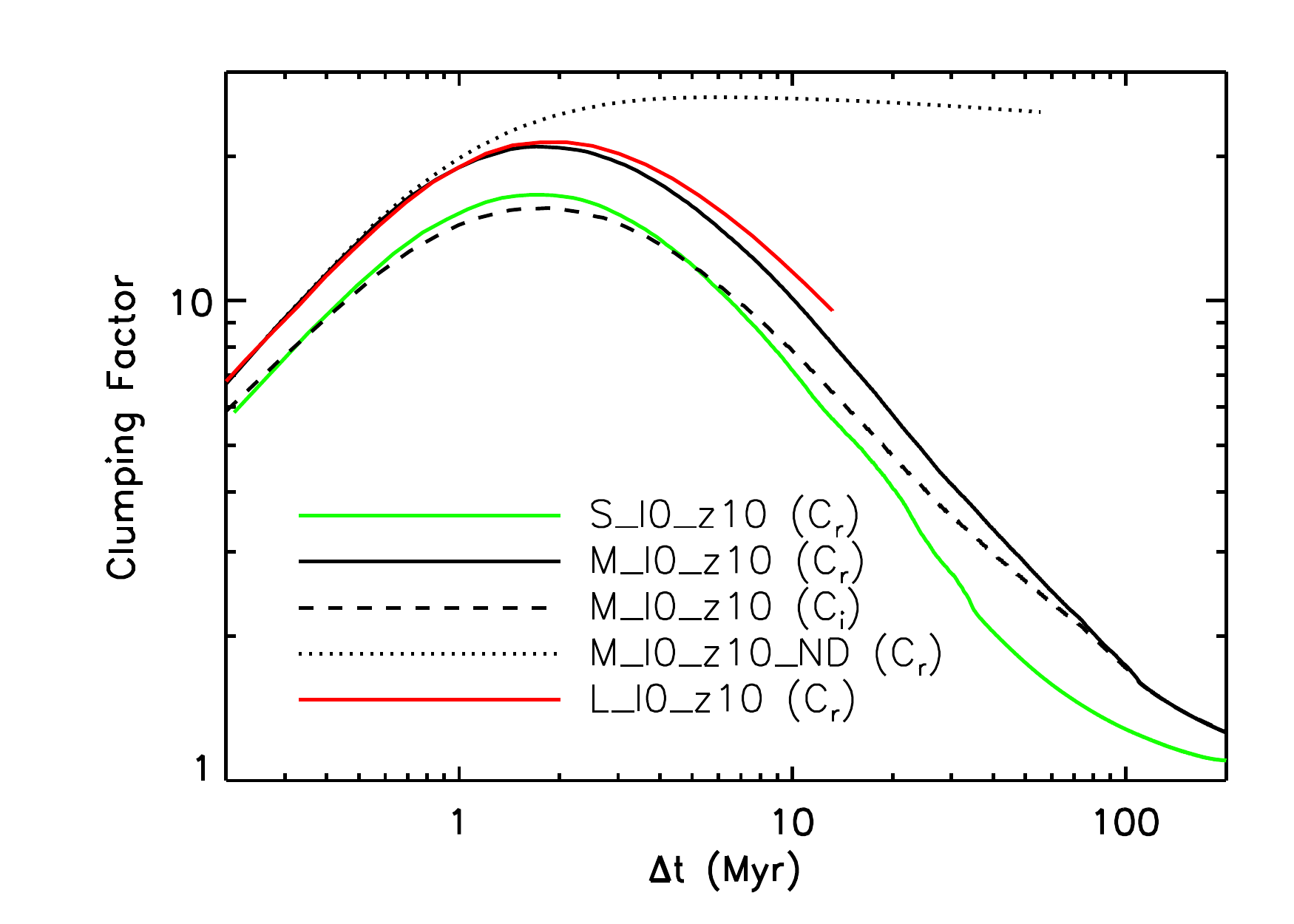}
  \caption{The clumping factor plotted as a function of time for S\_I0\_z10 (green solid), M\_I0\_z10 (black solid), L\_I0\_z10 (red solid), M\_I0\_z10\_NS (black dotted), \& M\_I0\_z10 (black dashed).}
     \label{fig:C_vs_t}
  \end{center}
\end{figure} 

\begin{figure*}
  \begin{center}  
    \includegraphics[scale=0.5]{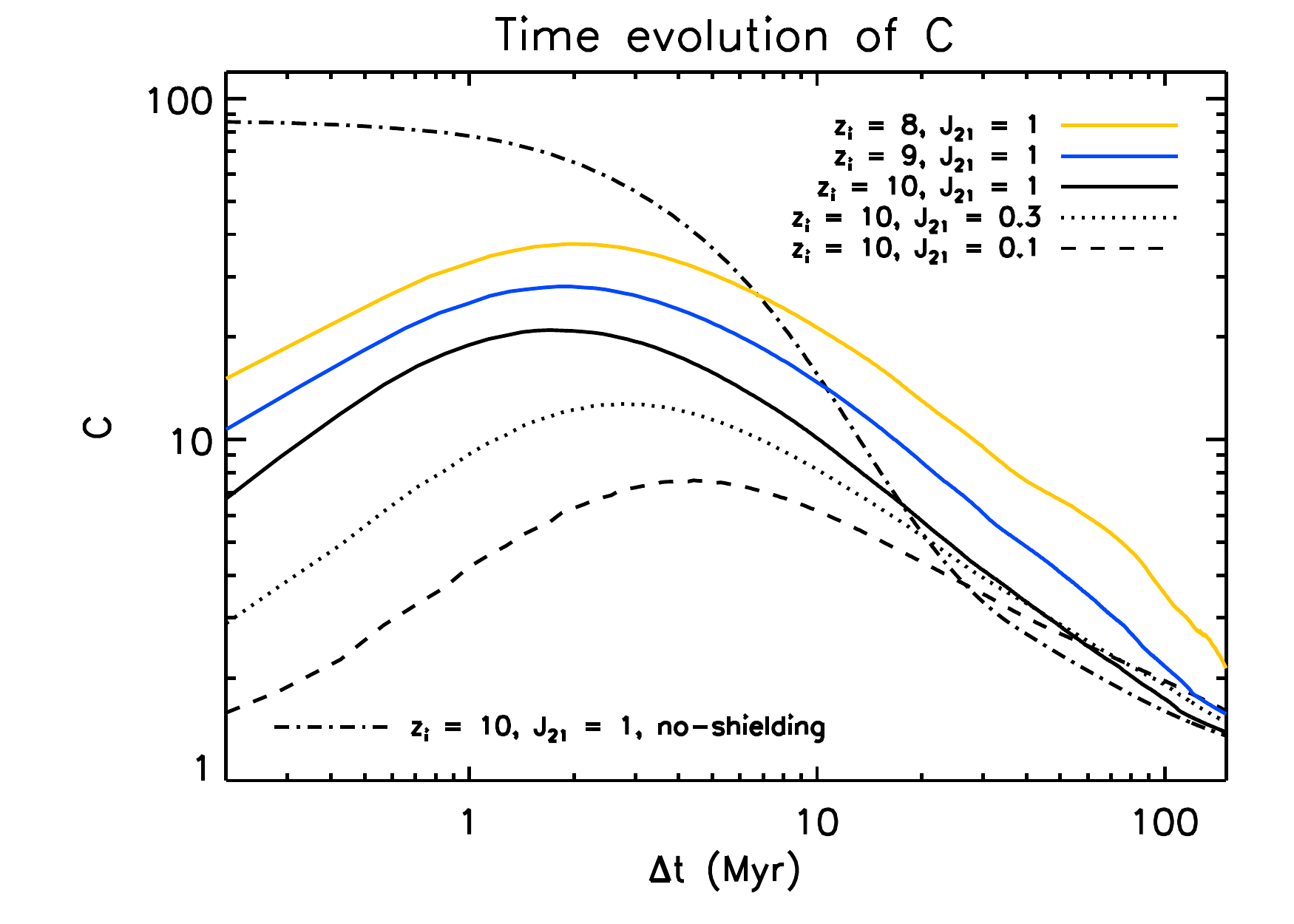}
    \includegraphics[scale=0.5]{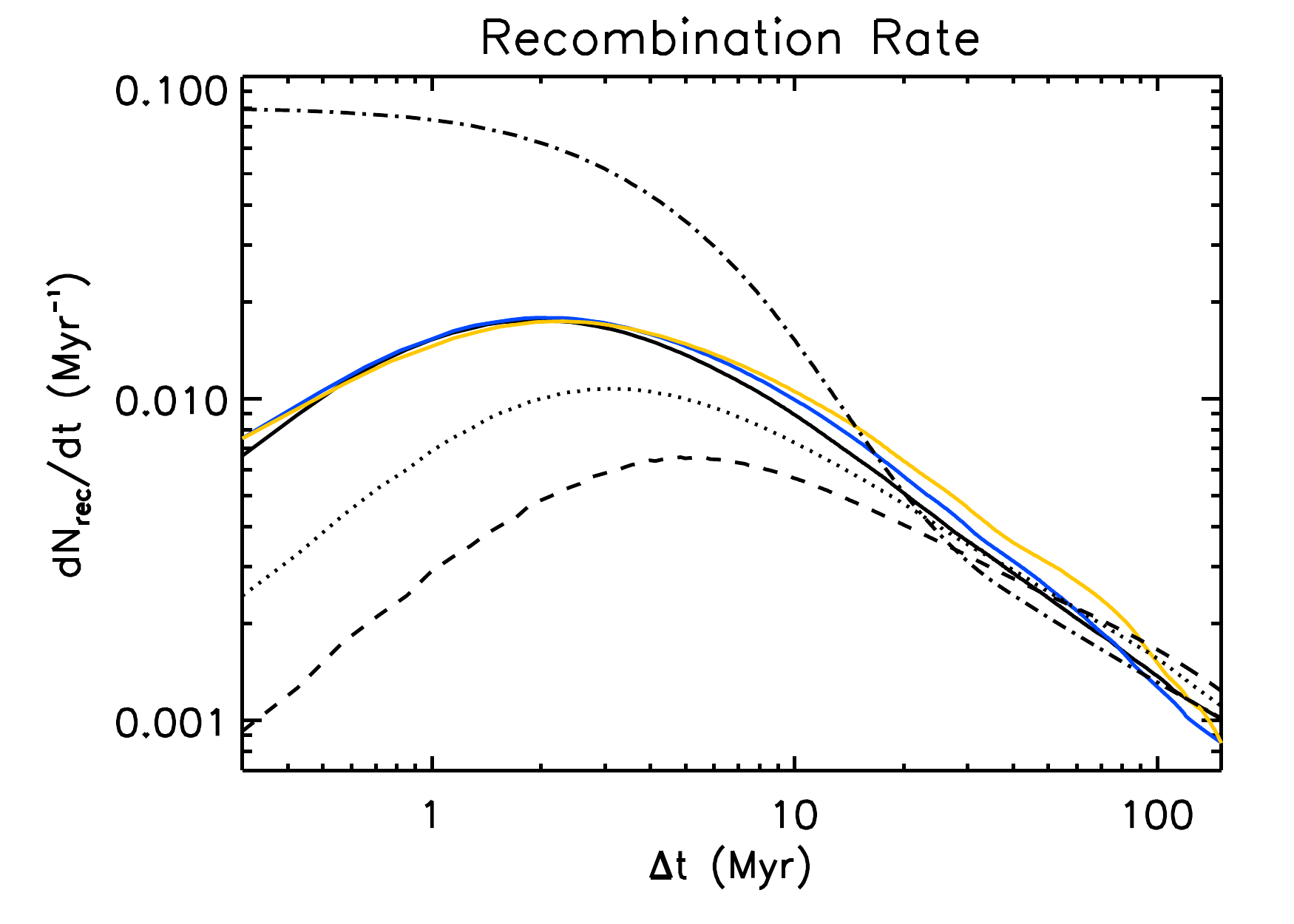}
        \caption{(left) The clumping factor as a function of time for M\_I0\_z10 (black solid), M\_I-0.5\_z10 (black dotted), M\_I-1\_z10 (black dashed), M\_I0\_z9 (blue solid), M\_I0\_z8 (yellow solid), and M\_I0\_z10\_NS (black dot-dashed). (right) The recombination rate as a function of time for the same runs with same line type as in the left panel.}
        \label{fig:C_vs_t_compare_z}
      \end{center} 
\end{figure*} 

\begin{figure*}
  \begin{center}
    \includegraphics[scale=0.55]{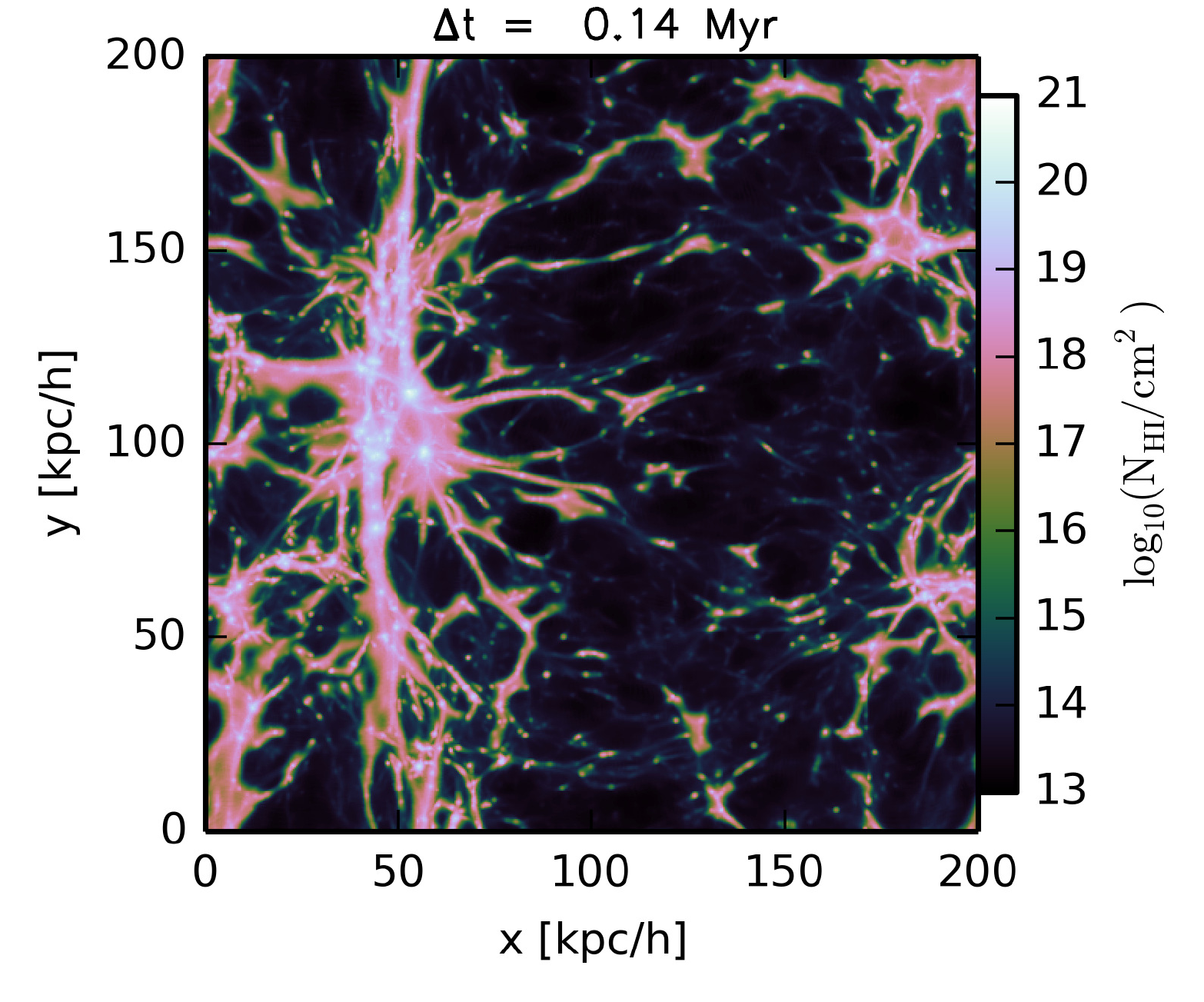}
    \includegraphics[scale=0.55]{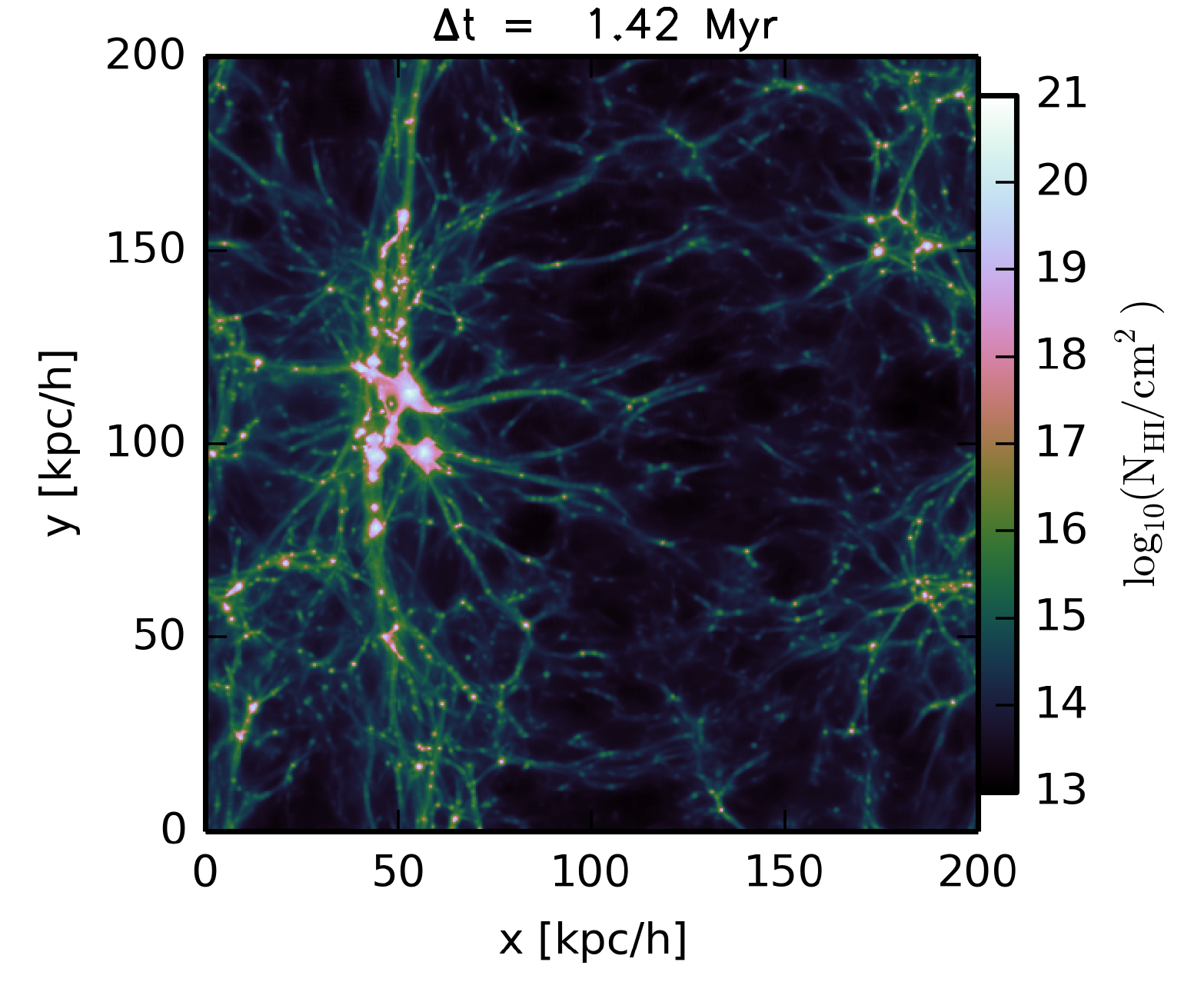}
    \includegraphics[scale=0.55]{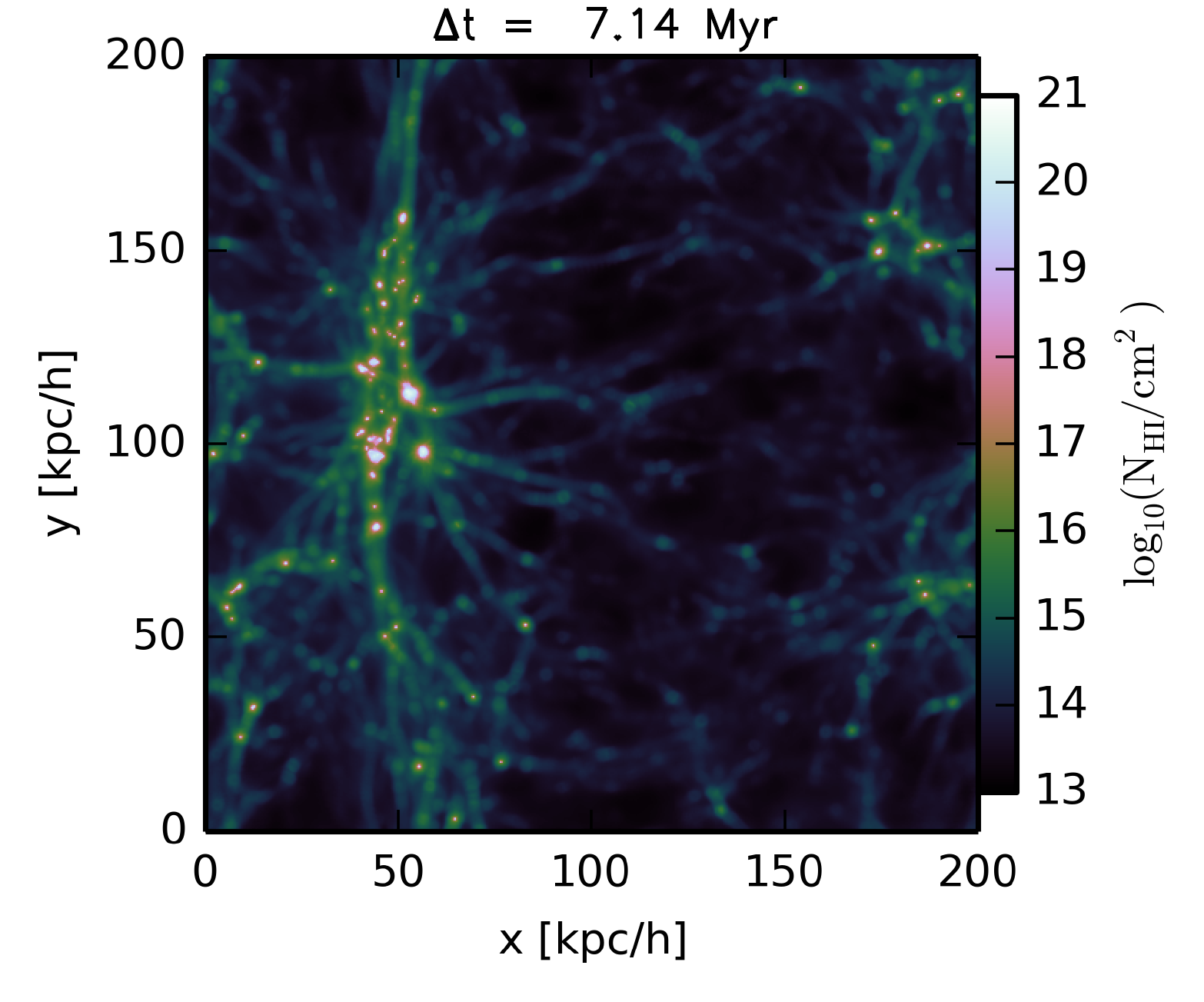}
    \includegraphics[scale=0.55]{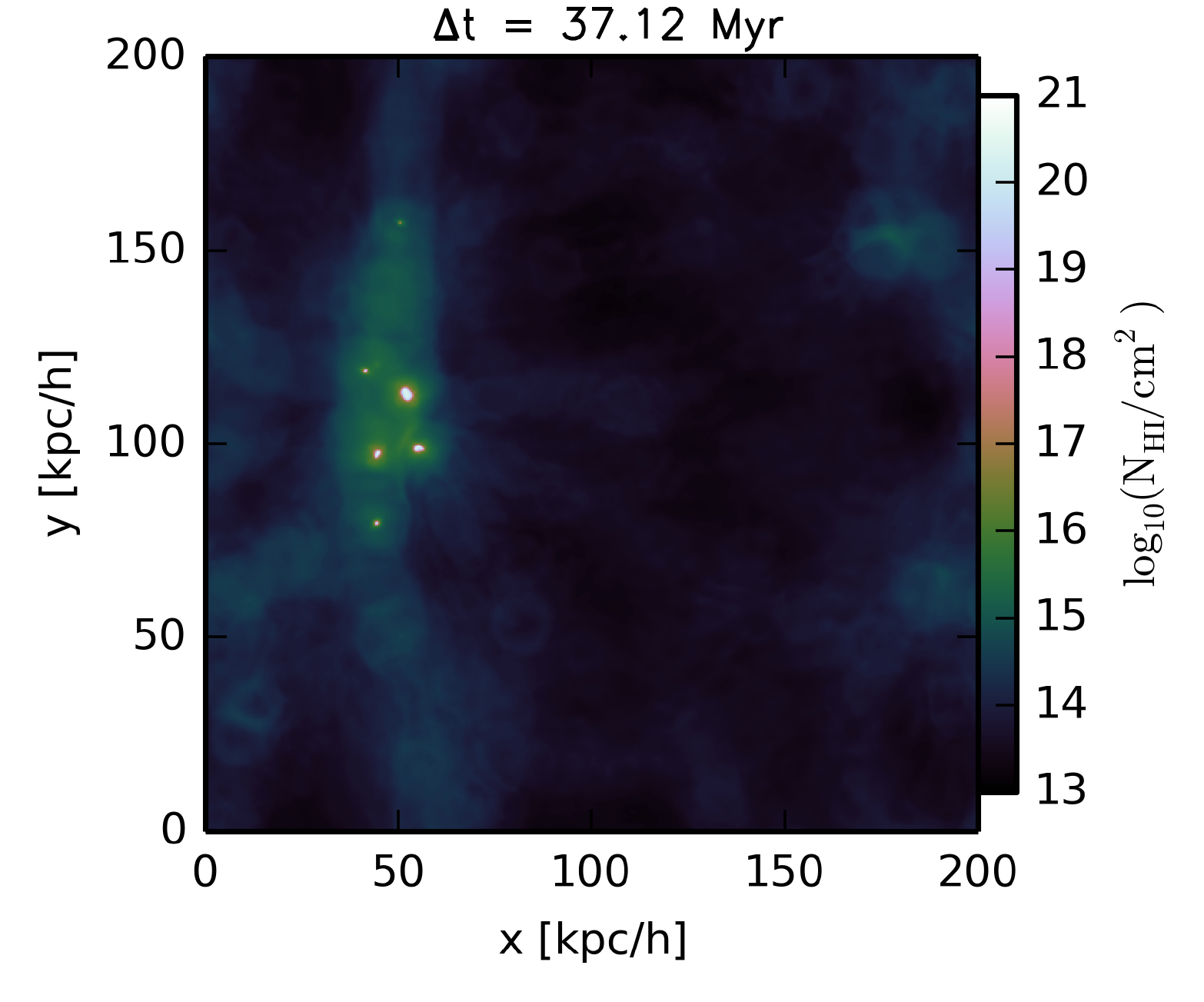}
    \caption{ The projected HI column density of M\_I0\_z10 at $\Delta t=$ 0.14 Myr (top left), 1.4 Myr (top right),  7.1 Myr (bottom left), and 37 Myr (bottom right). White and pink colors display the neutral regions, and green and black colors display the ionized regions. }
    \label{fig:ColumnDen}
  \end{center} 
\end{figure*} 

\subsection{Physical Quantities Relevant to the Clumping Factor} \label{sec:understanding_clumping_factor}

In Equations~(\ref{eq:C_i_int}) and (\ref{eq:C integral}), we express the clumping factor in terms of $\left<\chi \right>_n$, $P_M(n)$, $\left<f_e\right>_n$, and $\left<T \right>_n$. In this section, we demonstrate that the $n$-dependence of $dC_{\rm r}/d\log_{10}{n}$ (the integrand of the integral in Eq.~\ref{eq:C integral}) is practically dictated by $\left<\chi \right>_n$ and $P_M(n)$ going as $n^2 \left<\chi \right>_n^2P_M(n)$ to justify focusing only on $\left<\chi \right>_n$ and $P_M(n)$ to understand the time evolution of the clumping factor in the rest of this paper. 

We plot $\left<\chi \right>_n$, $P_M(n)$, $\left<T \right>_n$, and $dC_{\rm r}/d\log_{10}{n}$ in Figure~\ref{fig:X_vs_n} for M\_I0\_z10 at $\Delta t = 1.42$ Myr. Throughout our analysis, $f_e$ is found to be close to $0.82\chi$, which is consistent with helium being singly-ionized when hydrogen is ionized.
Due to shielding against the EIBR, a break appears in $\left<\chi \right>_n$ at a certain $n$, above which $\left<\chi \right>_n$ falls to zero. At $n\lesssim 0.3~\rm{cm}^{-3}$ where the gas is at least partially ionized ($\left<\chi \right>_n>0$), the gas temperature tends to anti-correlate with the density. $P_M$ is nearly unchanged from the initial conditions at the turn-on of EIBR. $P_M$ has a gaussian-like distribution around the cosmic mean ($n=3.3\times 10^{-4}~\rm{cm}^{-3}$) with an extended power-law-like tail between $n\approx 10^{-3}$ and $5\times10^{-1}~\rm{cm}^{-3}$ with the power-law index of $-1.5$. Above $n\sim 5\times10^{-1}~\rm{cm}^{-3}$, $P_M$ falls faster and eventually cuts off. This behavior of $P_M$ is consistent with what was reported in \citet{2000ApJ...530....1M}.

To assess the impact of each term, we create the following three hypothetical cases and plot $dC_{\rm r}/d\log_{10}{n}$ for those cases in the bottom panel of Figure~\ref{fig:X_vs_n}.
\begin{itemize}
\item[Case] 1: Assume a constant temperature $T=$ 20,000 K, in which  $C_{\rm r}$ becomes $C_{\rm i}$.
\item[Case] 2: On top of assuming $T=$ 20,000 K, set $c_r=1$ in Equation~(\ref{eq:C integral}). $dC_{\rm r}/d\log_{10}{n}$ goes precisely as $n^2 \left<\chi\right>_n^2P_M(n)$ in this case.
\item[Case] 3: Assume complete ionization of all the IGM by setting $\chi=1$.
\end{itemize}

Case 1 and Case 2 mostly reproduce the shape of $dC_{\rm r}/d\log_{10}{n}$ with a moderate underestimation at $n\gtrsim 10^{-2}~\rm{cm}^{-3}$. $dC_{\rm r}/d\log_{10}{n}$, which is roughly proportional to $n^{0.5}$ up to $n\approx 3\times 10^{-2}~\rm{cm}^{-3}$, where $\left<\chi\right>_n$ is almost unity and the $n^2 P_M$ term determines its behavior. At above $n\approx 3\times 10^{-2}~\rm{cm}^{-3}$, the break in $\left<\chi\right>_n$ suppresses $dC_{\rm r}/d\log_{10}{n}$. 

Around the break, $dC_{\rm r}/d\log_{10}{n}$ in Case 1 and Case 2 is lower then that in the original case due the effect of the gas temperature.
The actual gas temperature falls toward the high-$n$ direction intersecting 20,000 K at $n\approx  10^{-2}~\rm{cm}^{-3}$. The recombination coefficient goes as $T^{-0.7}$ and thus decreases with increasing density. So, $\alpha_B(T)$ is underestimated at $n\gtrsim 10^{-2}~\rm{cm}^{-3}$ and overestimated at $n\lesssim 10^{-2}~\rm{cm}^{-3}$ when assuming $T=$ 20,000 K. Nevertheless, only the underestimation stands out because $dC_{\rm r}/d\log_{10}{n}$ practically vanishes at $n\lesssim 10^{-2}~\rm{cm}^{-3}$ due to its $n^{0.5}$ scaling. Setting $c_r=1$ causes yet another underestimation at $n\gtrsim 3\times10^{-2}~\rm{cm}^{-3}$ in Case 2. In that density range, the gas is partially ionized with scattered values of $\chi$, $f_e$ and $\alpha_B(T)$ at a given density with $\chi$ and $f_e$ highly correlated, resulting in $c_r>1$.

%Although setting $c_r=1$ or $c_i=1$ does not lead to a large error, they are needed to be correctly calculated in order to precisely match the results of Equations~(\ref{eq:C_i}) \& (\ref{eq:C_r}) with those of Equations~(\ref{eq:C_i_int}) \& (\ref{eq:C integral}), respectively.

The break in $\left<\chi \right>_n$ is an important consequence of the shielding algorithm. Case 3 shows how drastically the clumping factor would be overestimated without the break. $dC_{\rm r}/d\log_{10}{n}$ keeps rising as $n^{0.5}$ up to $n\sim 1~\rm{cm}^{-3}$ in that case. The resulting clumping factor is about 70, which is much higher than the clumping factor 20 in M\_I0\_z10.

Case 1 corresponds to using $C_{\rm i}$ for the clumping factor as in most previous literature that did not keep track of the gas temperature. Despite the fact that the impact of the gas temperature on the clumping factor is relatively minor compared to those of $\left< \chi \right>_n$ and $P_M$, it still matters consider precise estimates of the recombination rate. The time evolutions of $C_{\rm i}$ and $C_{\rm r}$ are compared in Figure~\ref{fig:C_vs_t}. The difference between $C_{\rm i}$ and $C_{\rm r}$ peaks at $\Delta t = 1.42$ Myr, where $C_{\rm i}=16$ and $C_{\rm r}=21$. Later ($\Delta t\gtrsim $ 10 Myr), the difference between the two diminishes as they both asymptote to one.

\subsection{Time Evolution of the Clumping Factor: Dual Phase Evolution} \label{sec:Time Evolution of C}

\begin{figure*}
  \begin{center}
    \includegraphics[scale=0.58]{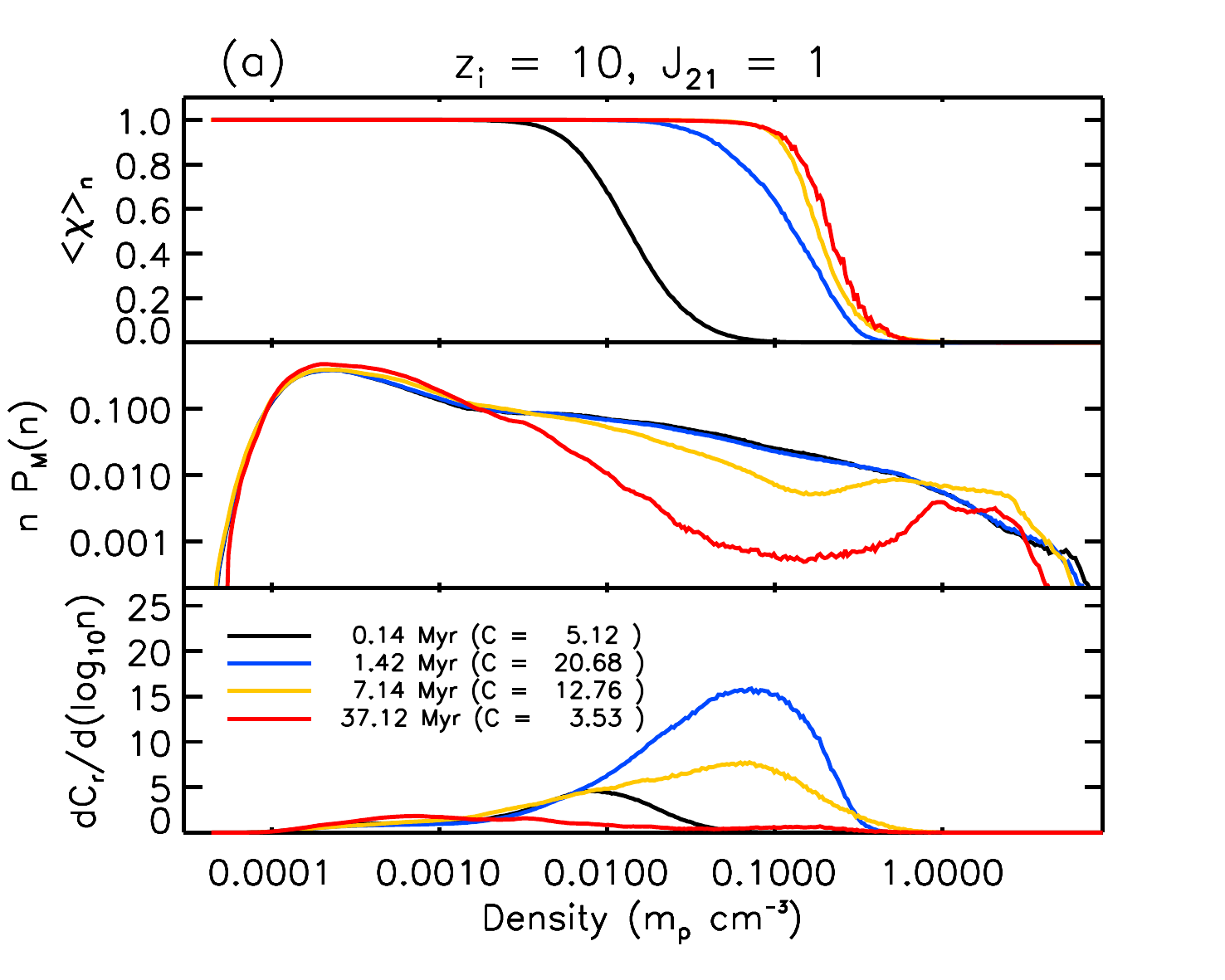}
    \includegraphics[scale=0.58]{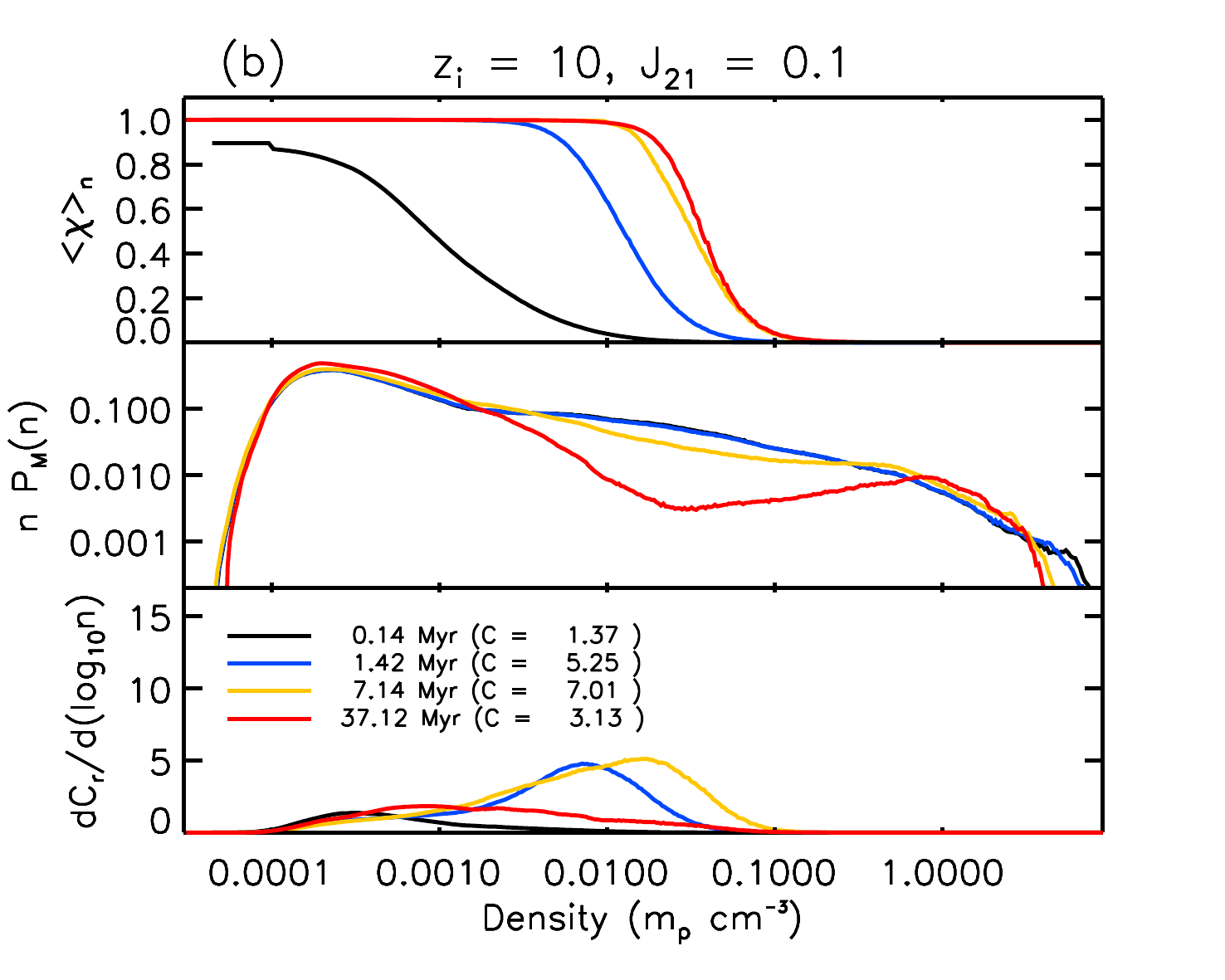}
    \includegraphics[scale=0.58]{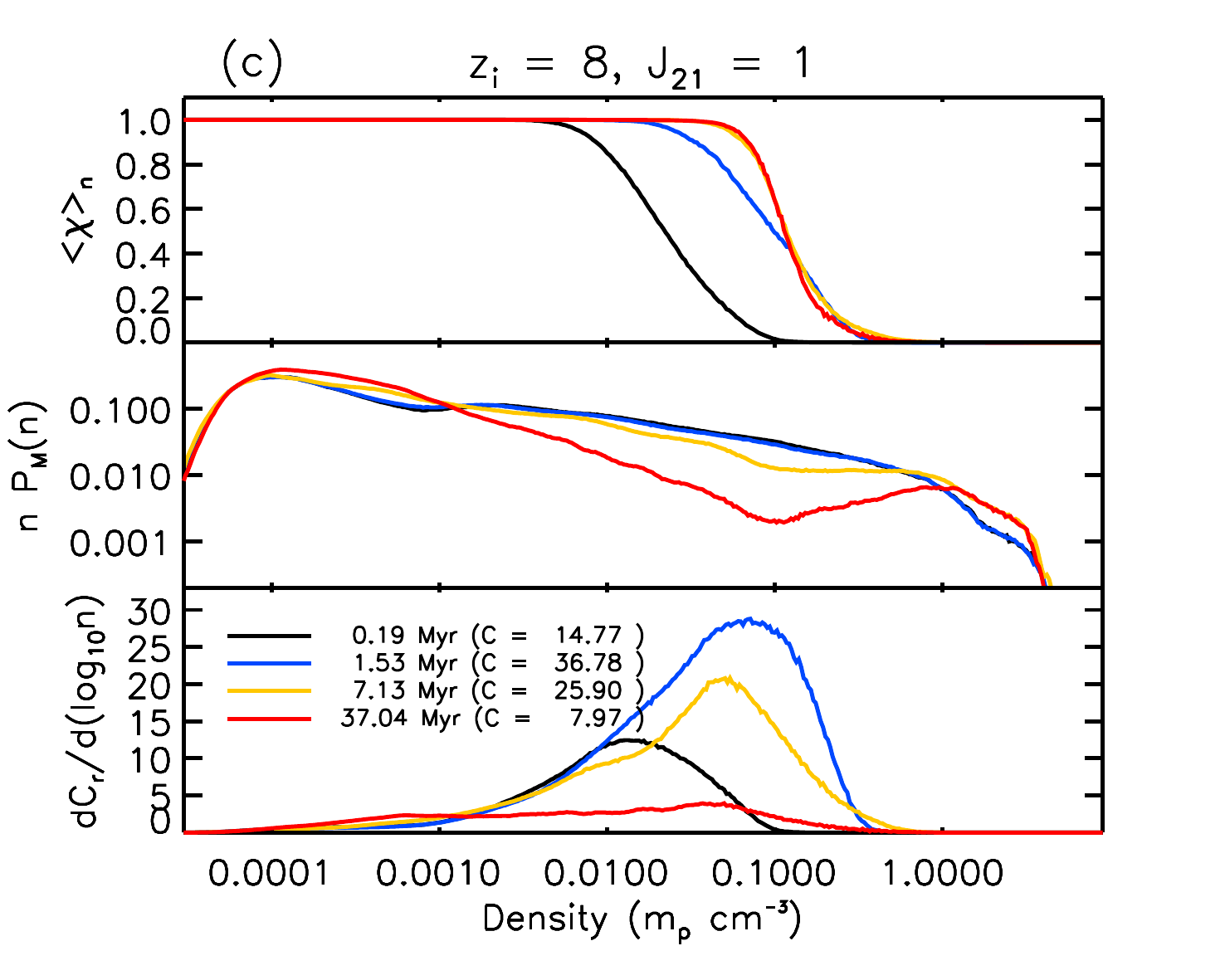}
    \includegraphics[scale=0.58]{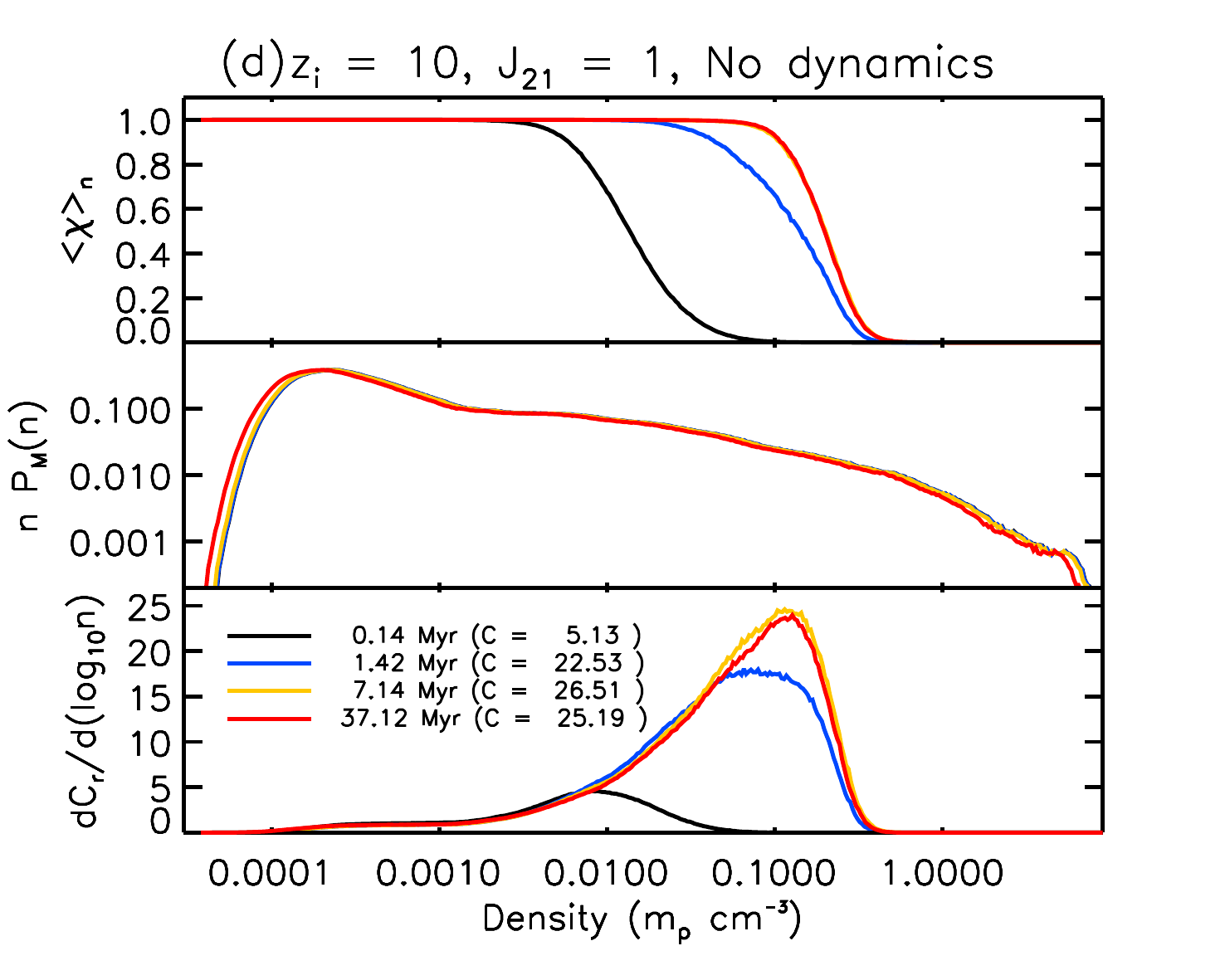}
    \includegraphics[scale=0.58]{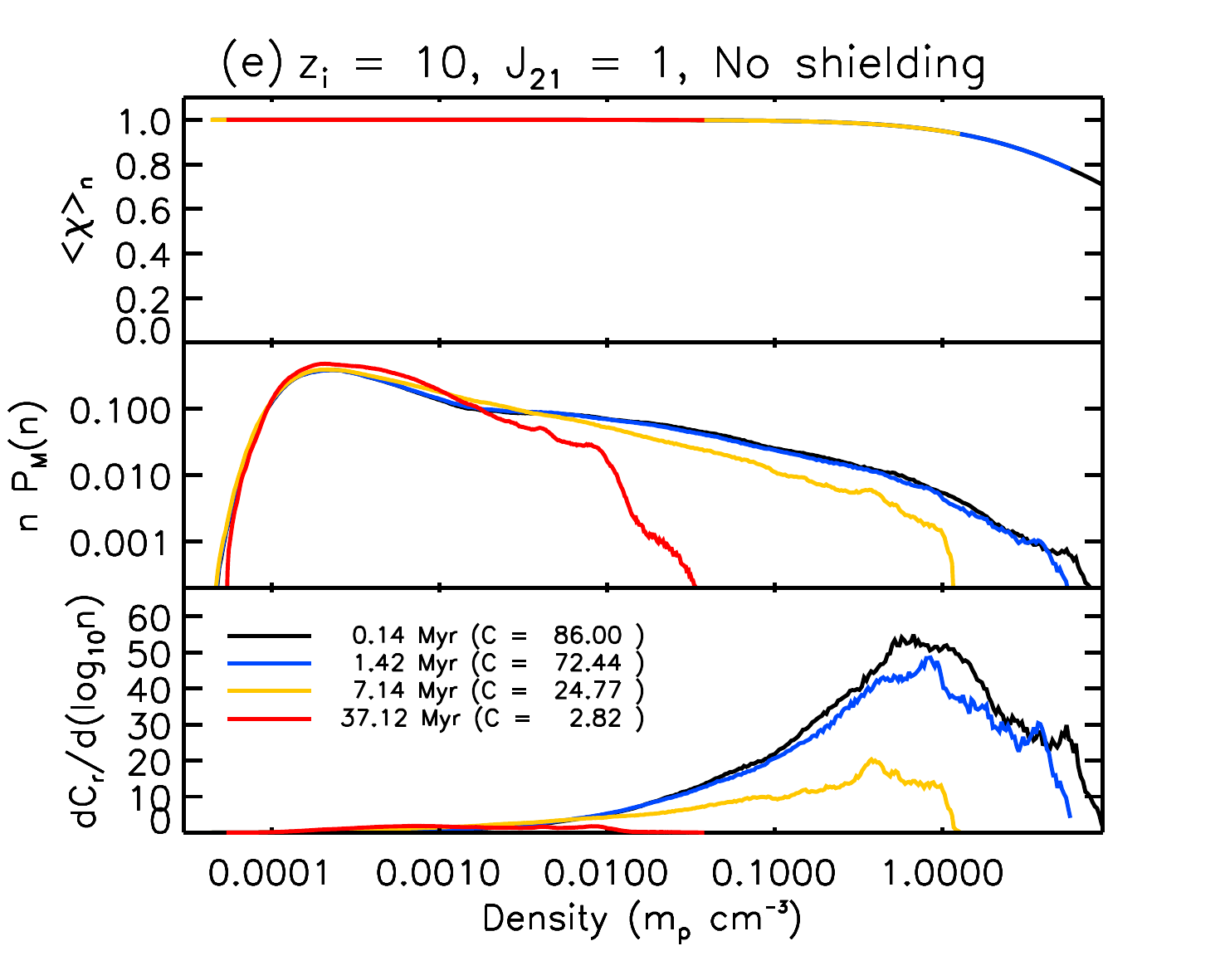}
    \caption{ The mean ionized fraction ($\left<\chi \right>_n$), probability density function of gas particle density ($P_M$), and clumping factor contribution ($dC_{\rm r}/d\log_{10}n$) at given densities are plotted in the top, middle, and bottom panels, respectively. The results are shown for M\_I0\_z10 (panel $a$), M\_I-1\_z10 (panel $b$), M\_I0\_z8 (panel $c$), M\_I0\_z10\_ND (panel $d$), and M\_I0\_z10\_NS (panel $e$). Except for M\_I0\_z10\_z8, the black, blue, yellow and red lines describe the results at $\Delta t=$ 0.14 Myr, 1.4 Myr,  7.1 Myr, and 37 Myr, respectively. For M\_I0\_z10\_z8, the same colors describe $\Delta t=$ 0.19 Myr, 1.5 Myr,  7.1 Myr, and 37 Myr, respectively.}
    \label{fig:X_vs_n2}
  \end{center} 
\end{figure*} 

The clumping factor is shown as a function of time for each model in Figure~\ref{fig:C_vs_t} and in the left panel of Figure~\ref{fig:C_vs_t_compare_z}. Except for the runs without shielding (M\_I0\_z10\_NS) or dynamics (M\_I0\_z10\_ND), we find that the clumping factor starts rising in the beginning, turns over at $\Delta t = $ 1 - 3 Myr, and falls afterwards eventually converging to one at $\Delta t \gtrsim 100$ Myr. We explain this behavior with two phases of ionization fronts (I-fronts) as explained in the following.
\begin{itemize}
\item[1.] {\bf R-type}: I-fronts propagate super-sonically through the low density IGM. They sweep gaseous structures without giving enough time for them to react to ionization.
\item[2.] {\bf D-type}: As I-fronts reach dense regions, they become sub-sonic and can no longer proceed before the hydrodynamic feedback begins to move the gas. The gas expands substantially due to increased pressure from photo-ionization.
\end{itemize}

\begin{figure}
  \begin{center}
    \includegraphics[scale=0.45]{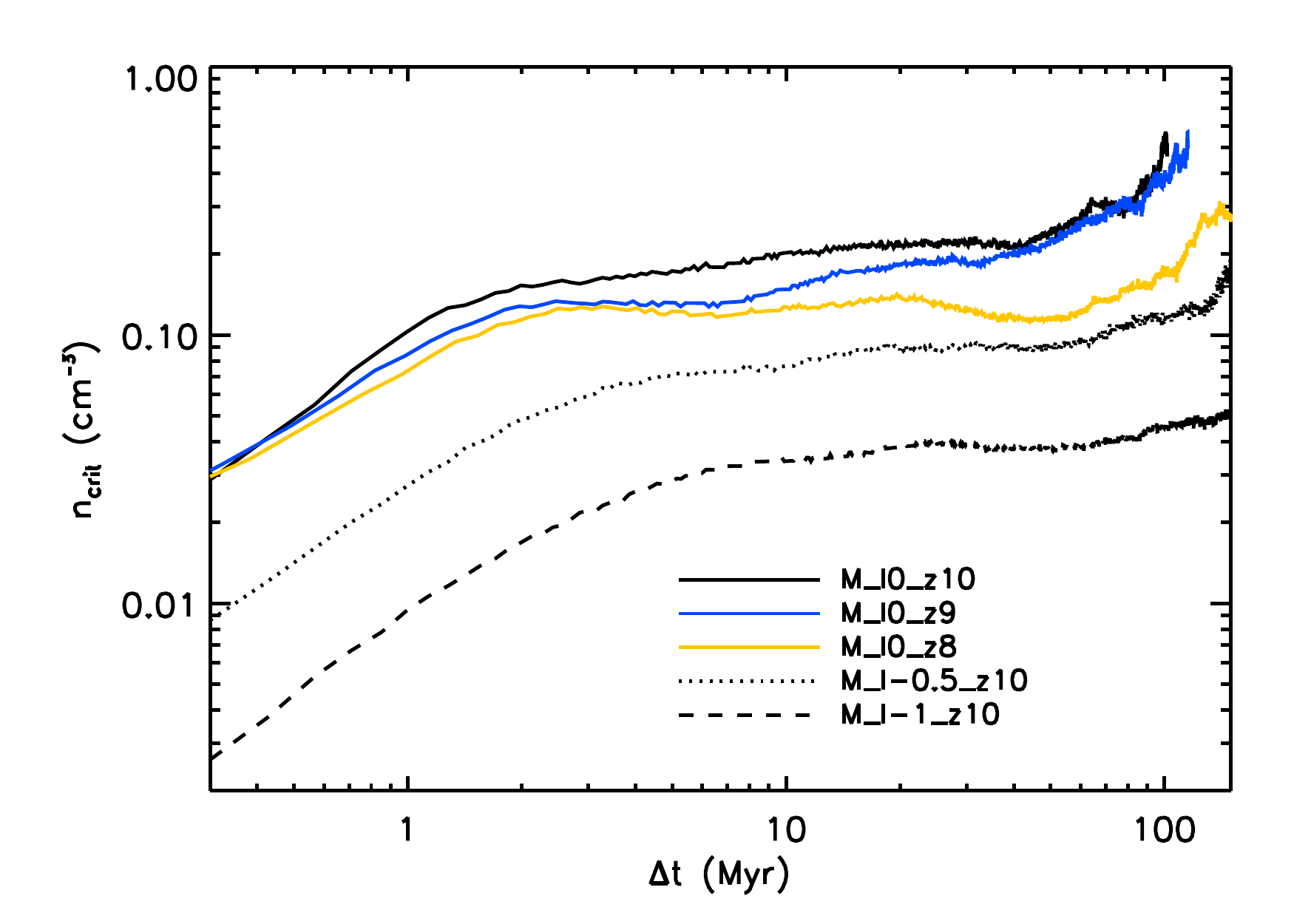}
   \caption{The critical density of ionization ($n_{\rm crit}$) as a function of time from the turn-on of EIBR ($\Delta t$). We show the result for M\_I0\_z10 (black solid), M\_I0\_z9 (blue solid), M\_I0\_z8 (yellow solid), M\_I-0.5\_z10 (black dotted), and M\_I-1\_z10 (black dashed). Above a certain density, all the gas is ionized and $n_{\rm crit}$ cannot be defined because $\left< \chi \right>_n >0.5$ for all $n$'s. That is where the curves end for M\_I0\_z10 and M\_I0\_z9. We also advice the reader to be cautious about the fact that $n_{\rm crit}$ is not statistically reliable near where it ends because there is very few neutral particles left there. }
    \label{fig:n_crit}
  \end{center} 
\end{figure} 

To explain the two phases, we shall look into the standard run (M\_I0\_z10).
A schematic description of the R-type phase is seen in Figure~\ref{fig:ColumnDen} for the snapshots at $\Delta t=$ 0.14 Myr and 1.42 Myr. During this time the neutral regions shrink while the density field remains nearly unchanged. The ionization profile and density PDF in Figure~\ref{fig:X_vs_n2}$a$ give more quantitative descriptions. The major change in the physical quantities during this phase is the shift of the break in $\left<\chi\right>_n$ from $n=0.03~{\rm cm^{-3}}$ to $0.2~{\rm cm^{-3}}$ between $\Delta t=$ 0.14 Myr and 1.42 Myr while the density PDF remains almost the same. To describe the location of the break, we define $n_{\rm crit}$ as the density that $\left< \chi \right>_{n=n_{\rm crit}} = 0.5$, which can be seen in Figure~\ref{fig:n_crit} plotted as a function of time. As $n_{\rm crit}$ rises, $dC_{\rm r}/d\log_{10}{n}$ picks up a contribution from the gas whose density satisfies $n < n_{\rm crit}$. Due to $\sqrt{n}$ scaling of $n^2 P_M$, this rise of $n_{\rm crit}$ adds progressively more to the clumping factor, which explains the rapid rise of the clumping factor during the R-type phase.

After $\Delta t \approx 2~\rm{Myr}$, I-fronts in the simulation transition to D-type. The column density maps (Fig.~\ref{fig:ColumnDen}) show disruption of ionized structures. Filamentary structures diminish as they expand and dilute with the background. $n_{\rm crit}$ in Figure~\ref{fig:n_crit} settles down at around $0.2~{\rm cm^{-3}}$ and no longer evolves substantially\footnote{$n_{\rm crit}$ is shown to rise slightly after $\Delta t \sim 40~\rm{Myr}$ in the figure. But, this is not a statistically meaningful feature as it is from very little gas left in a single evaporating minihalo.}. Neutral clumps that survived during the R-type phase slowly evaporate from their surfaces. At $\Delta t=$ 37 Myr, most of the structures are gone except for a few clumps that located in the most massive minihalos in the volume. In the density PDF (middle panel of Fig.~\ref{fig:X_vs_n2}$a$), this hydrodynamic feedback appears as a suppression of the PDF of the gas with $n<n_{\rm crit}$. Because high-density ionized gas gives the main contribution to $dC_{\rm r}/d\log_{10}{n}$ (bottom panel of Fig.~\ref{fig:X_vs_n2}$a$), the clumping factor decays during the R-type phase. 

The density and ionization fraction histories of individual particles support this dual phase picture as well. We sample ten SPH particles with difference initial densities and show how their densities and ionized fractions evolve over time in Figure~\ref{fig:pdt_den}. Shortly after the turn-on of EIBR, SPH particles above the asymptotic value of $n_{\rm crit}$ ($\sim 0.2~\rm cm^{-3}$) are ionized by R-type I-fronts and drop in their densities down close to the cosmic mean ($n_{\rm mean}= 3.3\times 10^{-4}~ \rm cm^{-3}$). The particles with $n> 0.2~\rm cm^{-3}$ are shielded in dense clumps and are not ionized immediately. But, they eventually get exposed to the radiation at the I-fronts, slowly process towards the center of the clumps and go through similar drops in their densities.

%$n_{\rm mean}= 3.3\times 10^{-4}~ \rm cm^{-3}$

\subsubsection{The effects of Shielding and Hydrodynamics}

The dual phase evolution described above is a consequence of implementing the self-shielding of dense neutral gas while simultaneously considering full hydrodynamic effects. To highlight the difference between the effect of self-shielding and that of hydrodynamics, we run one no-shielding run (M\_I0\_z10\_NS) and one no-dynamics run (M\_I0\_z10\_ND). For the no-shielding run, we simply turn off shielding and let all the SPH particles be exposed to the EIBR. In the no-dynamics run, we force particles to stay in their initial locations to mimic post-process radiative transfer.

Ionization in the no-shielding run happens everywhere from the beginning. I-fronts therefore do not exist in this run. The H I column density map for $\Delta t = 1.4$ Myr (left panel of Fig.~\ref{fig:ColumnDen_NS}) lacks most of the spurious high-column density regions with $N_{\rm H I}> 10^{19}~{\rm cm^2}$ present in M\_I0\_z10 (top right panel of Fig.~\ref{fig:ColumnDen}). The subsequent expansion of gas looks similar, but the no-shielding run lacks self-shielded cores as can be seen for $\Delta t = 37$ Myr in the right panel of Figure~\ref{fig:ColumnDen_NS}. The break in the ionization profile cannot exist in this case because the gas is nearly fully ionized at all densities. $dC_{\rm r}/d\log_{10}{n}$ at $\Delta t = 0.14~\&~1.4$ Myr (bottom panel of Fig.~\ref{fig:X_vs_n2}$e$) picks up a huge contribution from gas whose $n$ is greater than $n_{\rm crit}$ of the standard run. This is similar to the $\chi=1$ case of the standard run (bottom panel of Fig.~\ref{fig:X_vs_n}) discussed is Section~\ref{sec:understanding_clumping_factor}.
At $\Delta t = 37$ Myr, gas with $n\gtrsim 0.03~\rm{cm}^{-3}$ no longer exist in the no-shielding run while the standard run retains some amount of self-shielded gas in that range. The clumping factor in the no-shielding run therefore starts much higher ($\sim 100$; See Fig.~\ref{fig:C_vs_t_compare_z}) than in the standard run. Then, it declines rapidly even down to lower than in the standard run after $\Delta t \sim 20$ Myr. That is  because the no-shielding run lacks self-shielded clumps, of which a small amount of dense ionized gas from evaporation contribute slightly to the clumping factor.

The no-dynamics run on the other hand reproduces the R-type phase precisely, but not the subsequent D-type phase. Up to $\Delta t \sim $ 1.4 Myr, the clumping factor (See Fig.~\ref{fig:C_vs_t}) and $\left<\chi\right>_n$ (See Fig.~\ref{fig:ColumnDen_ND}$d$) evolve similarly to the standard case but, $P_M$ remains unchanged for all time. The expansion of the gas that is the main process in the D-type phase is completely suppressed in this no-dynamics run. Soon, I-fronts get to the point that dense ionized gas on the surfaces of neutral clumps completely absorbs EIBR and they can not proceed any more. For this reason, H I column density (Fig.~\ref{fig:ColumnDen_ND}$d$) show little evolution from $\Delta t = 1.4$ Myr to $37$ Myr and the clumping factor asymptotes to a value after $\Delta t \sim 1$ Myr.

The results in this section demonstrate the importance of shielding in reproducing R-type I-fronts in early times, and that of the dynamics in reproducing D-type I-fronts that come after. Neglecting the former hugely overestimates the clumping factor in the early times by not excluding the self-shielded high-density gas in the calculation. And, neglecting the latter would not reproduce the hydrodynamic feedback effect that strongly suppresses the clumping factor for ionized gas.

\subsection{Dependence of the Clumping Factor of Properties of Ionizing Radiation} \label{sec:Dependence of the Clumping Factor of Properties of Ionizing Radiation}

On large scales, there would be sub-Mpc volumes that are ionized at different times ($z_i$) by EIBR with different intensities ($J_{21}$) than in the standard run due to the variance in their environments. In order to cover all such cases, we create multiple runs, in which we change one of $J_{21}$ and $z_i$ from the parameter choice of the standard run ($z_i=10;J_{21}=1$). We have two runs, M\_I-0.5\_z10 and M\_I-1\_z10, with the EIBR intensities $J_{21}=0.3$ and 0.1, respectively, and another two runs, M\_I0\_z9 and M\_I0\_z8, that ionizes at $z_i=9$ and 8, respectively. The resulting clumping factors are shown in the left panel of Figure~\ref{fig:C_vs_t_compare_z}. While both the R-type and D-type phases appear as in the standard run, there are notable differences in some details. 

When $J_{21}$ is lower, the clumping factor starts lower and turns over later. This is because I-fronts with a lower intensity propagate more slowly and transition into D-type at a lower density. The column density map of M\_I-1\_z10 at $\Delta t = 1.42~\rm{Myr}$ in the left panel of Figure~\ref{fig:ColumnDen_I-1} shows that the high column density ($N_{\rm H I}\gtrsim 10^{19}~\rm{cm}^{-2}$) regions are more extended than in M\_I0\_z10 (upper right panel of Fig.~\ref{fig:ColumnDen}), indicating that the progress of ionization is slower in M\_I-1\_z10 than in M\_I0\_z10. At a later time $\Delta t = 37~\rm{Myr}$ (right panel of Fig.~\ref{fig:ColumnDen_I-1}), a much larger number of neutral clumps are still observed than in the standard run.
$n_{\rm crit}$ in M\_I-1\_z10 (Fig.~\ref{fig:n_crit}) asymptotes to $\sim 0.04~\rm{cm}^{-3}$ that is 5 times lower than it does in the standard run. 
Also, it takes $\sim 10~$ Myr to asymptote, taking about 7 times longer than in the standard run (See also the evolution of the ionization profile in the top panel of Fig.~\ref{fig:X_vs_n2}$b$). 
With $dC_{\rm r}/d\log_{10}{n}$ suppressed from the lower density $(\sim0.04~\rm{cm}^{-3})$, the resulting clumping factor is also lower. The slower evaporation leads to a larger amount of high-density self-shielded gas remaining at late time as can be seen in the density PDF at $\Delta t = 37~\rm{Myr}$ (middle panel of Fig.~\ref{fig:X_vs_n2}$b$). This delay in evaporation causes the clumping factor to decay more slowly and eventually result in M\_I-1\_z10 having a slightly higher clumping factor at around $\Delta t = 50~\rm{Myr}$ than in the standard run.

For lower $z_i$'s, the reaction to the EIBR is similar to the standard case. The evolution of H I density in M\_I0\_z8 (Fig.~\ref{fig:ColumnDen_z8}), for example, is quite similar to M\_I0\_z10 (Fig.~\ref{fig:ColumnDen}). The time dependences of the clumping factor are similar, too, but the overall magnitudes are higher for the lower $z_i$ cases (See Fig.~\ref{fig:C_vs_t_compare_z}). The peak clumping factor ($C^{\rm peak}_{\rm r}$) listed in Table~\ref{simulation parameters} can be taken as the reference for the relative magnitude of each case.

We find $C^{\rm peak}_{\rm r}$ scales nearly as $(1+z_i)^{-3}$, which is the inverse of the cosmic mean density. Noting that the gas is nearly fully ionized in the simulation roughly satisfying $\bar{n}_{\rm H II}\bar{n}_{e}\propto \bar{n}^2$, where $\bar{n}$ is the average density of the simulation box divided by $m_p$, we have the following relation for recombination rate per hydrogen ($dN_{\rm rec}/dt$).
\bea \label{eq:dNdt}
\frac{dN_{\rm rec}}{dt}\equiv \frac{\left<\mathcal{R}\right>_V}{f_{\rm H}\bar{n}}=C_{\rm r}\frac{\alpha_B(\bar{T})\bar{n}_{\rm H II}\bar{n}_{e}}{f_{\rm H}\bar{n}}
\propto C_{\rm r} \bar{n}
\eea
Since we are considering cosmic mean density volume here, $C^{\rm peak}_{\rm r}$ and $\bar{n}$ cancel out, resulting in $dN_{\rm rec}/dt$ remaining constant for changing $z_i$.

Plotting $dN_{\rm rec}/dt$ directly (right panel of Figure~\ref{fig:C_vs_t_compare_z}), we find that $dN_{\rm rec}/dt$ starts almost the same up to $\Delta t \sim 3$ Myr for the cases with different $z_i$'s. But, $dN_{\rm rec}/dt$ falls more slowly for the lower $z_i$ cases later on. The density PDF of M\_I0\_z10 (middle panel of Fig.~\ref{fig:X_vs_n2}$a$) and M\_I0\_z8 (middle panel of Fig.~\ref{fig:X_vs_n2}$c$) at $\Delta t = 37$ Myr shows that M\_I0\_z8 has more gas remaining at $n>10^{-1}~\rm{cm}^{-3}$ shielded from the EIBR. This is due to M\_I0\_z8 starting with more collapsed structures due to structure growth from $z=10$ to 8. Those structures can contribute to the clumping factor from their evaporation in the late time. $dC_{\rm r}/d\log_{10}{n}$ in M\_I0\_z8 shows a significant contribution from $n>10^{-1}~\rm{cm}^{-3}$ while there is almost none in M\_I0\_z10 indicating that the collapsed structures are indeed responsible for higher recombination rate in lower $z_i$ cases.

\begin{figure}
  \begin{center}
    \includegraphics[scale=0.45]{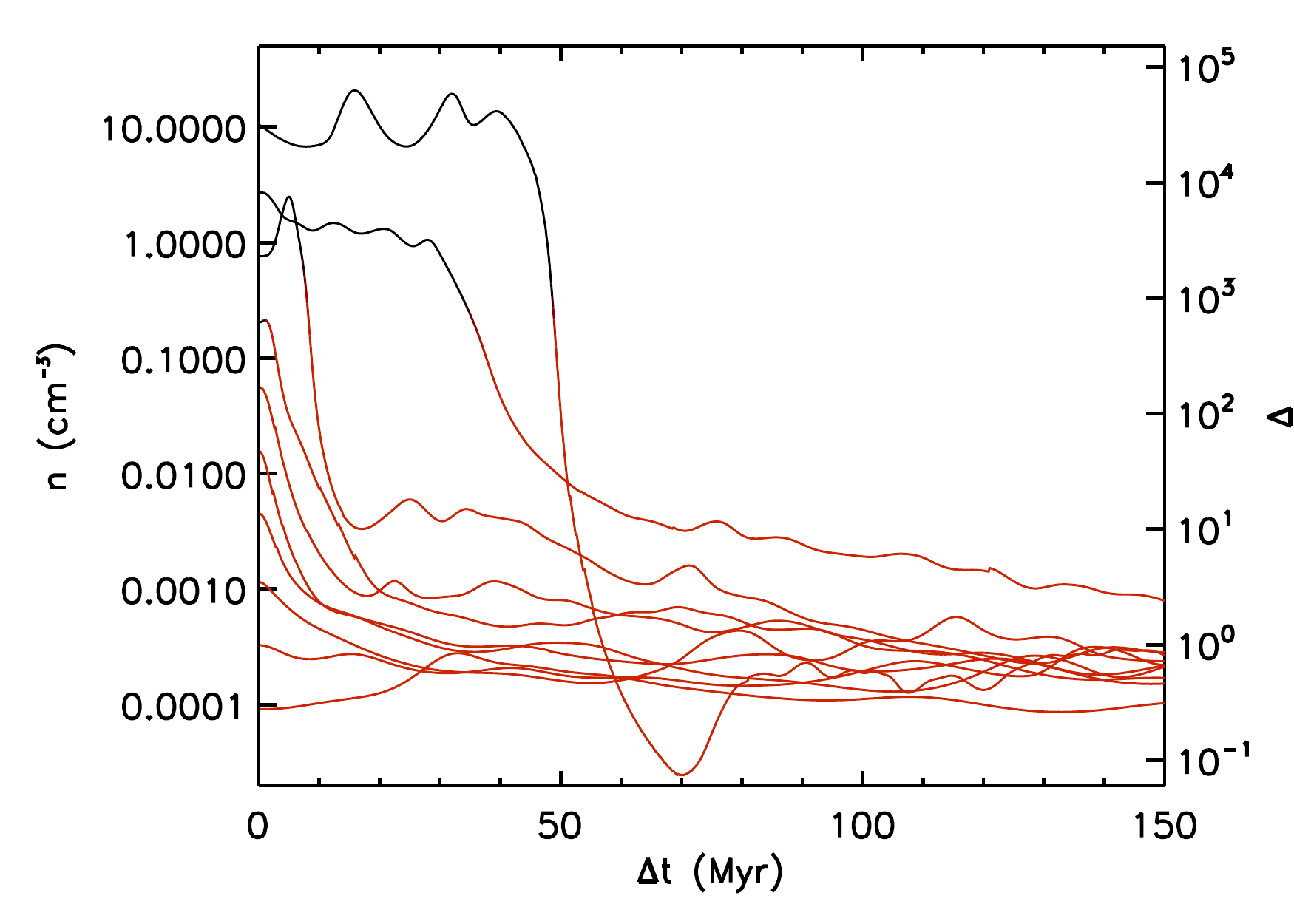}
          \caption{The densities of ten SPH particles in M\_I0\_z10 as functions of time. Red/Black color denotes that the particle is ionized/neutral.} 
          \label{fig:pdt_den}
        \end{center} 
\end{figure} 

\begin{figure*}
  \begin{center}
    \includegraphics[scale=0.55]{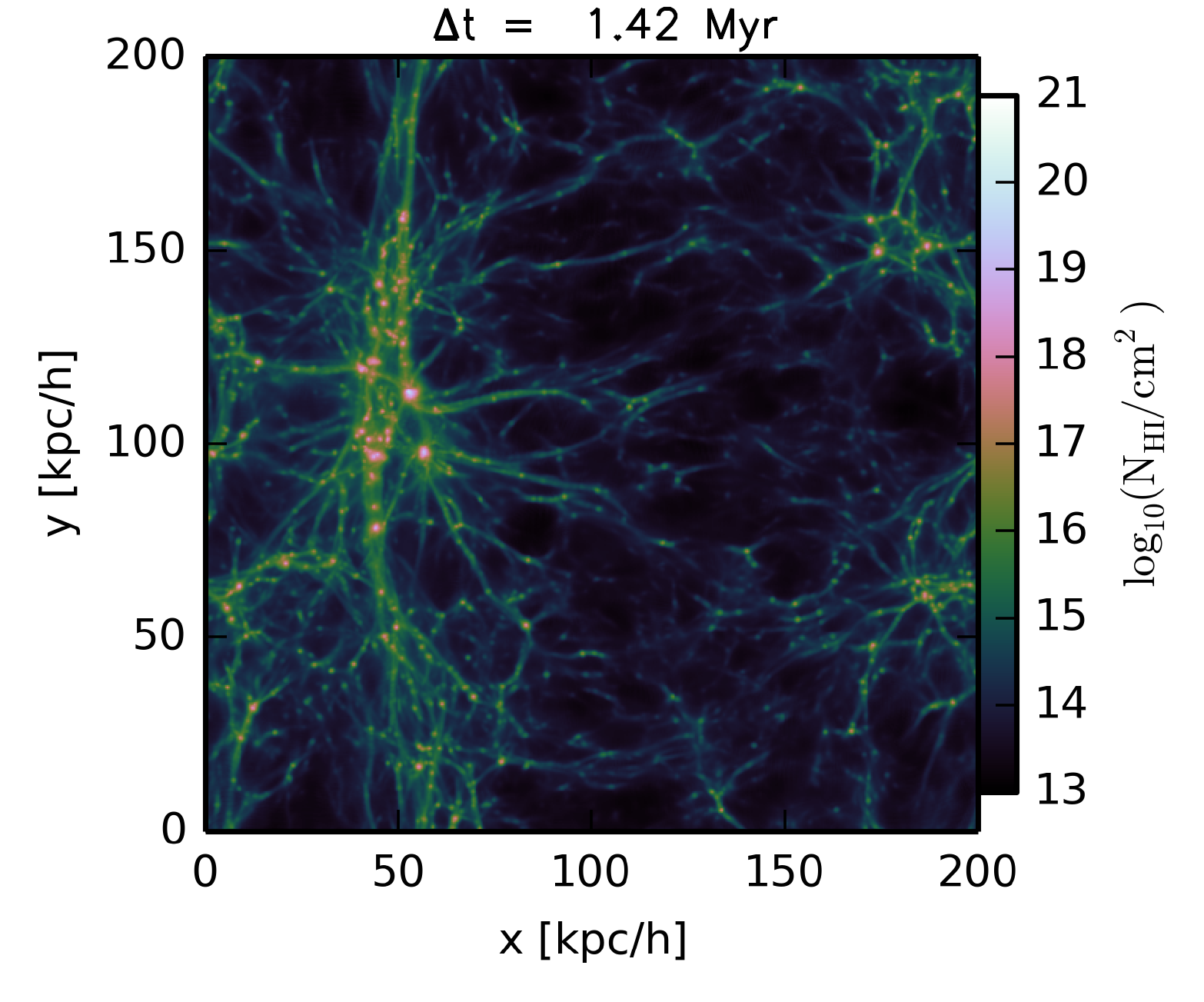}
    \includegraphics[scale=0.55]{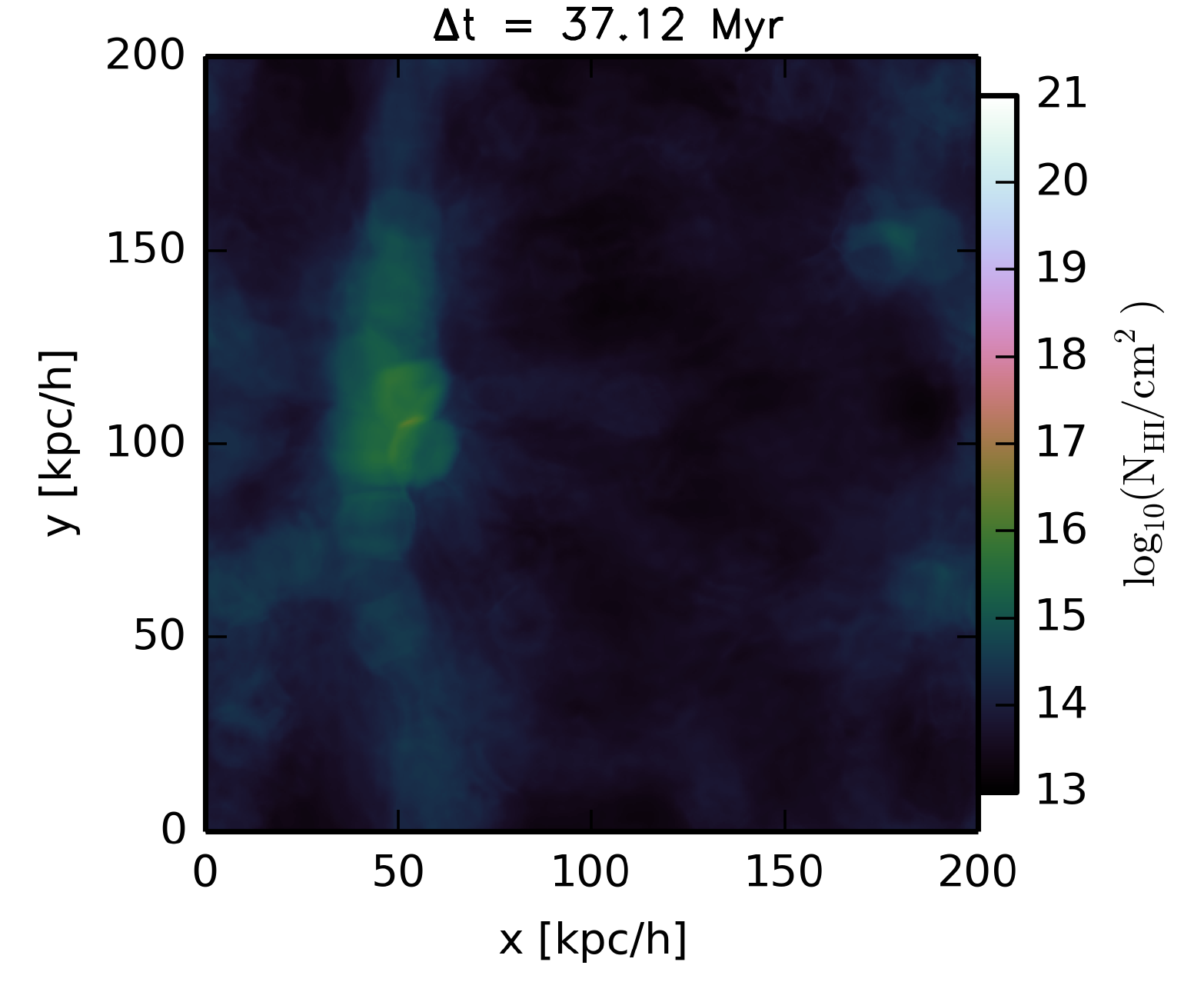}
    \caption{ The projected H I column density of M\_I0\_z10\_NS at $\Delta t=$ 1.4 Myr (left) and 37 Myr (right).}
        \label{fig:ColumnDen_NS}
  \end{center} 
\end{figure*} 

\begin{figure*}
  \begin{center}
    \includegraphics[scale=0.55]{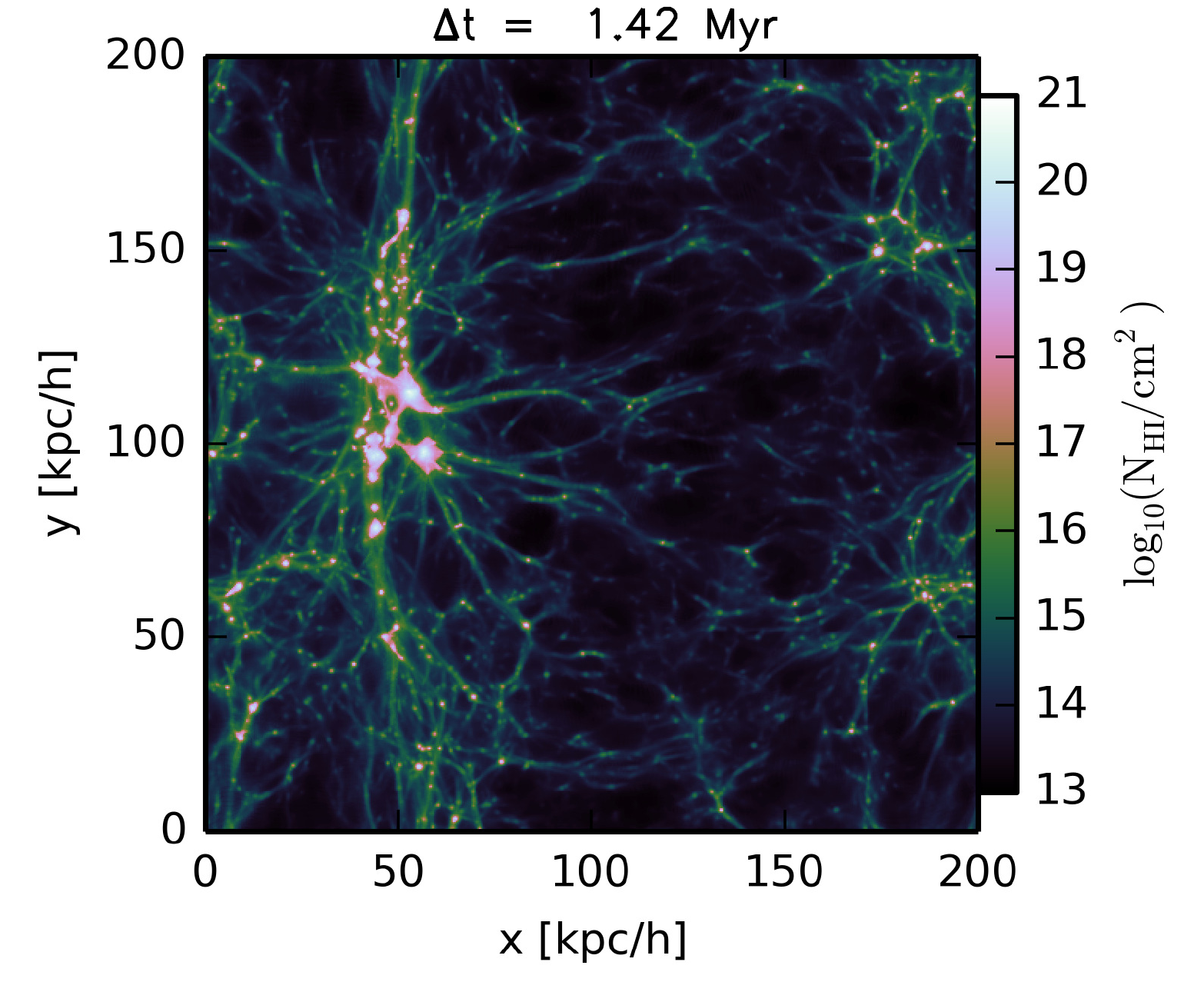}
    \includegraphics[scale=0.55]{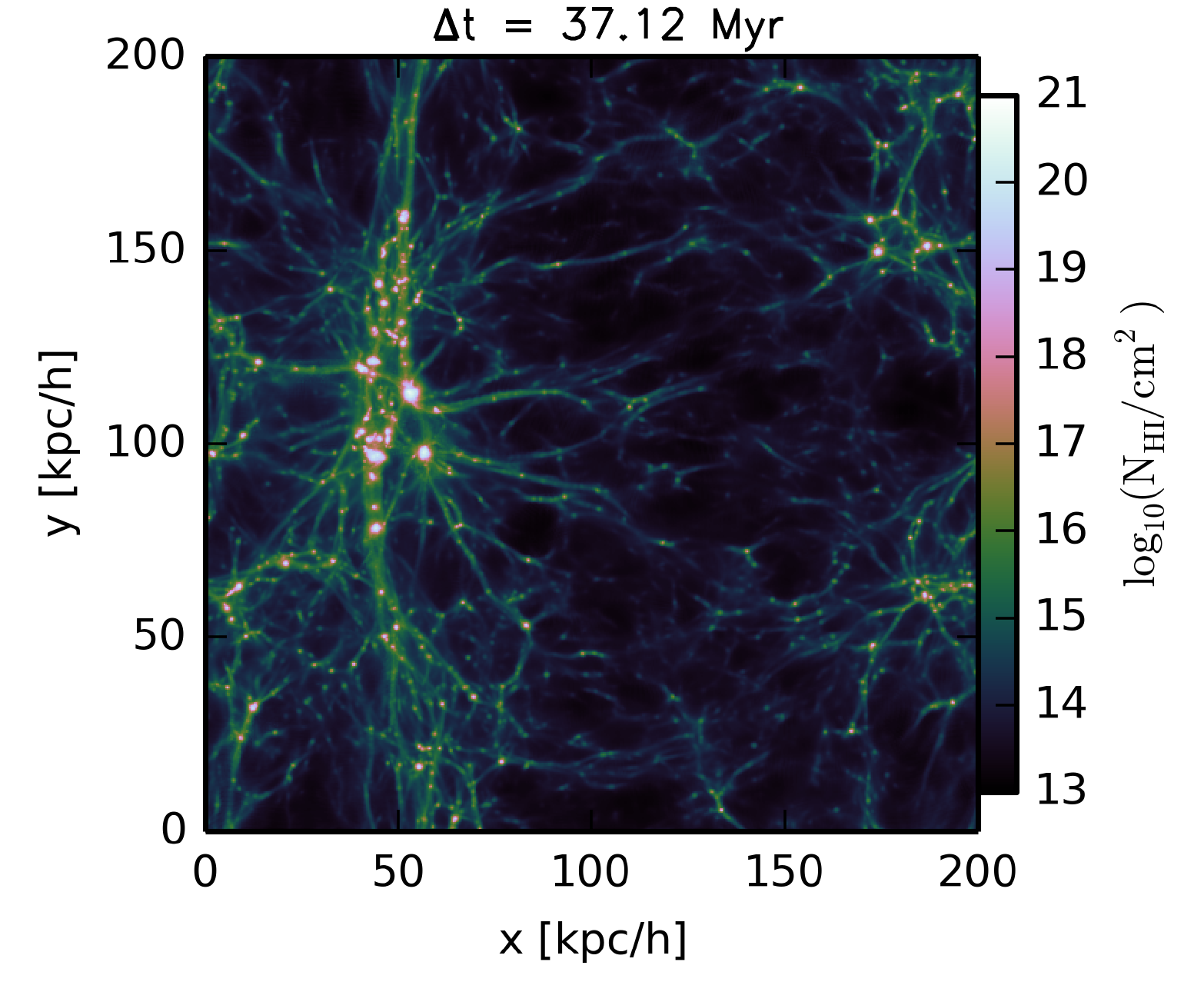}
    \caption{ The projected H I column density of M\_I0\_z10\_ND at $\Delta t=$ 1.4 Myr (left) and 37 Myr (right).}
        \label{fig:ColumnDen_ND}
  \end{center} 
\end{figure*} 

\begin{figure*}
  \begin{center}
    \includegraphics[scale=0.55]{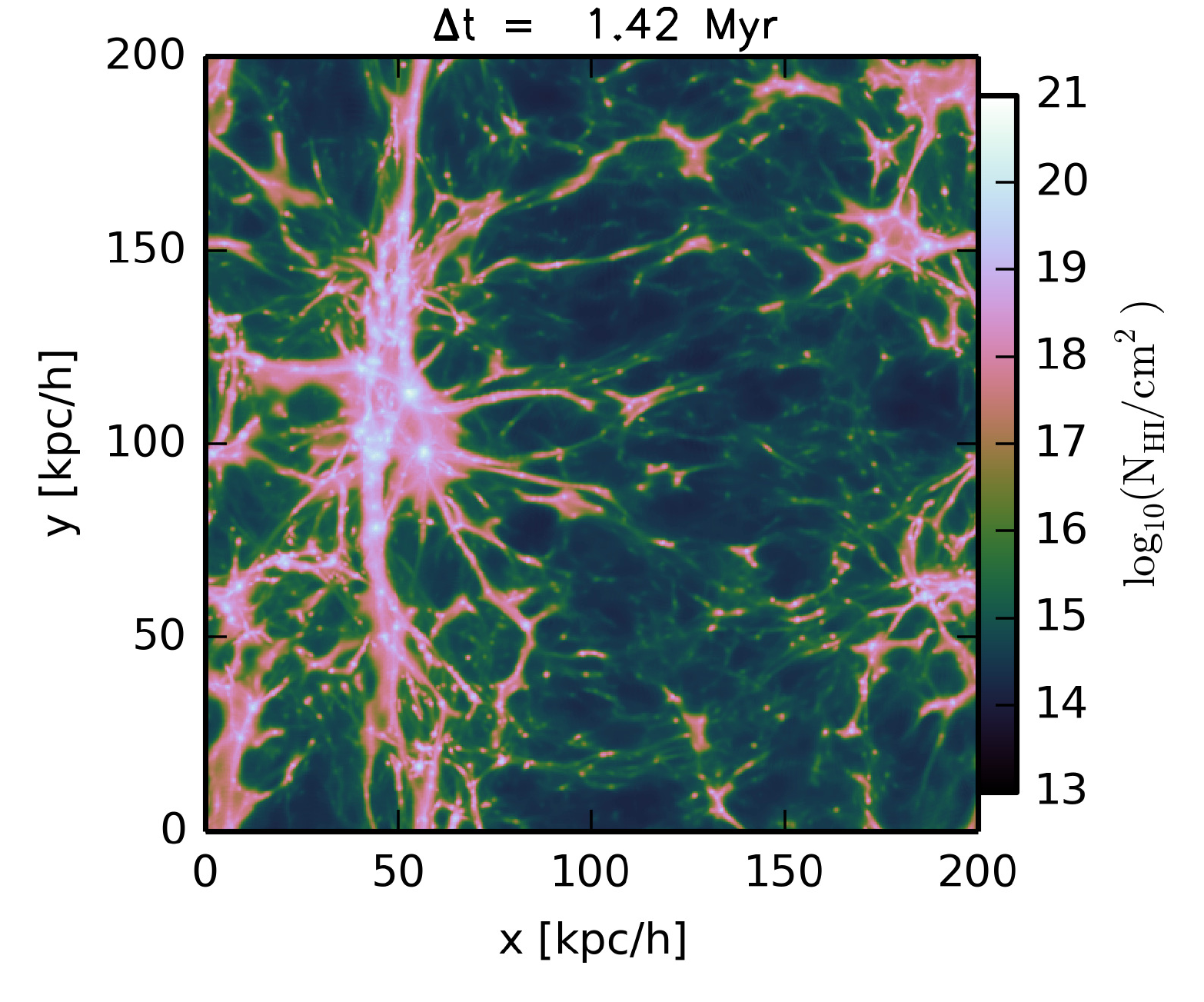}
    \includegraphics[scale=0.55]{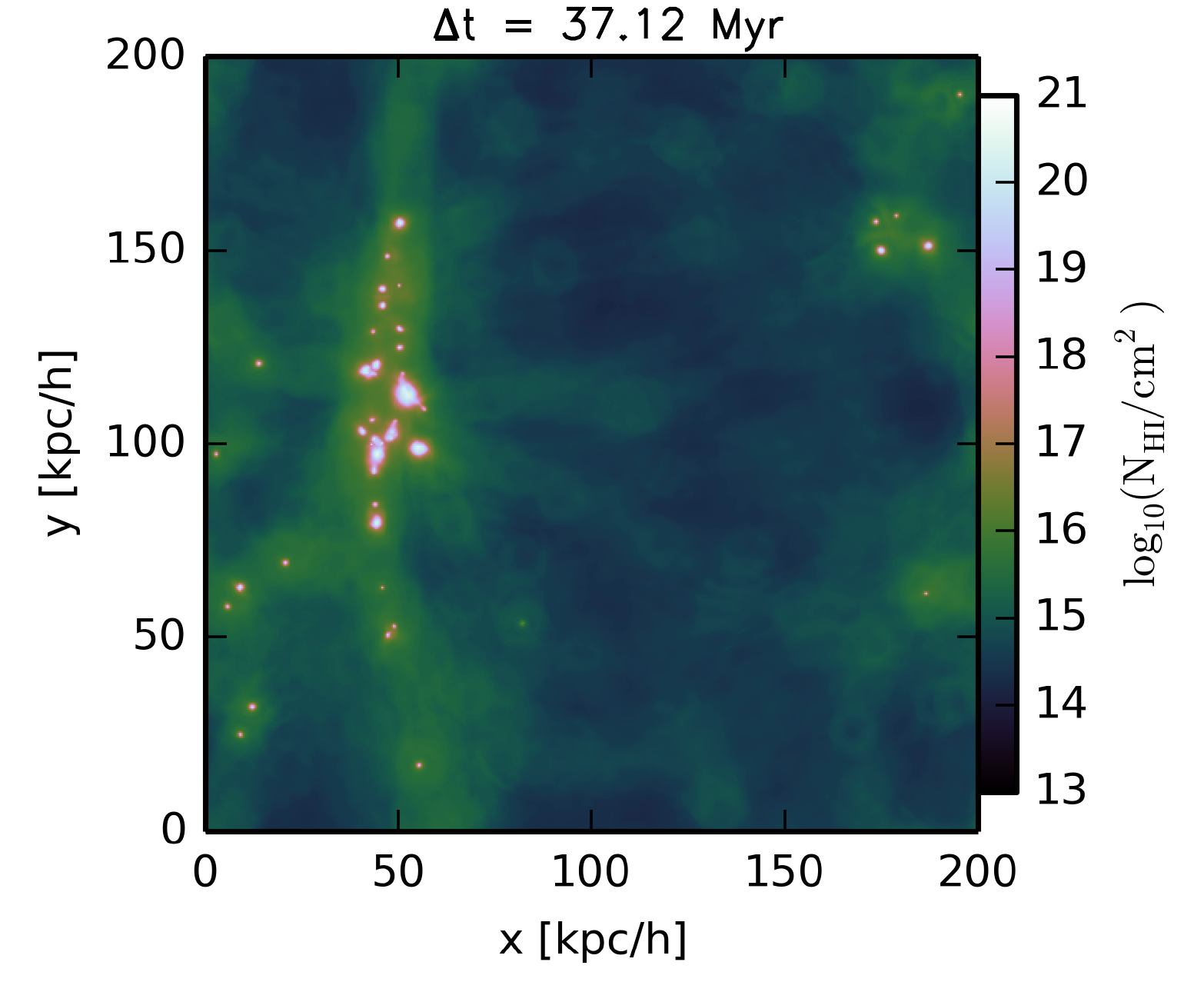}
    \caption{The projected H I column density of M\_I-1\_z10 at $\Delta t=$ 1.4 Myr (left) and 37 Myr (right).}
  \label{fig:ColumnDen_I-1}
  \end{center} 
\end{figure*} 

\begin{figure*}
  \begin{center}
    \includegraphics[scale=0.55]{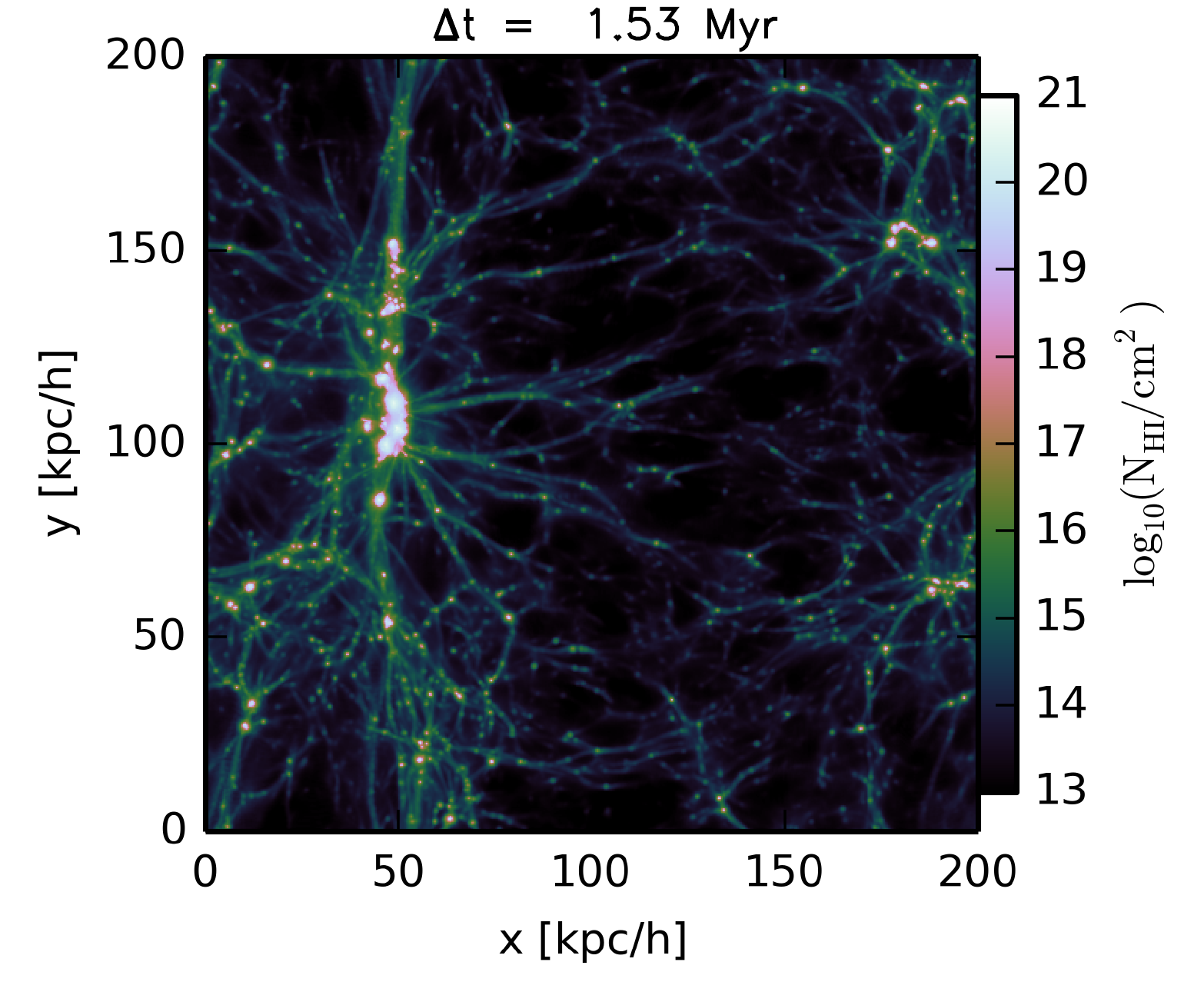}
    \includegraphics[scale=0.55]{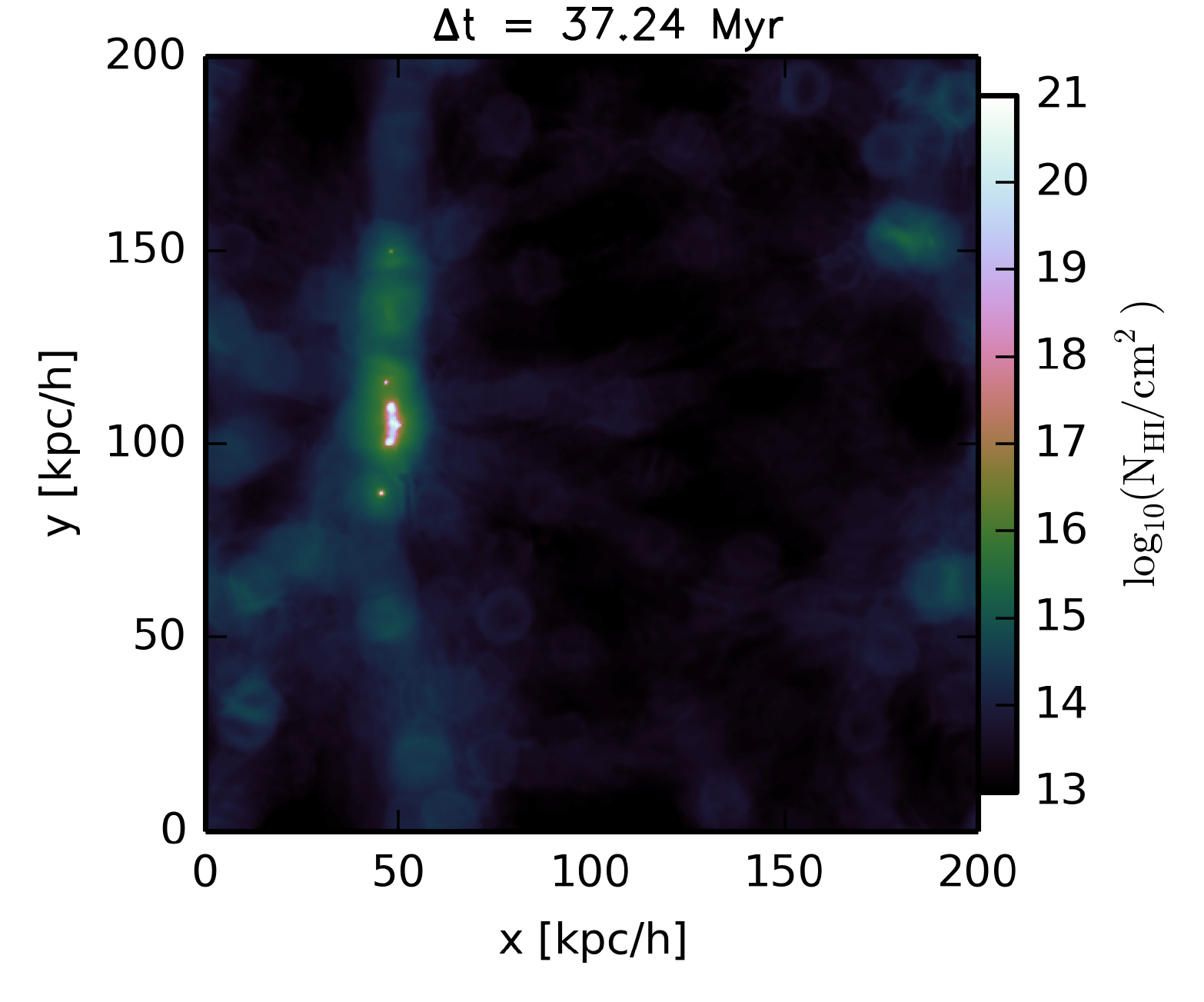}
    \caption{ The projected H I column density of M\_I0\_z8 at $\Delta t=$ 1.5 Myr (left) and 37 Myr (right).}
  \label{fig:ColumnDen_z8}
  \end{center} 
\end{figure*}

\subsection{Interpretation of High Clumping Factor in ETA13} 

ETA13 used their post-processed radiative-transfer simulations to explore the dependence of the clumping factor on $z_i$ and $\Gamma_{-12}$. Here
\bea
\Gamma_{-12}\equiv \left(\frac{0.3}{10^5~{\rm cm}^{-2}~{\rm s}^{-1}}\right)\int \Omega \int^{54.4~{\rm eV}}_{13.6~{\rm eV}} \frac{I_\nu}{h\nu}d\nu,
\eea
where $I_\nu$ is the intensity of the EIBR at the frequency $\nu$. Their main result for the clumping factor is in Figure 4 of their work\footnote{ Both left and right panels of the figure give the clumping factor for given $z_i$ and $\Gamma_{-12}$. But, each panel gives a slightly different clumping factor for the parameters of our interest. We shall adopt the left panel in our discussion.}. Their reported clumping factor was substantially larger than what recent works reported ($\sim 3$) in most of their parameter space. For example, their clumping factor is well above 10 for $z_i <10$ and $\Gamma_{-12}> 1$. We list $z_i$ and $\Gamma_{-12}$ for our simulations in Table~\ref{simulation parameters} to allow reading out their version of the clumping factor.

Our no-dynamics run (M\_I0\_z10\_ND) mimics their simulation by activating EIBR while fixing the locations of particles. We suppose the asymptotic state of the no-dynamics run corresponds to the simulation result of ETA13. The no-dynamics run uses $[z_i,\Gamma_{-12}]=[10,9.2]$ that corresponds to the clumping factor of 21 according to Figure 4 of ETA13. The asymptotic value of $C_{\rm r}$ in the no-dynamics run on the other hand is 26. Here we note that the clumping factor ETA13 calculates corresponds to $C_{\rm i}$ in this work. As we find in Section~\ref{sec:understanding_clumping_factor}, $C_{\rm i}$ is lower than $C_{\rm r}$ by 10 - 30 percents. In the asymptotic state, the no-dynamics run gives $C_{\rm i}=21$  that agrees very well with ETA13.

In other runs that we allow particles to move, the peak value of $C_{\rm i}$ tends to be lower than in ETA13's results as can be expected from the effects of hydrodynamic feedback. For M\_I0\_z10, M\_I0\_z9, and M\_I0\_z8, $C^{\rm peak}_{\rm i}$ = 16, 21, \& 29 while ETA13 reports 21,30, \& 45, respectively. Here, simulations in this work gives roughly two third of that ETA13 give. For M\_I0\_z10, M\_I-0.5\_z10, \& M\_I-1\_z10, $C^{\rm peak}_{\rm i}$ = 16, 9, \& 6 while ETA13 reports 21, 15, \& 12, respectively. Here the difference grows toward lower $J_{21}$ case. This is because it takes more time for the clumping factor to reach the peak when $J_{21}$ is lower allowing the hydrodynamic feedback to suppress the clumpiness more.

To summarize, the high ($>10$) clumping factor reported by ETA13 does appear in the early phase of our simulation. But, it lasts only for a few Myr until the hydrodynamic feedback effects of photo-ionization wipes out gas density structure. We shall discuss its impact on the UV photon consumption during EoR in Section~\ref{sec:Recombination Rate}.

\subsection{Clumping Factor in Non-cosmic Mean Density Regions} 

\begin{figure*}
  \begin{center}  
    \includegraphics[scale=0.46]{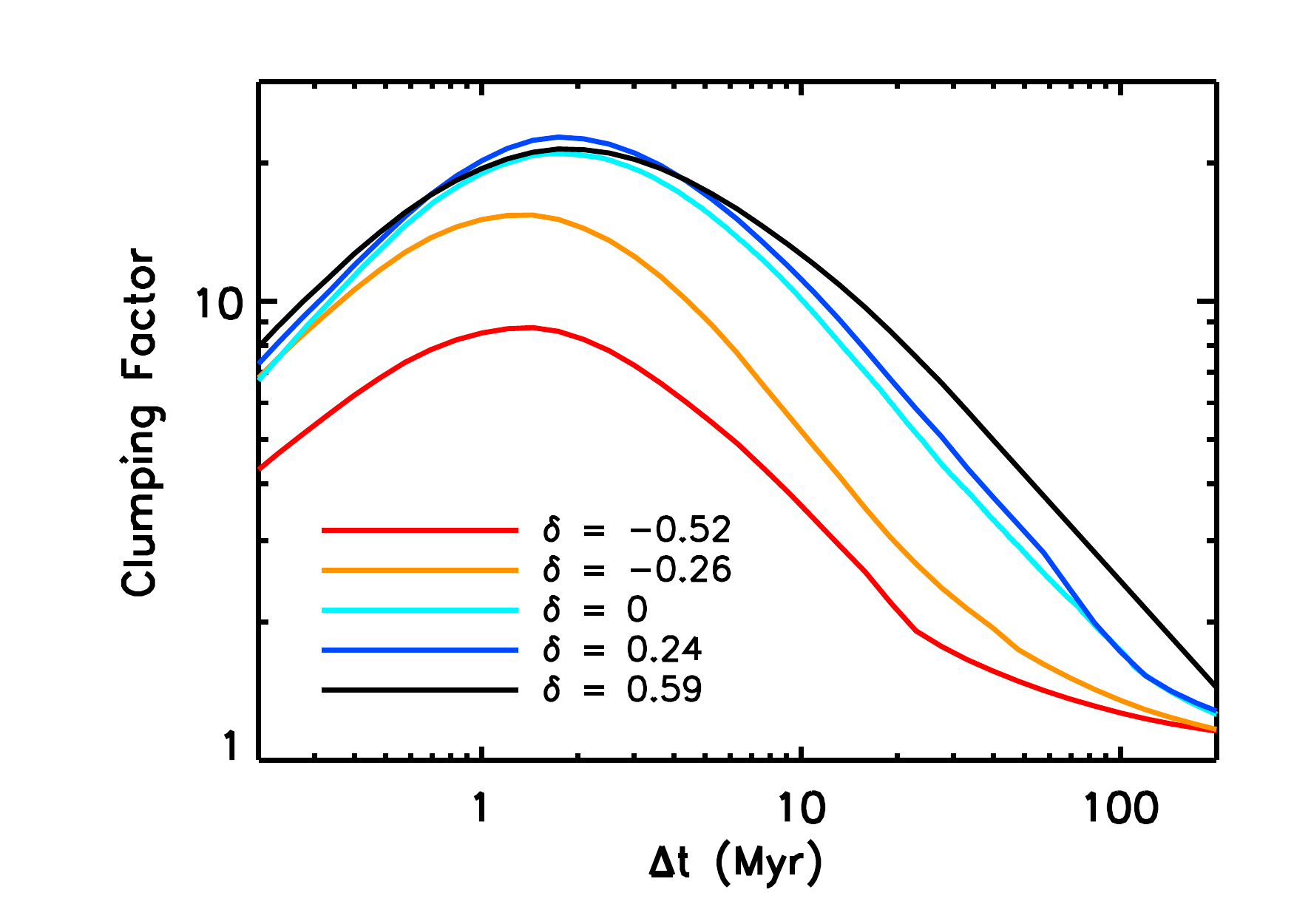}
    \includegraphics[scale=0.50]{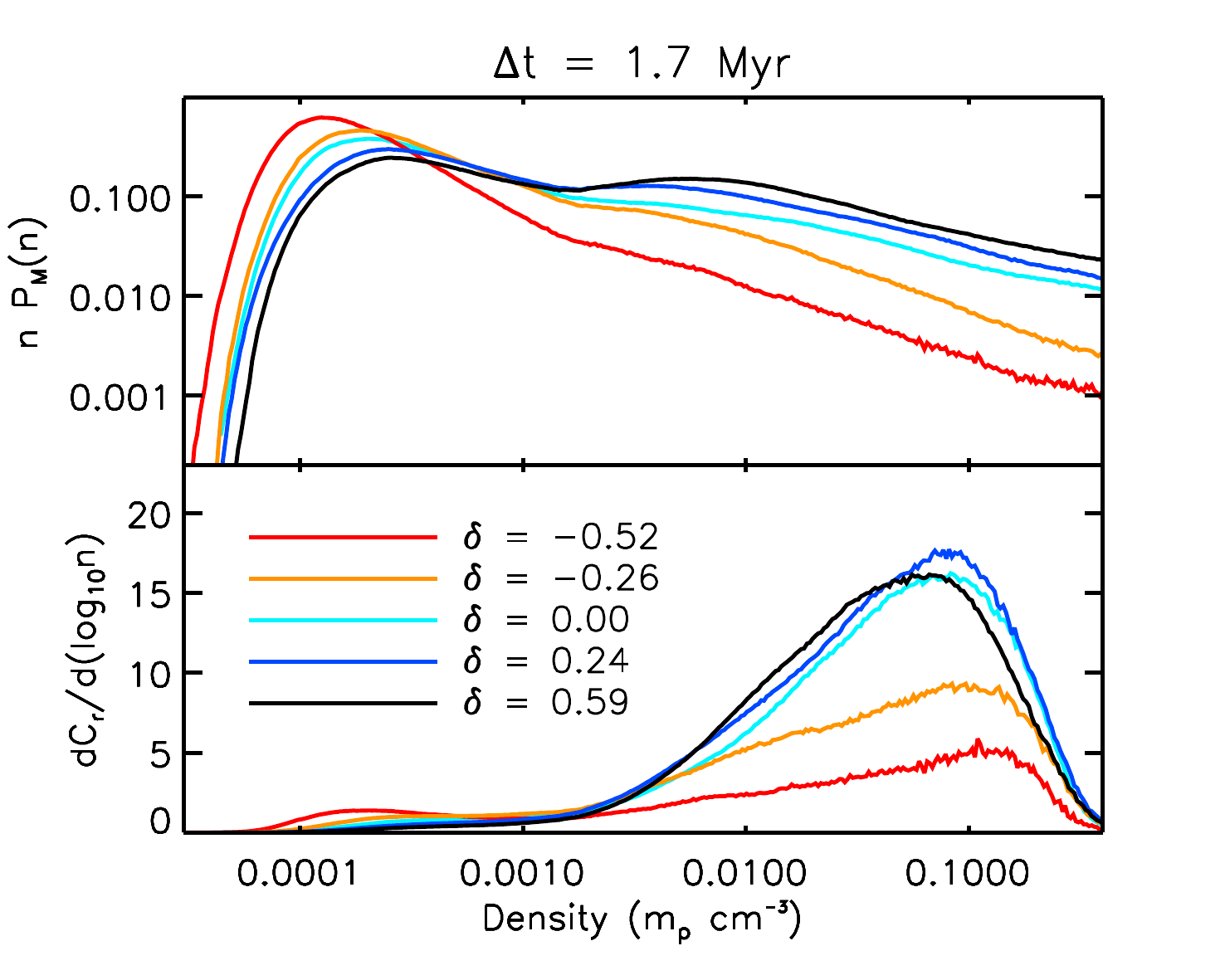}
        \caption{(left) The clumping factor ($C_{\rm r}$) versus time for five runs with a same box-size and EIBR intensity, but different mean densities. Red, orange, cyan, blue, \& black lines correspond to M\_I0\_z10\_VL$\delta$, M\_I0\_z10\_L$\delta$, M\_I0\_z10, M\_I0\_z10\_H$\delta$, \& M\_I0\_z10\_VH$\delta$ with $\bar\delta$ =  -0.52, -0.26, 0, 0.24, \& 0.59, respectively. (right) The probability density function of gas particle density ($P_M$, upper panel) and clumping factor contribution ($dC_{\rm r}/d\log_{10}n$, lower panel) as functions of $n$ at $\Delta t = 1.7$ Myr for the same runs considered in the left panel. The correspondence between line colors and the runs is same as in the left panel as well.}
        \label{fig:C_vs_t_del}
      \end{center} 
\end{figure*} 

In the analyses above, we have only considered sample sub-Mpc volumes, in which the mean density equals the cosmic mean. In reality, such volumes in a cosmological environment should have a substantial variation in their mean densities at $z\lesssim10$. At $z=10$, the variation in the mean density of a $200~h^{-1}~\rm{kpc}$ box can be given roughly by
\bea
\int d^3k ~W(kR)~P_{\delta\delta}(k,z=10) \approx (0.6)^2,
\eea
where $P_{\delta\delta}$ is the density power spectrum and $W(x=kR)=[3/x][\sin(x)/x^2-\cos(x)/x]$ is the window function for a spherical top-hap with radius $R$, which we set to be the size of the box, $200~h^{-1}~\rm{kpc}$. To cover roughly the one sigma ($\sim$0.6) range of the mean density contrast, we simulate EIBR in four more $200~h^{-1}~\rm{kpc}$ boxes with the mean density contrasts $\bar\delta$ = -0.52, -0.26, 0.24, \& 0.59. They are sub-samples of the $800~h^{-1}~\rm{kpc}$ box introduced in Section~\ref{sec:simulation}. From low to high $\bar\delta$, we name them as M\_I0\_z10\_VL$\delta$, M\_I0\_z10\_L$\delta$, M\_I0\_z10\_H$\delta$, and M\_I0\_z10\_VH$\delta$. 

The clumping factor result for different $\bar\delta$'s are compared in the left panel of Figure~\ref{fig:C_vs_t_del}. $P_M$ and $dC_{\rm r}/d\log_{10}n$ are shown for those runs in the right panel of Figure~\ref{fig:C_vs_t_del}. Also, $C^{\rm peak}_{\rm r}$ values for those runs can be found at Table~\ref{simulation parameters}. Before reading the clumping factor values, it is important to remember that the recombination rate goes as both the clumping factor and the mean density of the box as described in Equation~(\ref{eq:dNdt}). Thus, one needs to multiply $C_r[1+\bar\delta]$ to the background recombination rate to get the net recombination rate.

Up to $\bar\delta = 0$, the clumping factor clearly correlates with $\bar\delta$ being higher for higher $\bar\delta$ at all $\Delta t$. For $\bar\delta$ above 0, the early time ($\Delta t\lesssim$  2 Myr) clumping factor insensitive to $\bar\delta$, but the speed that the clumping factor decays is slower after $\Delta t=$  2 Myr for higher $\bar\delta$. Thus, the late time ($\Delta t\gtrsim$  2 Myr) clumping factor still correlates with $\bar\delta$ above 0. The peak clumping factor $C^{\rm peak}_{\rm r}$ represents the dependence of the early clumping factor on $\bar\delta$. It rises from 8.8 to 21 as we increase $\bar\delta$ from -0.52 to 0, but stays around 21 when increasing $\bar\delta$ from 0 to 0.59. 

The $\bar\delta$-dependence of the early time clumping factor is determined mainly by $P_M$ shown in the right panel of Figure~\ref{fig:C_vs_t_del}. At $0.01\lesssim n\lesssim 0.1~\rm{cm}^{-3}$, where most of the clumping factor contribution comes from, $P_M$ highly correlates with $\bar\delta$ up to $\bar\delta=0$. The correlation gets weaker for $\bar\delta$ above 0, resulting in a saturation of the early-stage clumping factor. Considering the extra $[1+\delta]$ factor on top of the clumping factor for the recombination rate, the recombination rate should keep increasing with $\bar\delta$ for $\bar\delta>0$. Thus, the recombination rate correlates with $\bar\delta$ for all $\bar\delta$'s.

%As we have seen from Equation~(\ref{eq:dNdt}), a fixed clumping factor leads to the mass-weighted recombination rate going as the mean density $\bar{n}$ or $[1+\bar\delta]$. Thus the recombination rate positively correlates with $\bar\delta$ for any value of $\bar\delta$.

\subsection{Ionizing Photon Budget for Small-scale Structure} \label{sec:Recombination Rate} 

\begin{table*} 
\caption{Fitting Parameters for $dN^{\rm add}_{\rm rec}/dt$ and $d\bar\tau/ds$}
\begin{center} 
\begin{tabular}{@{}llllllllllllllllllll||}  
\hline
label & $z_i$  &  $J_{21}~(\Gamma_{-12})$ & $\bar\delta$ &$a_0$ & $a_1$& $a_2$&$a_3$& $A$& $\gamma$ \\
\hline
M\_I0\_z10                         & 10 & 1~(9.2) & 0 &  -3.8&-0.18&-0.097&-0.0078& 0.025&-0.82\\
M\_I-0.5\_z10                     & 10 & 0.3~(2.8) & 0 &-3.9&-0.67&0.14&-0.034&0.076&-0.82\\
M\_I-1\_z10                         & 10 & 0.1~(0.92) & 0 &  -5.2 &0.094&-0.056&-0.012&0.16&-0.75\\
M\_I0\_z9                            & 9  & 1~(9.2)& 0 & -3.5 &-0.56&0.079&-0.030&0.026&-0.76\\
M\_I0\_z8                            & 8  & 1~(9.2) & 0 & -4.0 &-0.16&0.019& -0.027&0.024&-0.61\\
M\_I0\_z10\_VH$\delta$     & 10 & 1~(9.2)& 0.59 & -4.6 &1.04&-0.41& 0.023&0.069&-0.62\\
M\_I0\_z10\_H$\delta$       & 10 & 1~(9.2)& 0.24 & -12.2 &6.4&-1.71& 0.120&0.051&-0.75\\
M\_I0\_z10\_L$\delta$        & 10 & 1~(9.2)& -0.26 & -0.118 &-3.7&0.69& -0.061&0.0039&-1.07\\
M\_I0\_z10\_VL$\delta$     & 10 & 1~(9.2)& -0.52 & 0.089 &-4.7&0.94& -0.075&0.00083&-1.27\\
\hline
\end{tabular}  
\end{center}
\label{table: fitting}
\end{table*}

\begin{figure}
  \begin{center}  
    \includegraphics[scale=0.5]{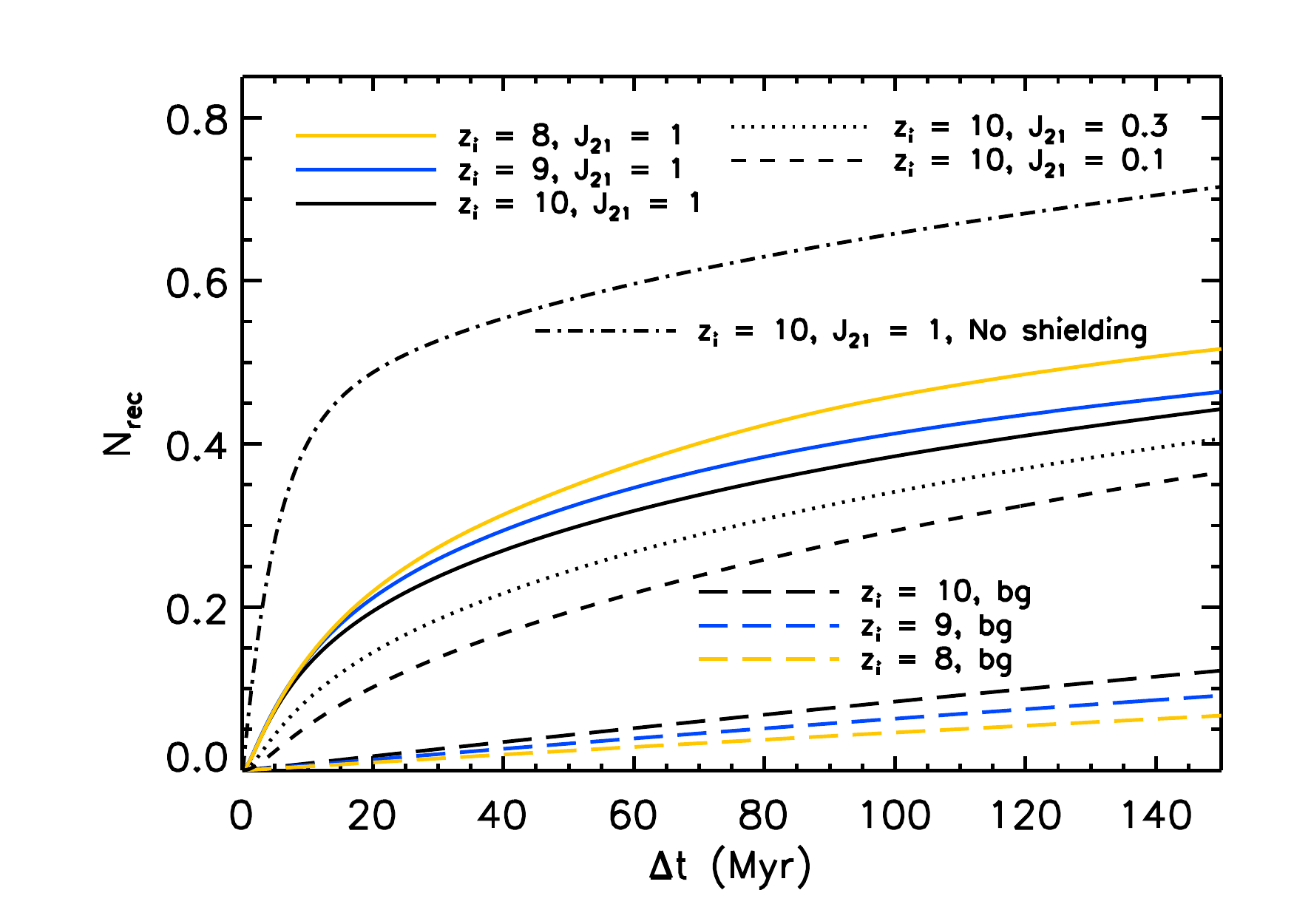}
        \caption{The accumulated recombination per $H$ atom, $N_{\rm rec}$, for M\_I0\_z10 (black solid), M\_I0\_z9 (blue solid), M\_I0\_z8 (yellow solid), M\_I-0.5\_z10 (black dotted), M\_I-1\_z10 (black dashed), and M\_I0\_z10\_NS (black dot-dashed). The black, blue, and yellow long dashed lines describe the background recombination rate calculated from the average gas density and the temperature at each time. }
        \label{fig:Rec_acc}
      \end{center} 
\end{figure}

 To tell the significance of the temporarily high clumping factor at the early time due to small-scale structure, we need to assess the recombination accumulated over time. For that purpose, we obtain the accumulated recombination per hydrogen atom by integrating Equation~(\ref{eq:dNdt}) w.r.t. time: 
\bea
N_{\rm rec}|_{\Delta t}\equiv \int_{t=0}^{\Delta t} \frac{dN_{\rm rec}}{dt}dt = \int_{t=0}^{\Delta t}C_{\rm r}\frac{\alpha_B(\bar{T})\bar{n}_{\rm H II}\bar{n}_{e}}{f_{\rm H}\bar{n}}dt.~~~~
\eea
To separate out the base amount expected from the case that the simulation volume is homogenous without any structure, we define the ``background recombination count" as $N^{\rm bg}_{\rm rec} \equiv C_{\rm r}^{-1}N_{\rm rec}$. Then, the rest, $N^{\rm add}_{\rm rec}\equiv(1-C_{\rm r}^{-1})N_{\rm rec}$, can be interpreted as the additional due to the structures on top of the background.
We plot $N_{\rm rec}$ and $N^{\rm bg}_{\rm rec}$ as functions of time for M\_I0\_z10, M\_I0\_z9, M\_I0\_z8, M\_I-0.5\_z10, M\_I-1\_z10, and M\_I0\_z10\_NS in Figure~\ref{fig:Rec_acc}. In this case, $C_{\rm r}$ is the ratio of the slope of $N_{\rm rec}$ to that of $N^{\rm bg}_{\rm rec}$. 

At $\Delta t \lesssim 20$ Myr, the boost of the clumping factor makes $N_{\rm rec}$ accumulate much faster than $N^{\rm bg}_{\rm rec}$ does. Later ($\Delta t \gtrsim 20$ Myr), the slope of $N_{\rm rec}$ asymptotes to that of $N^{\rm bg}_{\rm rec}$ as $C_{\rm r}$ decays toward unity. As a result, $N_{\rm rec}$ minus $N^{\rm bg}_{\rm rec}$ becomes nearly a fixed quantity after $\Delta t= 150$ Myr. Thus, we interpret $N^{\rm add}_{\rm rec,150}\equiv N^{\rm add}_{\rm rec}|_{\Delta t = 150~\rm{Myr}}$ as {\it the ionizing photon budget for the pre-ionization IGM}. We list $N^{\rm add}_{\rm rec,150}$ and $N^{\rm bg}_{\rm rec,150}\equiv N^{\rm bg}_{\rm rec}|_{\Delta t = 150~\rm{Myr}}$ in Table~\ref{simulation parameters} for each run. %Note that $N^{\rm bg}_{\rm rec,150}$ is {\it not} a fixed quantity and it grows in time. 

Similarly to the clumping factor, $N^{\rm add}_{\rm rec,150}$ is larger for lower $z_i$, higher $J_{21}$, and higher $\bar\delta$. 
We fit the result with the following scaling relation.
\bea \label{eq:Naddrec150}
N^{\rm add}_{\rm rec,150} \approx 0.32\times [J_{21}]^{0.12}\left[\frac{1+z_i}{11}\right]^{-1.7}[1+\bar\delta]^{2.5}
\eea
The relation above shows that the ionization budget scales very weakly with the EIBR intensity, and much more strongly with $z_i$ and $\bar\delta$ that are closely related to the abundance of structure. It is notable how strongly $N^{\rm add}_{\rm rec,150}$ scales with $\bar\delta$ even at $\bar\delta>0$ while the $C^{\rm peak}_{\rm r}$ value remains nearly unchanged. This highlights the impact of slower decaying clumping factor in $\bar\delta>0$ cases in Figure~\ref{fig:C_vs_t_del}. The lesson here is that the magnitude of clumping factor in early time cannot fully describe the ionization photon budget and one has to seriously take into account the subsequent evolution of structure with hydrodynamic feedback.

Applying the relation in Equation~(\ref{eq:Naddrec150}) for all the $200~h^{-1}~\rm{kpc}$ sub-boxes taken from the $z=10$ snapshot of the $(800~h^{-1}~\rm{kpc})^3$ volume, we find the average of $N^{\rm add}_{\rm rec,150}$ over the whole $(800~h^{-1}~\rm{kpc})^3$ volume for $z_i=10$ \& $J_{21}=1$ is
\bea \label{eq:meanN_rec}
\left< N^{\rm add}_{\rm rec,150} \right>  &=& \frac{1}{64}\sum^{64}_{i=1}[1+\bar\delta_i]N^{\rm add}_{\rm rec,150,i} \nonumber\\
&=&\frac{1}{64}\sum^{64}_{i=1}0.32[1+\bar\delta_i]^{3.5}\nonumber\\
&=&0.67,
\eea
where the index $i$ in the subscripts denotes the $i$th of the 64 sub-cubes from the $800~h^{-1}~\rm{kpc}$ box. The reason for the ionization budget over the whole $800~h^{-1}~\rm{kpc}$ box being larger than for the $200~h^{-1}~\rm{kpc}$ box the cosmic mean density is the strong power-law scaling of $N^{\rm add}_{\rm rec,150}$ with $[1+\bar\delta]$ and nonlinear growth of structure enhancing probability for the high-$\bar\delta$ end above what is expected from the pure gaussian distribution. For reionization models that ionize most of its volume toward the end of the EoR near $z=6$ \cite[e.g., the model of][]{2014MNRAS.439..725I}, we can get an additional factor of two enhancement in the ionization budget according to the scaling. Given that 2 - 3 per H atom have been considered for the ionization budget for the entire EoR, this level of extra recombination can potentially require a huge change for our current estimate.

We provide log-log 3rd order polynomial fitting functions for $dN^{\rm add}_{\rm rec}/dt$ as a function of $\Delta t$ as the following.
\bea \label{eq:Recomb_fit}
\log\left(\frac{dN^{\rm add}_{\rm rec}}{dt}\right) &=& a_0 + a_1(\log\Delta t) + a_2(\log\Delta t)^2
\nonumber\\
&+& a_3(\log\Delta t)^3~~~~~(\Delta t>\rm 2 ~Myr)
\nonumber\\
\nonumber\\
&=& 0 ~~~~~~~~~~~~~~~~~~(\Delta t<\rm 2 ~Myr)
\eea
In Figure~\ref{fig:Rec_add}, we display both the actual rate (left panel) and the fitted result (right panel).
For simplicity of fitting, we do not fit for $\Delta t<\rm 2~Myr$ when $dN^{\rm add}_{\rm rec}/dt$ is rising rapidly. $dN^{\rm add}_{\rm rec}/dt$ during that time can be ignored with a small error for $\Delta t\gtrsim 10~\rm{Myr}$. The fitting result is provided in Table~\ref{table: fitting} for M\_I0\_z10, M\_I0\_z9, M\_I0\_z8, M\_I-0.5\_z10, M\_I-1\_z10, M\_I0\_z10\_VH$\delta$, M\_I0\_z10\_H$\delta$, M\_I0\_z10\_L$\delta$, \& M\_I0\_z10\_VL$\delta$.

\begin{figure*}
  \begin{center}
    \includegraphics[scale=0.59]{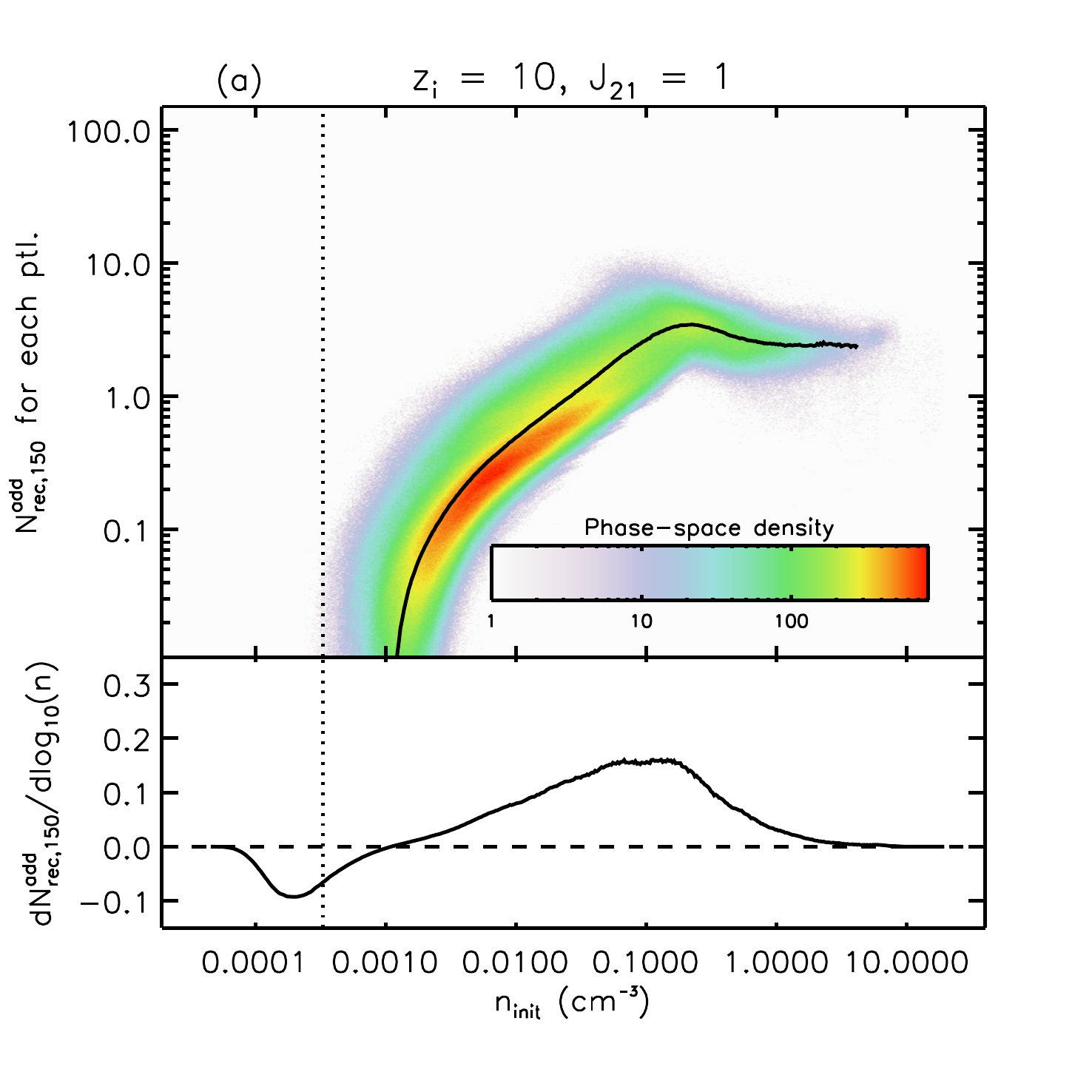}
    \includegraphics[scale=0.59]{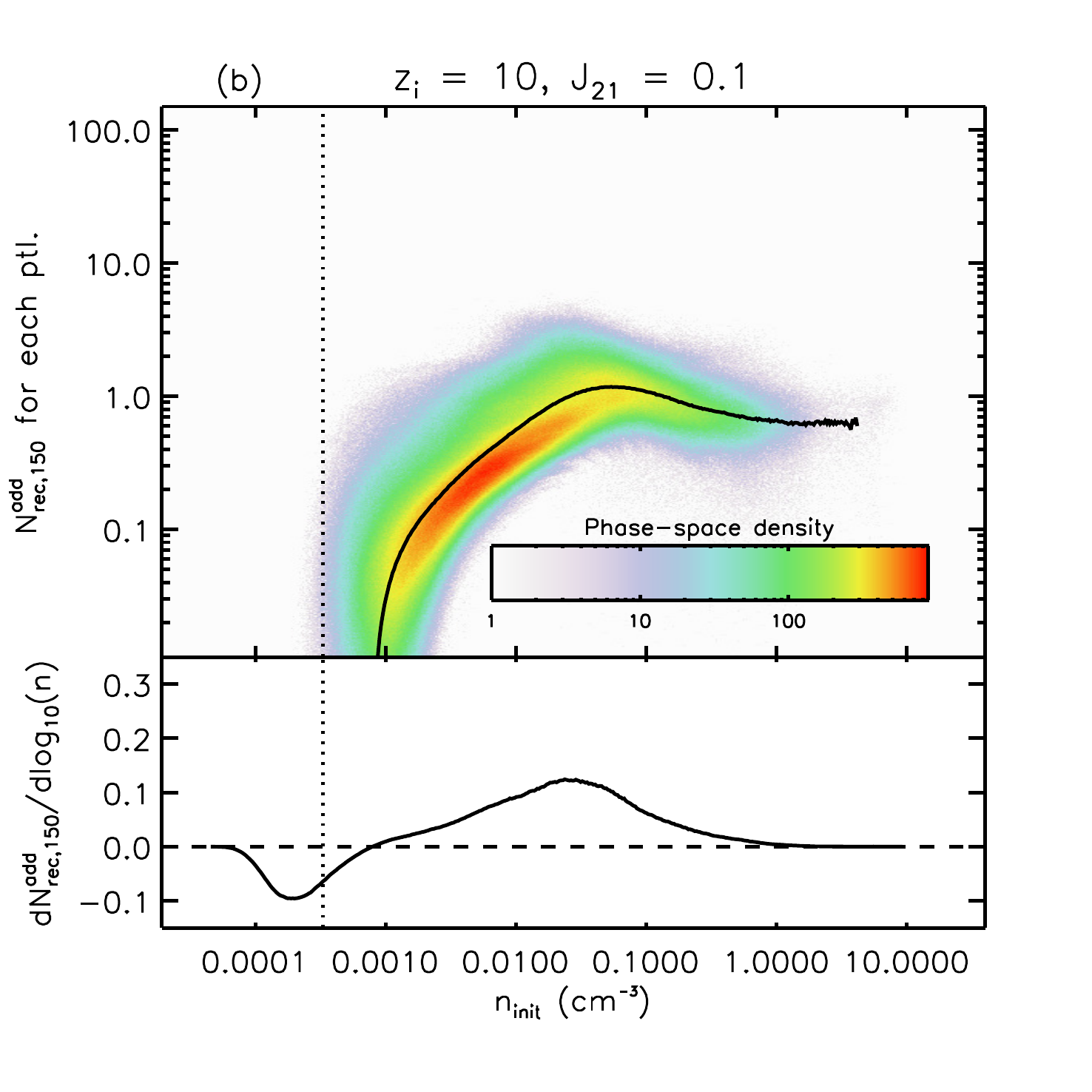}
    \includegraphics[scale=0.59]{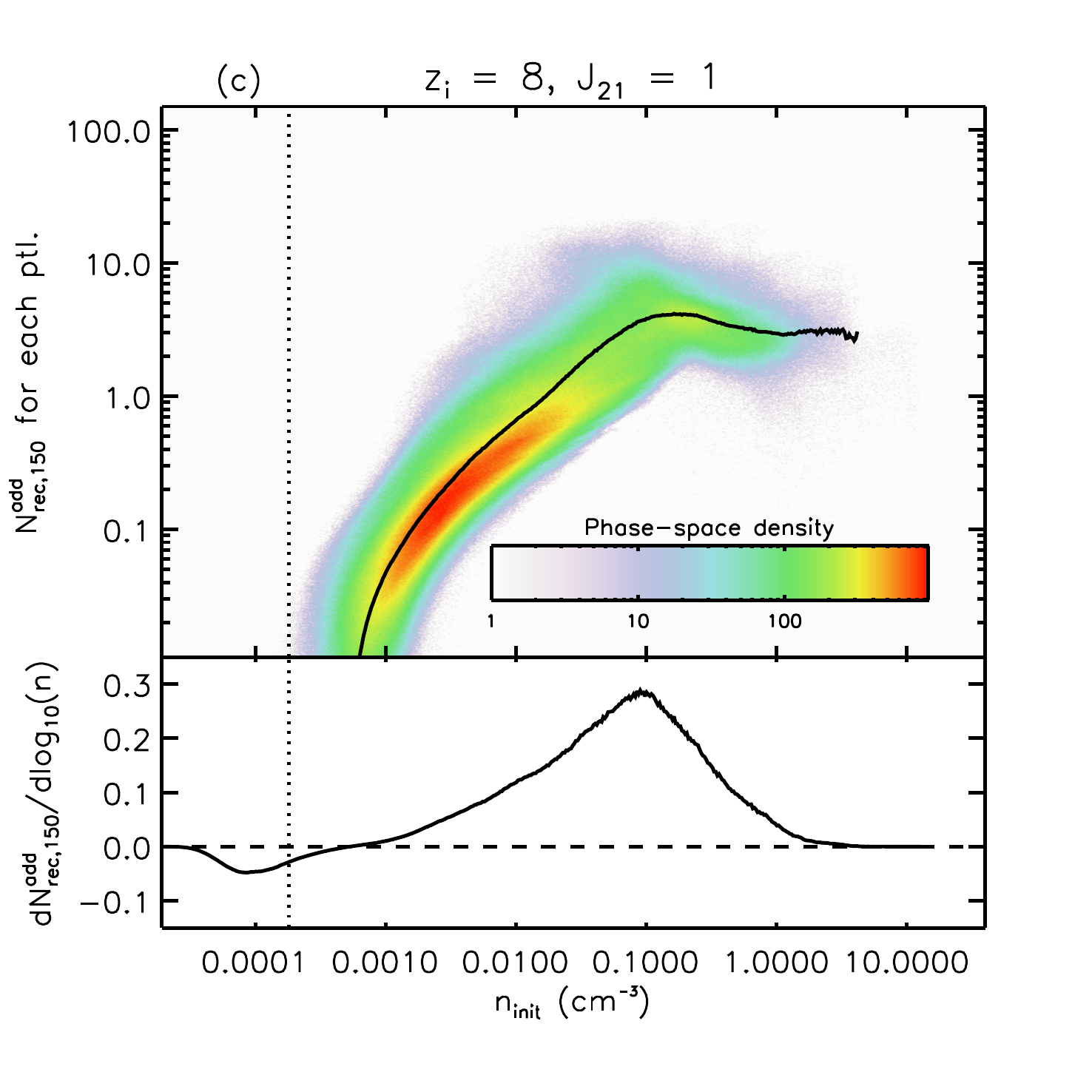}
    \includegraphics[scale=0.59]{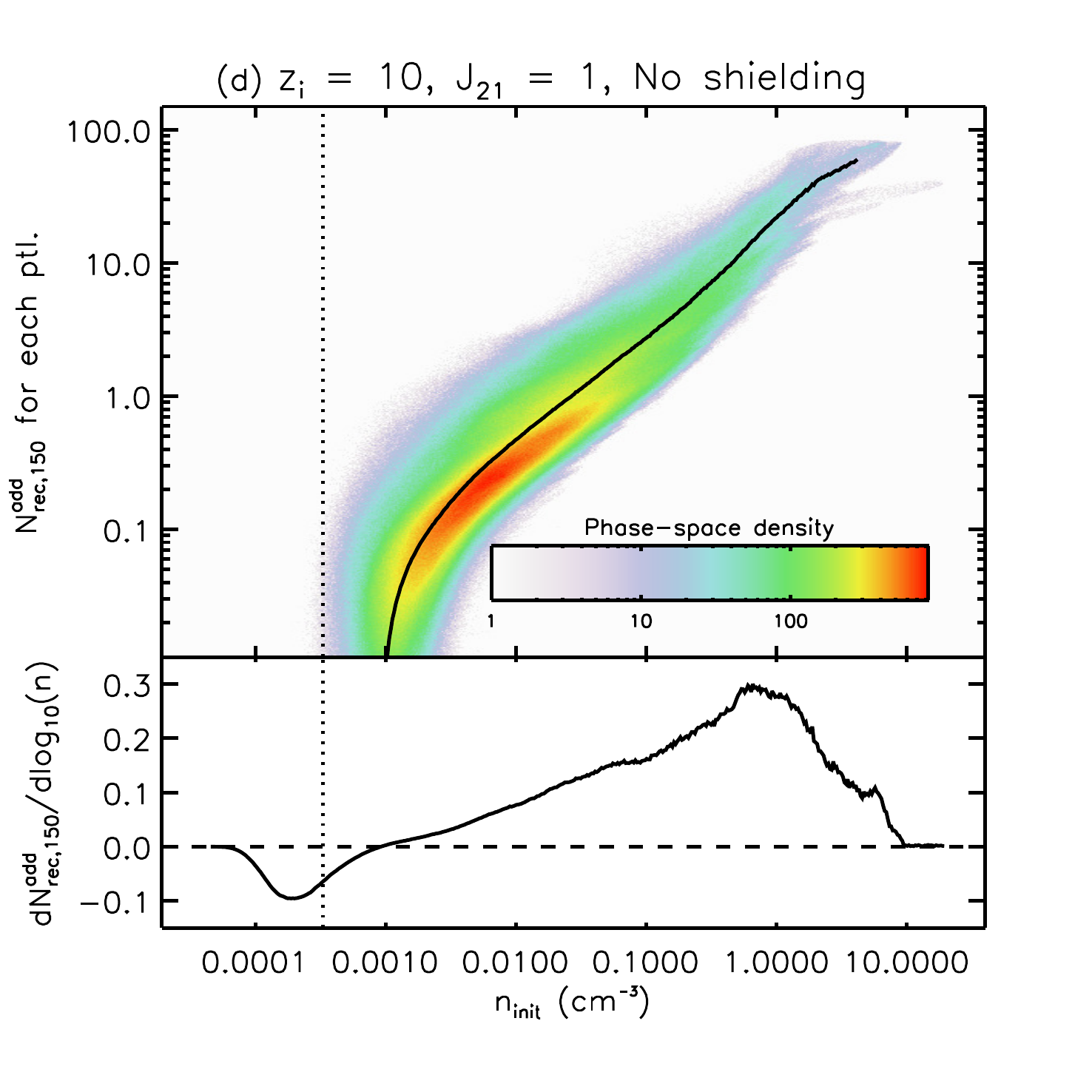}
    \caption{Upper panels show the scatter plot of the additional recombination ($N^{\rm add}_{\rm rec,150}$) versus initial particle density ($n_{\rm init}$) for M\_I0\_z10 (figure $a$), M\_I-1\_z10 (figure $b$), M\_I0\_z8 (figure $c$), and M\_I0\_z10\_NS (figure $d$). The vertical dotted line marks the cosmic mean density. Lower panels show $N^{\rm add}_{\rm rec,150}/d\log_{10}n$ to describe the contribution to $N^{\rm add}_{\rm rec,150}$ from each $n_{\rm init}$. This is obtained by binning and adding up $N^{\rm add}_{\rm rec,150}$ of individual particles in the log-space of $n_{\rm init}$.}
        \label{fig:r_vs_n_c}
     \end{center}
\end{figure*} 

\begin{figure*}
  \begin{center}  
    \includegraphics[scale=0.5]{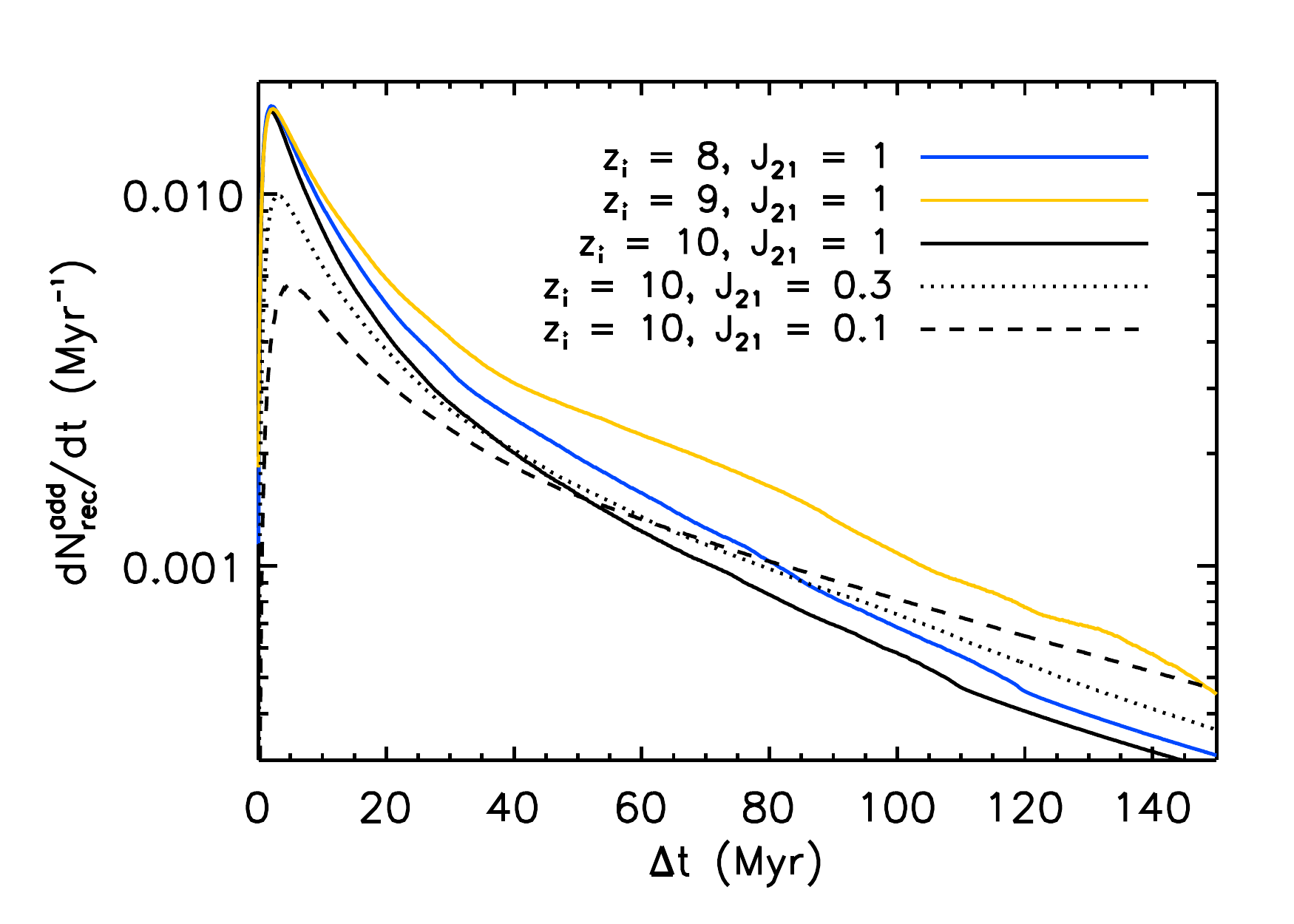}
    \includegraphics[scale=0.5]{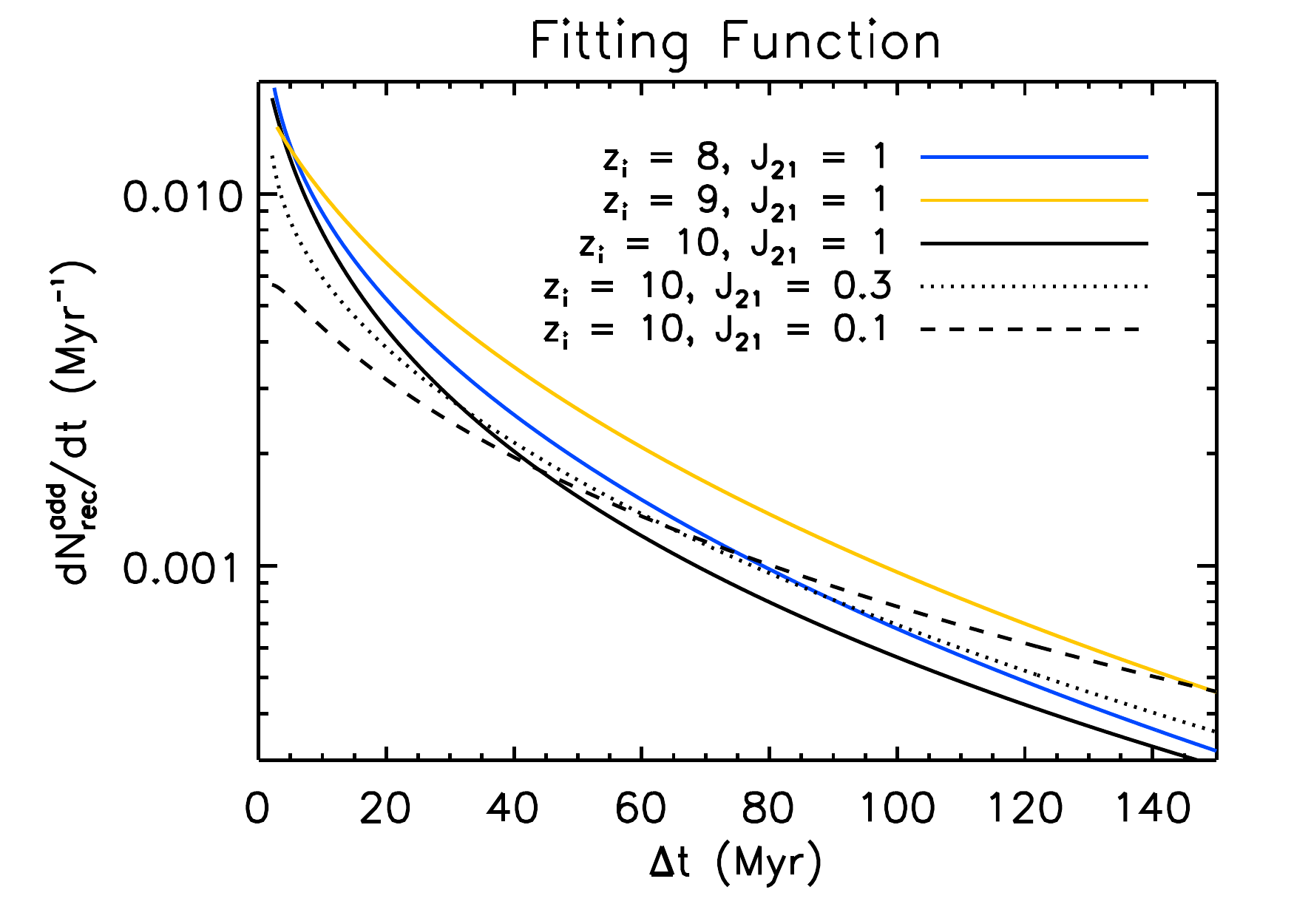}
        \caption{ (left) The recombination rate subtracted by the background rate, $dN^{\rm add}_{\rm rec}/dt$, plotted for M\_I0\_z10 (black solid), M\_I0\_z9 (blue solid), M\_I0\_z8 (yellow solid), M\_I-0.5\_z10 (black dotted), M\_I-1\_z10 (black dashed), and M\_I0\_z10\_NS (black dot-dashed). (right) The fitted results of $dN^{\rm add}_{\rm rec}/dt$ for the same runs considered in the left panel. The line types and colors correspond to the same runs as in the left panel.}
        \label{fig:Rec_add}
      \end{center} 
\end{figure*} 

\subsubsection{Initial Density of Gas Parcel and Ionizing Photon Budget}

We find it helpful to look into $N^{\rm add}_{\rm rec,150}$ for individual SPH particles in understanding the dependance of the global $N^{\rm add}_{\rm rec,150}$ on $J_{21}$ \& $z_i$. In the upper panels of Figures~\ref{fig:r_vs_n_c}$a$, \ref{fig:r_vs_n_c}$b$, \ref{fig:r_vs_n_c}$c$, \& \ref{fig:r_vs_n_c}$d$, we scatter-plot $N^{\rm add}_{\rm rec,150}$ of each particle versus its SPH density at the turn-on of the EIBR ($n_{\rm init}$) for M\_I0\_z10, M\_I-1\_z10, M\_I0\_z8, and M\_I0\_z10\_NS, respectively. We also bin all the particles in $n_{\rm init}$-space to examine how much is contribute to $N^{\rm add}_{\rm rec,150}$ from given $n_{\rm init}$, which is written as $dN^{\rm add}_{\rm rec,150}/d\log_{10}n_{\rm init}$.

Except for the no-shielding run, $n_{\rm init}$ correlates with $N^{\rm add}_{\rm rec,150}$ up to a certain density and the correlation saturates above that density. The $N^{\rm add}_{\rm rec,150}-n_{\rm init}$ relations are almost the same across the different runs up to the saturation density, and the $N^{\rm add}_{\rm rec,150}$ is fixed above the saturation density that is not always the same for different runs. For M\_I0\_z10 and M\_I0\_z8, the saturations happen at almost the same density at $n_{\rm init}\sim 0.2~\rm{cm}^{-3}$. But, the saturation happens at $n_{\rm init}\sim 0.04~\rm{cm}^{-3}$ in M\_I-1\_z10. 

This saturation density is similar to the asymptotic value of $n_{\rm crit}$ when the R-type phase ends (See Fig.~\ref{fig:n_crit} for the behavior of $n_{\rm crit}$). According to the density and ionization histories of individual particles in Figure~\ref{fig:pdt_den}, particles below the threshold density will ionize almost immediately at their initial densities and will expand until its density drops close to the cosmic mean density. In this case, particles that started with higher $n_{\rm init}$ will achieve more recombination. In contrast, particles with their densities above the threshold experience expansion before ionization, get ionized at the threshold density, and go through density drops similar to one that started from the threshold density. The particles that started from higher than the threshold therefore ends up with similar amounts of recombination to those that started at the threshold do. In the no-shielding run, there is no such a threshold density because all the particles are ionized instantly at their initial densities. So $N^{\rm add}_{\rm rec,150}$ keeps correlating with $n_{\rm init}$ no matter how high $n_{\rm init}$ is.

Despite the small difference in the $N^{\rm add}_{\rm rec,150}$-$n_{\rm init}$ relation between M\_I0\_z10 and M\_I0\_z8, the global average of $N^{\rm add}_{\rm rec,150}$ is significantly higher in M\_I0\_z8. This difference comes from the difference in the initial density PDF. M\_I0\_z8 has a larger number of high-$n_{\rm init}$ particles that contribute highly to the global $N^{\rm add}_{\rm rec,150}$. That is seen by $N^{\rm add}_{\rm rec,150}/d\log_{10}n$ being higher at $n_{\rm init} \gtrsim 0.1~{\rm cm}^{-3}$ in M\_I0\_z8 (lower panel of Figure~\ref{fig:r_vs_n_c}$c$) than in M\_I0\_z10 (lower panel of Fig.~\ref{fig:r_vs_n_c}$a$). This is expected because growth of structure would put more particles in high-density end in lower redshifts and it also explains the negative scaling of $N^{\rm add}_{\rm rec,150}$ with $[1+z_i]$ in Equation~(\ref{eq:Naddrec150}).

The $N^{\rm add}_{\rm rec,150}$-$n_{\rm init}$ relation is quite different between M\_I0\_z10 and M\_I-1\_z10. The relation starts similarly in the low-density end in both cases, but it saturates at a lower density in M\_I-1\_z10. The saturation happens at $n_{\rm init}\approx 0.2~{\rm cm}^{-3}$ in M\_I0\_z10 and at $n_{\rm init}\approx 0.04~{\rm cm}^{-3}$ in M\_I-1\_z10. This is because the I-fronts settle down at a lower density for lower EIBR intensity. Despite the fact that both cases have the same initial density PDF, particles with $0.04 \lesssim n_{\rm init}\lesssim 0.2~{\rm cm}^{-3}$ only contributed to $N^{\rm add}_{\rm rec,150}$ in M\_I0\_z10 and M\_I-1\_z10 as can be seen by comparing $N^{\rm add}_{\rm rec,150}/d\log_{10}{n}$ in M\_I0\_z10 (lower panel of Fig.~\ref{fig:r_vs_n_c}$a$) and M\_I-1\_z10 (lower panel of Fig.~\ref{fig:r_vs_n_c}$b$). This is responsible for the positive scaling of $N^{\rm add}_{\rm rec,150}$ with $J_{21}$.

%The saturation at a lower $N^{\rm add}_{\rm rec,150}$ in M\_I-1\_z10 weakens $N^{\rm add}_{\rm rec,150}/d\log_{10}{n}$ above the threshold. This is responsible for the positive scaling to $N^{\rm add}_{\rm rec,150}$ for $J_{21}$.

%{\color{blue}
%\subsubsection{Fitting for Time-Dependent Extra Recombination}
%}

\subsection{Opacity of the IGM due to Small-scale Structure}\label{sec: Opacity} 

\begin{figure*}
  \begin{center}
    \includegraphics[scale=0.5]{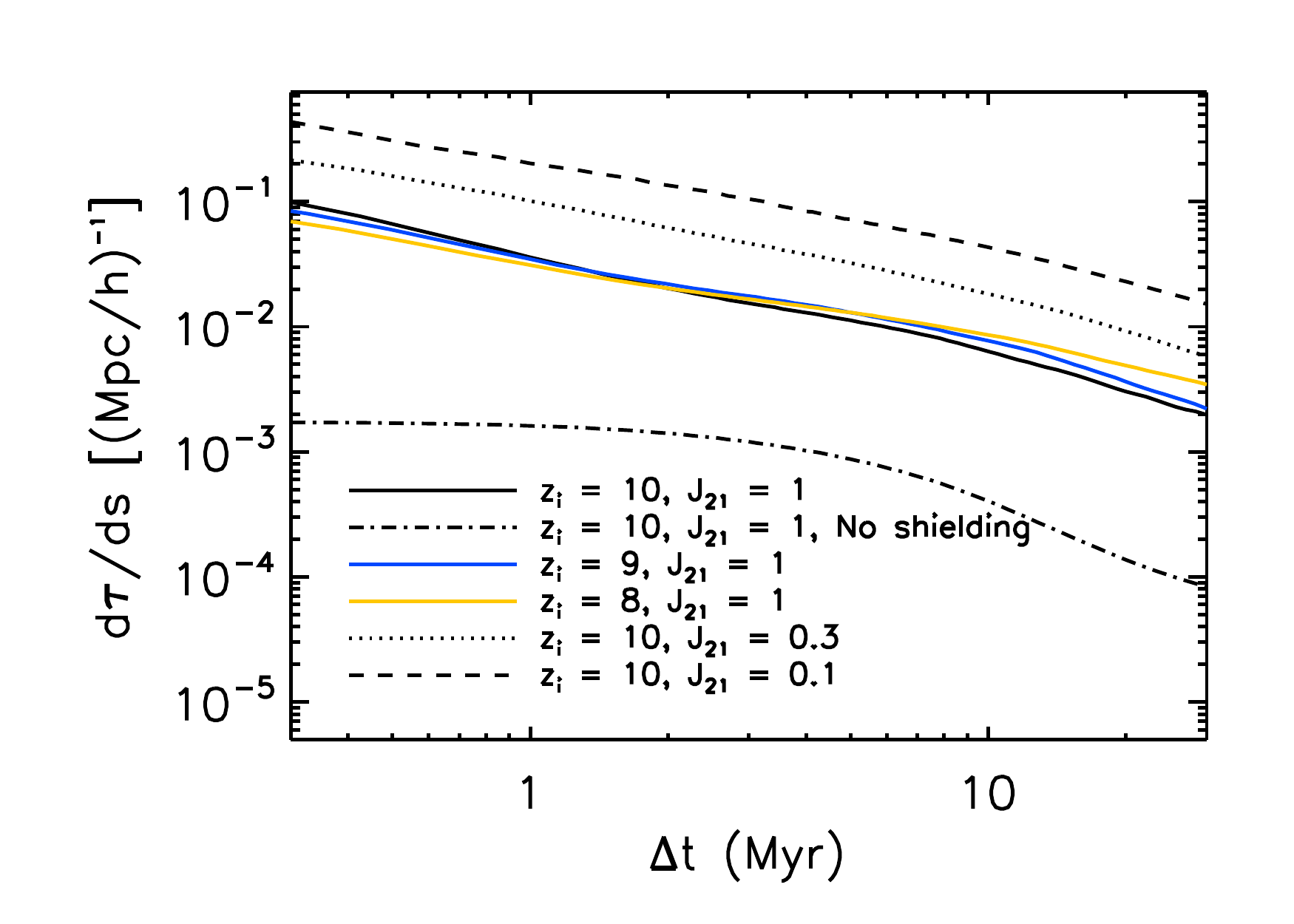}
     \includegraphics[scale=0.5]{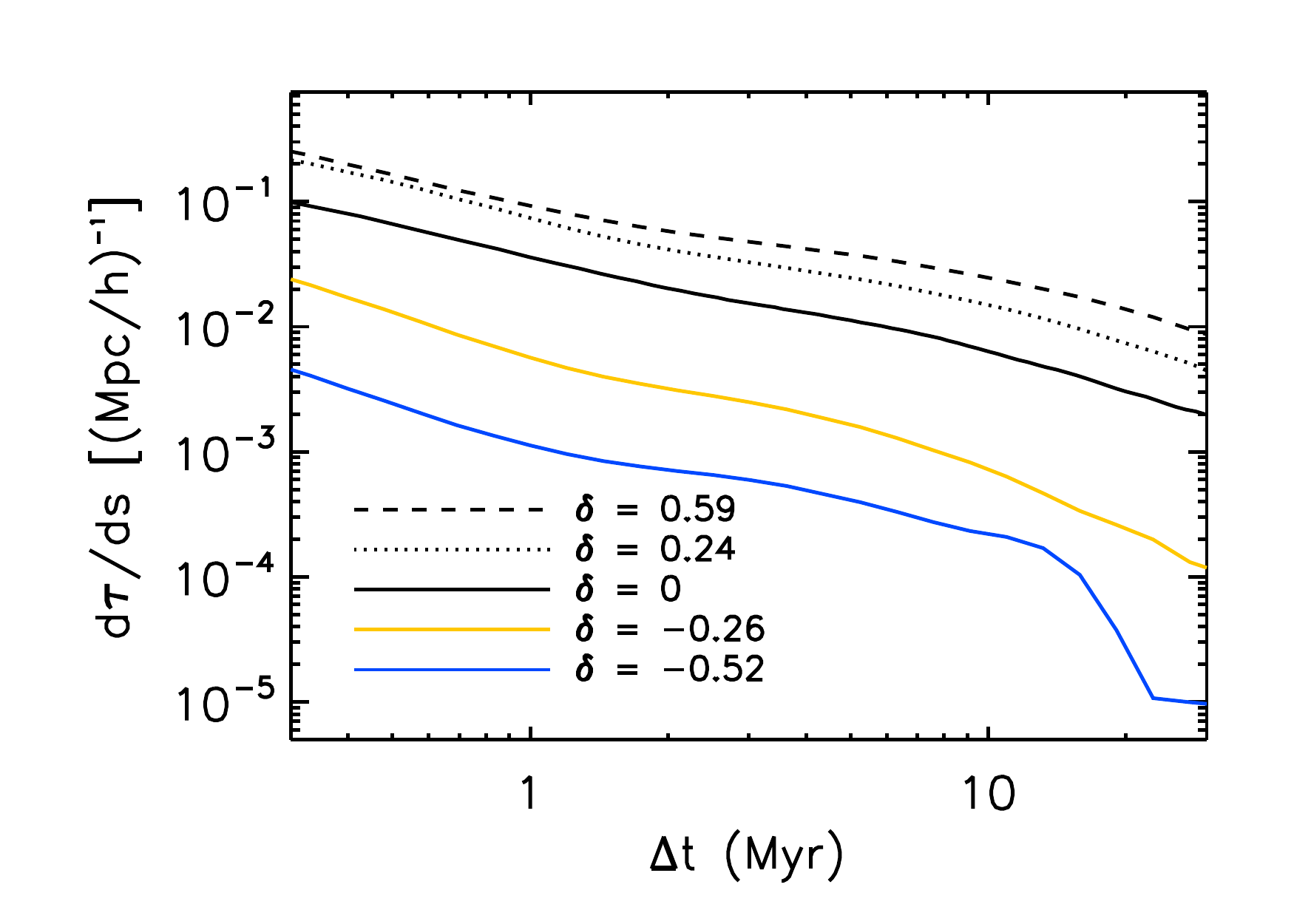}
          \caption{(left) The mean opacity per distance $d\bar\tau/ds$ as a function of time $\Delta t$ for M\_I0\_z10 (black solid), M\_I0\_z9 (blue solid), M\_I0\_z8 (yellow solid), M\_I-0.5\_z10 (black dotted), M\_I-1\_z10 (black dashed), \& M\_I0\_z10\_NS (black dot-dashed). (right) Similar to the left, but for M\_I0\_z10\_VH$\delta$ (black dashed), M\_I0\_z10\_H$\delta$ (black dotted), M\_I0\_z10 (black solid), M\_I0\_z10\_L$\delta$ (yellow solid), \& M\_I0\_z10\_VL$\delta$  (blue solid).} 
        \label{fig:bartau}
        \end{center} 
\end{figure*} 

%Lyman-limit opacity from small-scale structure will be missed in large-scale simulations used for comparison against the Lyman alpha forest measurement. In this section, we present the Lyman-limit opacity from our sample volume with evaporating small-scale structures. 

Small-scale structures add to the Lyman-limit opacity, potentially explaining the absence of it in the large-scale simulations compared to what is observed after the end of reionization in the Lyman alpha forest, and extrapolated to earlier redshifts from that post-reionization observation. In this section, we present the Lyman-limit opacity from our sample volume with evaporating small-scale structures. 

The Lyman-limit cross-section for the EIBR used in this work is given by
\bea
\bar{\sigma}=\frac{\int^{\infty}_{13.6{\rm eV}/h_p} \frac{I_\nu}{h_p\nu} \sigma_\nu~d\nu}{\int^{\infty}_{13.6{\rm eV}/h_p} \frac{I_\nu}{h_p\nu}~d\nu}=1.62\times 10^{-18}~{\rm cm}^2,
\eea
where $h_p$ in the Planck's constant and $I_\nu$ is the intensity of EIBR. Using the projected 2D H I column density $N_{\rm H I}$ shown in Figures~\ref{fig:ColumnDen}, \ref{fig:ColumnDen_NS}, \ref{fig:ColumnDen_ND}, \ref{fig:ColumnDen_I-1}, \& \ref{fig:ColumnDen_z8}, we calculate the transmissivity by taking $e^{-N_{\rm H I}\bar{\sigma}}$. Then, we take the log of the average transmissivity on the map to calculate the opacity for the sample volume with $200~h^{-1}~\rm{kpc}$ depth:
\bea
\left< e^{-N_{\rm H I}\bar{\sigma}} \right>_{\rm{sim}} = \frac{d\bar\tau}{ds}(200~h^{-1}~\rm{kpc} ),
\eea
where $d\bar\tau/ds$ denotes the opacity per comoving distance. By multiplying 5 to the above, we obtain $d\bar\tau/ds$ per $h^{-1}~\rm{Mpc}$. 
We plot the result for M\_I0\_z10, M\_I0\_z9, M\_I0\_z8, M\_I-0.5\_z10, M\_I-1\_z10, M\_I0\_z10\_NS, M\_I0\_z10\_VH$\delta$, M\_I0\_z10\_H$\delta$, M\_I0\_z10\_L$\delta$, \& M\_I0\_z10\_VL$\delta$ in Figure~\ref{fig:bartau}.

In all cases, $d\bar\tau/ds$ falls monotonically over time as can be expected from decreasing H I fraction due to photo-evaporation. In all cases, $d\bar\tau/ds$ falls close to or below $0.01~({\rm Mpc}/h)^{-1}$ in 30 Myr. In reality, there should be large-scale structures preventing the mean free path from growing above $100~h^{-1}~\rm{Mpc}$. This shows that the opacity from small-scale structures is unlikely to last more than $\sim 30$ Myr in most cases.

The opacity is quite sensitive to $J_{21}$ and $\bar\delta$, but not to $z_i$. For the $J_{21}=0.1$ case (M\_I-1\_z10), small-scale structure can limit the mean free path within $10~h^{-1}~\rm{Mpc}$ for $\sim$10 Myr on its own. Depending on how much of the entire universe is filled with volumes like this one, small-scale structure can be a substantial source of opacity. The $\bar\delta = -0.52$ case, M\_I0\_z10\_VL$\delta$, in contrast has negligibly small opacity ($<10^{-2}~({\rm Mpc}/h)^{-1}$) at all time. Combining the result here with probabilistic distribution of $J_{21}$, $z_i$, and $\bar\delta$ in large scale EoR simulations will verify how much small-scale structure can contribute to the opacity.

Noting that $d\bar\tau/ds$ appears nearly as straight lines in the log-log plots of Figure~\ref{fig:bartau}, we fit the result between $\Delta t =$ 1.5 Myr and 20 Myr with a power-law using $\Delta t =$ 1.5 Myr as the pivot point:
\bea
\frac{d\bar\tau}{ds} = A \left[\frac{\Delta t}{1.5~\rm{Myr}}\right]^\gamma ({\rm Mpc }/h)^{-1}.
\eea
We list the fitted values for $A$ and $\gamma$ in Table~\ref{table: fitting}. M\_I0\_z10\_L$\delta$ and M\_I0\_z10\_VL$\delta$ have somewhat irregular behaviors and are not well-described by the power-law fitting above. But, their opacities are practically zero at all time anyway.

\section{Box-size effect} \label{sec:convergence}

Both of the main contributors of the ionization budget, filamentary structures and minihalos, have huge variation in their populations according to their local density environments \citep[e.g.,][]{2015MNRAS.450.1486A}. Thus, it is important to have enough samples of structure to make sure the result is applicable to global cosmic environment. Toward this end, ETA13 reported their result for a convergence test of the clumping factor for box size in Figure 9 of their work. While they give $1~\rm{Mpc}$ as the converging box size, their result for $500~\rm{kpc}$ is not much different. But, $500~\rm{kpc}$ is still about twice bigger than $200~h^{-1}~\rm{kpc}$ that we use for our main analysis. 

In this work, we do our own convergence test by comparing the results from three runs with different box sizes and same $z_i$, $J_{21}$, and $\bar\delta$. For this, we use S\_I0\_z10, M\_I0\_z10, and L\_I0\_z10 that are $100~h^{-1}~\rm{kpc}$, $200~h^{-1}~\rm{kpc}$, and $400~h^{-1}~\rm{kpc}$ in a side, respectively. Due to the excessive computational expense for running L\_I0\_z10, we run it until $\Delta t = $ 15 Myr while the other two cases are run down to $\Delta t = $ 150 Myr. We show the clumping factor result for those three runs in Figure~\ref{fig:C_vs_t}. 

The clumping factor in S\_I0\_z10 is about $20\%$ smaller than in the other two cases for all time suggesting $100~h^{-1}~\rm{kpc}$ is too small for modeling the clumping factor. In M\_I0\_z10 and L\_I0\_z10, the clumping factor evolves identically up to $\Delta t = 2~\rm{Myr}$, but falls slowly in L\_I0\_z10. At $\Delta t = 15~\rm{Myr}$, the difference grows to $\sim 10\%$ at $\Delta t = 15~\rm{Myr}$. This is because L\_I0\_z10 has more high-mass minihalos that take a long time to evaporate. Considering that this evaporation process is not included in ETA13, the converging box-size for the ionization budget may be even larger than what they find.

 We partially overcome this limitation by considering sub-sample volumes with several different overdensities from a bigger volume and providing the scaling of the ionization budget with overdensity of the volume (Eq.~\ref{eq:Naddrec150}). Simply applying the overdensity variation in our $800~h^{-1}~\rm{kpc}$ box to the scaling relation doubles the ionization budget at $z=10$ (See Eq.~\ref{eq:meanN_rec}), highlighting the importance of considering different density environments. Toward this end, it is crucial to apply our scaling result to large-scale EoR simulations that capture all the density environments to truly assess the ionization budget.
 
The strong dependence of the ionization budget on the overdensity also implies that there is a room for improvement in the scaling relation that can be achieved by having more samples with different overdensities and redshifts. In this work, we consider five samples with different overdensities at $z=10$ and one cosmic mean density sample for each of $z=8$ and 9. Accommodating late reionization scenarios where most of the volume is ionized at around $z\sim6$, for example, would require extrapolating our results at $z=8-10$ down to 6. Thus, it is preferable to cover a wider range of overdensities and redshifts to improve the result quantitatively.

The main goal of this work is to point out the significance of the small-scale structure contribution in the ionization budget. While the result might change at quantitative level in subsequent studies, the qualitative understanding about small-scale structure from this work should remain valid.

%The most massive ones in the box should be the ones that are the most effected by the the box size effect in their formation. L\_I0\_z10 is able to keep more self-shielded clumps that last until late time than M\_I0\_z10 does and this is making difference in the late time clumping factor. $200~h^{-1}~\rm{kpc}$ seems to be large enough for modeling the clumping factor up to $\Delta t \sim 1~\rm{Myr}$, but a larger box is needed to be more precise about the subsequent evolution. 

%We also expect the box size effect to be an issue for the mean free path. After $\Delta t = 10$ Myr, only handful of neutral clumps are determining the mean free path of the simulation. For example, we can see in the column density map of M\_I0\_z10 in Figure~\ref{fig:ColumnDen} that we have only six self-shielded clouds setting the mean free path to be $\sim1$ Gpc at $\Delta t = 37$ Myr. In this case, it is unclear that a $200~h^{-1}~\rm{kpc}$ box would actually represent hundreds of Mpcs that the light would go through in the universe.

%Given that $200~h^{-1}~\rm{kpc}$ is well into the nonlinear regime at $z_i$'s considered in this work, it is not surprising to have this limitation. But, the qualitative picture about the evolution of structures after ionization should remain valid. Also, having a larger box size would likely to enhance the effects of small-scale structures by allowing more structures to form. 

\section{Summary and Discussion} \label{sec:summary}

We have simulated the clumpiness of ionized IGM during the EoR while resolving structures down to the Jeans scale of the pre-ionization IGM, aiming to estimate the the ionizing photon budget for reionization and provide a sub-grid prescription for the recombination rate in large-scale EoR simulations. Our target volumes are sub-Mpc non-star-forming regions that are ionized externally by distant ionizing sources. Such regions act as the sinks of ionizing photons and are much more commons than regions that host the sources of ionizing radiation like star-forming galaxies.

To achieve this, we have developed the GADGET-RT code that fully couples hydrodynamics to a reasonably accurate prescription for EIBR. This unveils the subsequent evolution of high clumping factor in the early stage of reionization found in ETA13. Also, this work is a 3-dimensional extension for the halo evaporation simulations of \cite{2004MNRAS.348..753S}, \cite{2005MNRAS.361..405I}, and \citet{2007MNRAS.375..881A}. GADGET-RT has been tested against a well tested 1D code from \citet{2007MNRAS.375..881A} for a spherically symmetric halo evaporation problem. We have run simulations with different $J_{21}$'s, $z_i$'s, and $\bar\delta$'s to explore the dependence of the clumping factor and the resulting ionization budget on these parameters. In the following, we summarize our main results.

\vspace{\bulletskip}
{\bf Evolution of the clumping factor:} When EIBR arrives the target volume, R-type I-fronts start to sweep structures super-sonically from low density regions, during which the clumping factor grows to a large ($>10$) value. This phase comes to an end in a few megayears when the I-fronts reach dense parts of the structures and transition to D-type. Then, the hydrodynamical back-reaction on ionized gas destroys the structures over tens of megayears, causing the clumping factor to decay. 

\vspace{\bulletskip}
{\bf Photon budget for the pre-reionization IGM:} The enhanced clumping factor during the R-type phase adds substantially to the ionizing budget for the reionization, which is neglected in previous works. The resulting extra recombination per H atom due to small-scale structure in a $200~h^{-1}~\rm{kpc}$ box with the mean density contrast $\bar\delta$ ionized by EIBR with the intensity $J_{21}$ at redshift of $z_i$ is $0.32[J_{21}]^{0.12}\left[(1+z_i)/11\right]^{-1.7}[1+\bar\delta]^{2.5}$. Using a distribution of $\bar\delta$ obtained from a $800~h^{-1}~\rm{kpc}$ at $z_i = 10$ gives 0.67 extra recombination per H atom for $J_{21}=1$.

\vspace{\bulletskip}
{\bf Lyman-limit Opacity:} Photo-evaporation quickly suppresses the opacity contributed by small-scale structures, but some cases with high overdensity ($\bar\delta=0.59$) or low EIBR intensity ($J_{21}=0.1$) are found to be able to limit the mean free path within $100~h^{-1}~\rm{Mpc}$ for more than 10 megayears only with small-scale structures. It is to be verified with large-scale EoR simulations whether such high opacity cases do have a significant impact on the global mean free path during the EoR.

%\vspace{\bulletskip}
%{\bf Mean free path of H-ionizing photons:} The mean free path grows rapidly, cautioning about using it as a constant value. When structures are ionized later, more self-shielded clumps exist to keep the mean free path shorter. When the intensity of EIBR is higher, self-shielded clumps are evaporated faster, leading to more rapid growth of the mean free path.

\vspace{\bulletskip}

It is meaningful to confirm that the high clumping factor of the ionized IGM found in ETA13 does occur in simulations with coupled hydrodynamics, and it does contribute significantly to the ionization budget for the reionization even under the hydrodynamic feedback of ionization suppressing the clumpiness of the IGM. For late reionization scenarios that most of the universe gets ionized toward the end of reionization ($z\sim6$), we can have a factor of two increase from two third per H atom we found for $z_i=10$ and $J_{21}=1$ due to the $\left[1+z_i\right]^{1.7}$ scaling. This is substantial considering that $\sim2-3$ is usually considered as the reionization budget in the literature \cite[e.g.,][]{2014ApJ...789..149S}. For a more definitive conclusion, we need to apply the scaling relation to existing EoR models.

The isotropy of ionizing background is a powerful assumption that allowed us to make the shielding algorithm efficient enough to be coupled to the hydrodynamics. Yet, one needs to be careful about interpreting the results as the angular distribution of incoming radiation would be more complex in reality. When a small ($< 1~\rm{Mpc}$) volume is exposed to the ionizing background, it is likely that a large-scale I-front would be sweeping the entire volume uni-directionally from one side. The radiation would isotropizes later as that volume is exposed to from more and more ionizing sources from diverse directions. At the early time when the radiation is close to being uni-directional, the geometry of H II regions in reality might differ significantly from what we see in our simulation. However, we note that the details of how early R-type I-fronts go is rather unimportant for the recombination accumulated in the time scales of $\sim10$ Myr or longer. Whichever direction the R-type I-fronts sweep across the box, they will eventually get trapped at density peaks and transition to D-type. At this point, the intensity of the radiation will determine up to what density the gas would be ionized. And, the subsequent hydrodynamics feedback would make the gas expand from the density peaks that has nothing to do with the direction of EIBR. Here shadows behind self-shielded clumps in the uni-directional case make some difference by leaving some low density gas neutral. But, the column density maps (e.g., Fig.~\ref{fig:ColumnDen}) show that the shielded (white and pink) part of the volume is only a tiny fraction, suggesting that it is not so significant.

There are a number of EoR physics not included in this work that can potentially affect the results. A drift velocity between baryon and dark matter would hinder structure formation in small scales \citep{2010PhRvD..82h3520T} and pre-reionization heating from X-ray sources \citep{2004MNRAS.352..547R} would hinder the accumulation of high-density gas in minihalos that can achieve a large number of recombination, thereby reducing the global recombination rate during the EoR. Such physics will be explored using the GADGET-RT code as the extra parameters of the ionization budget in our subsequent studies and the results here will serve as the foundation.

\section{Acknowledgement}
Authors thank S. Finkelstein, M. Milosavljevic, E. Komatsu, E. L. Robinson, M. Alvarez, and the referee, N. Gnedin, for their helpful comments on this work.
This material is based upon a project supported by the National Science Foundation East Asian and Pacific Summer Institute Program under Grant No. SP13041. PRS. It was also supported in part by U.S. NSF grant AST-1009799, NASA grant NNX11AE09G, NASA/JPL grant RSA Nos. 1492788 and 1515294, and supercomputer resources from NSF XSEDE grant TG-AST090005 and the Texas Advanced Computing Center (TACC) at the University of Texas at Austin. K.A. was supported by NRF-2012K1A3A7A03049606 and NRF-2014R1A1A2059811.

\appendix

\section{Test Problem : Evaporation of a Spherical Halo} \label{sec:Test_Problem}

\begin{figure}
  \begin{center}
    \includegraphics[scale=0.45]{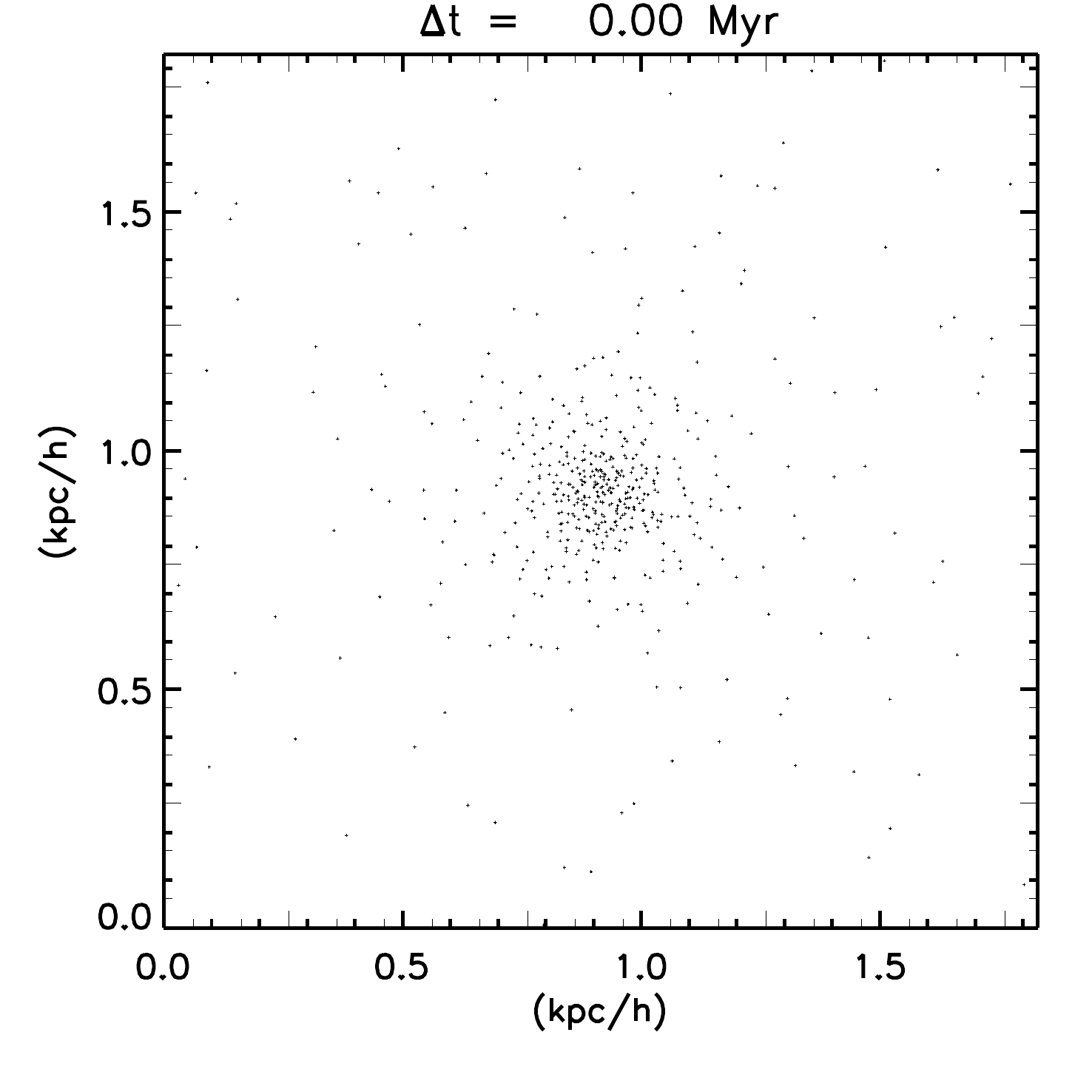}
    \includegraphics[scale=0.45]{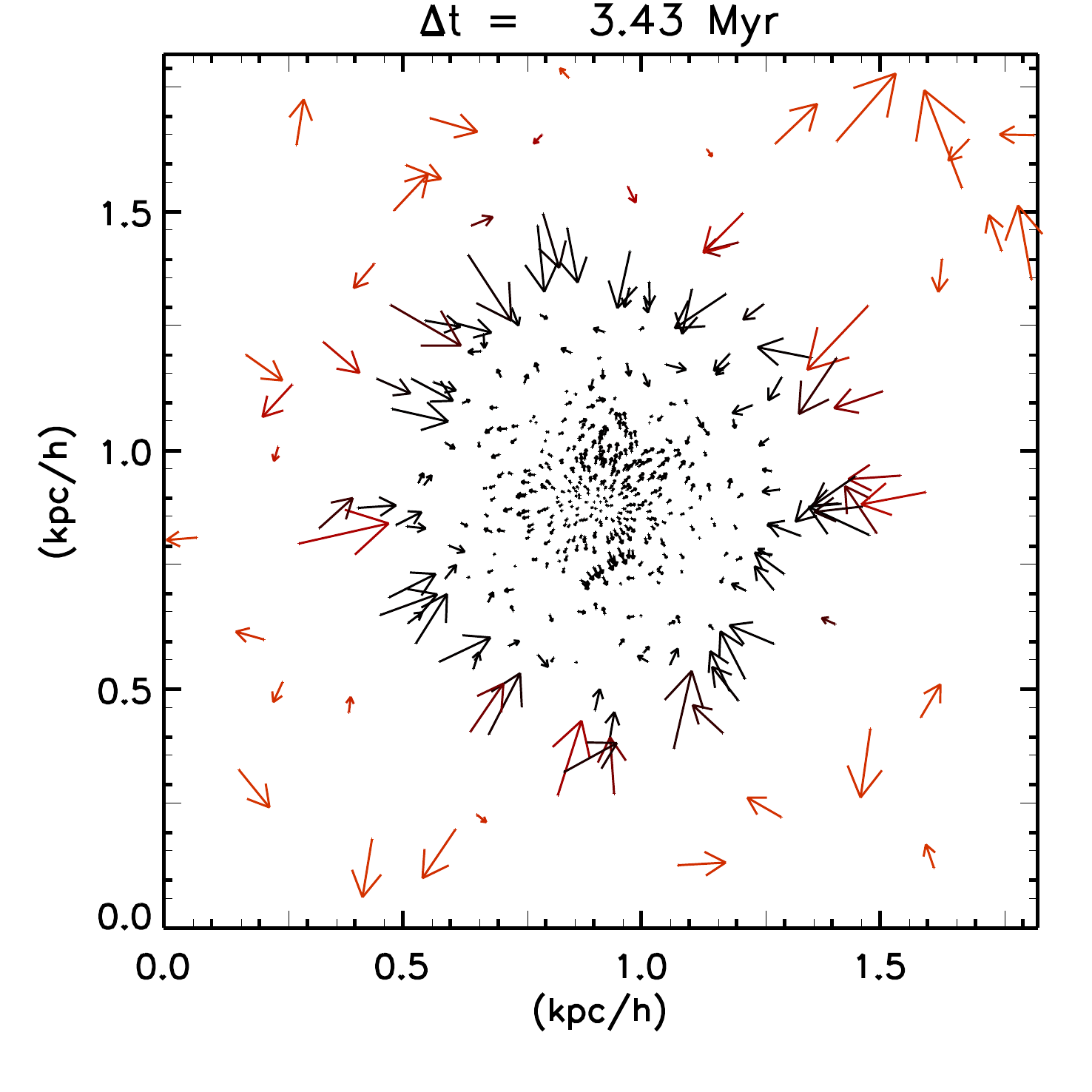}
    \includegraphics[scale=0.45]{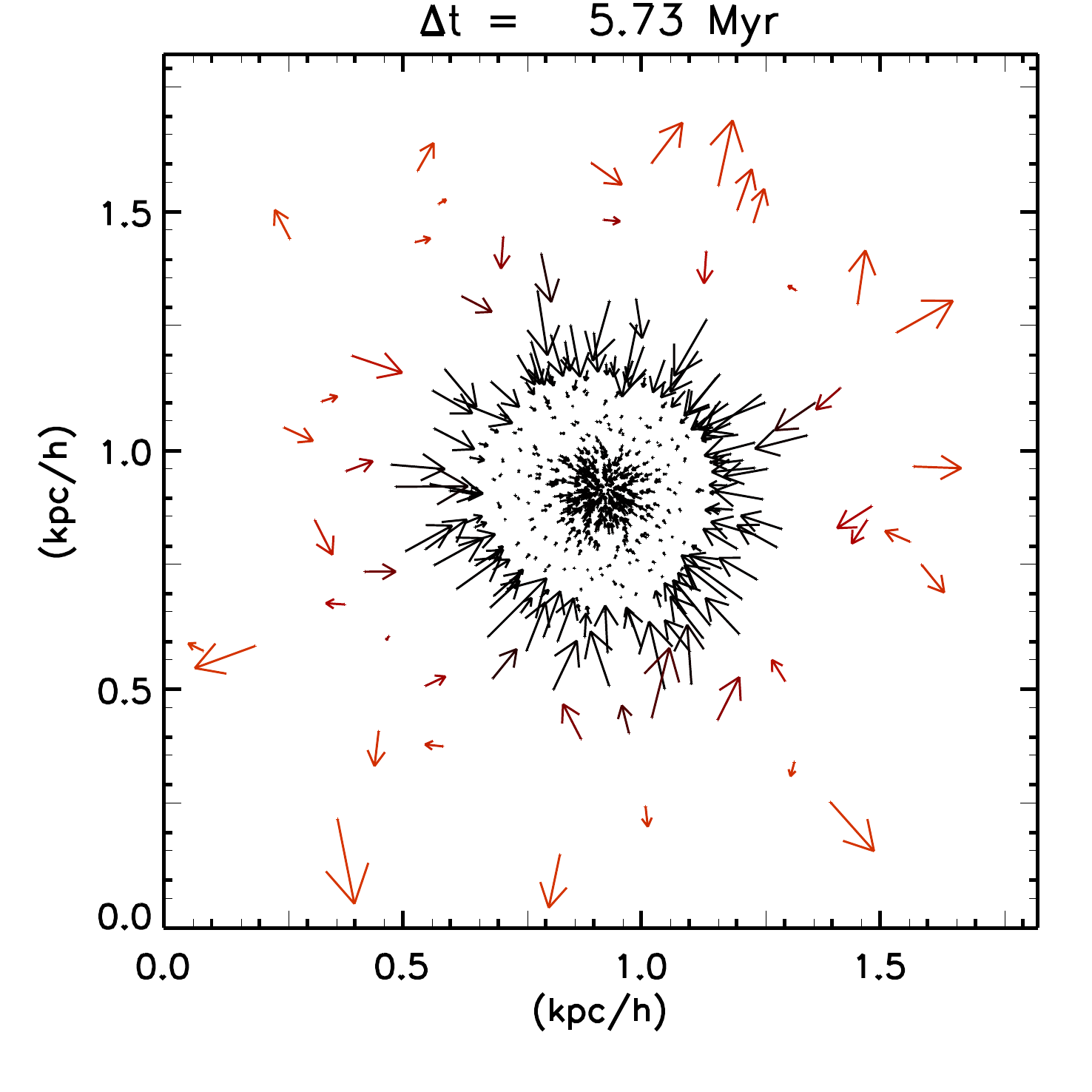}
    \includegraphics[scale=0.45]{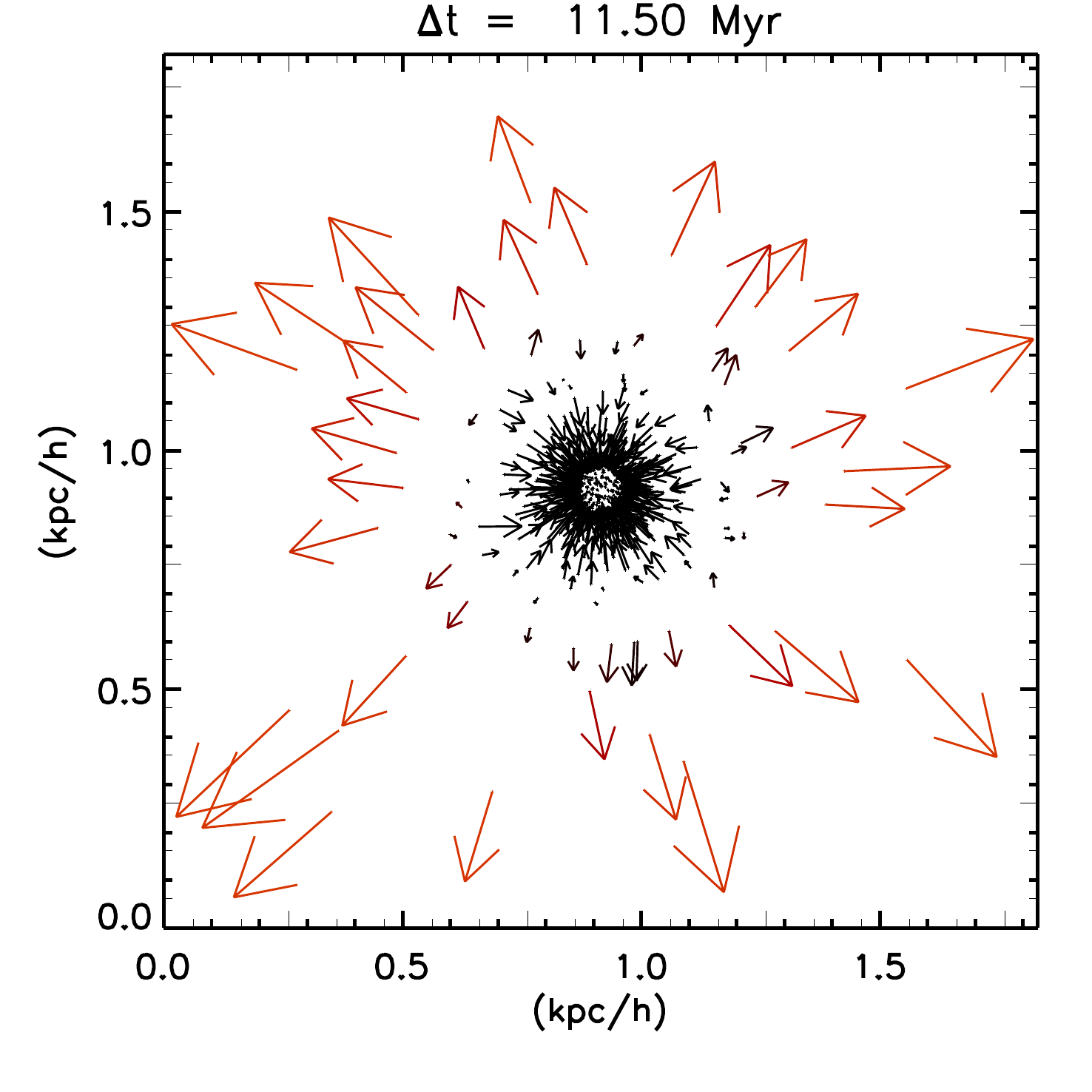}
  \caption{Spatial distribution of SPH particles in the GADGET-RT simulation with particle velocity and ionization status at $\Delta t =$ 0 (upper left), 3.43 (upper right), 5.73 (lower left), and 11.5 Myr (lower right). For visual convenience, only 20\% of the particles in a thin (0.2\% of the simulation box) slab that goes through the center of the halo are plotted. The arrows describe the projected particle velocities with the positions of their heads giving the linearly extrapolated positions after 5 Myr. Red/Black color indicates that the particle is ionized/neutral.}
   \label{fig:ptl_map}
  \end{center}
\end{figure}

\begin{figure*}
  \begin{center}
    \includegraphics[scale=0.5]{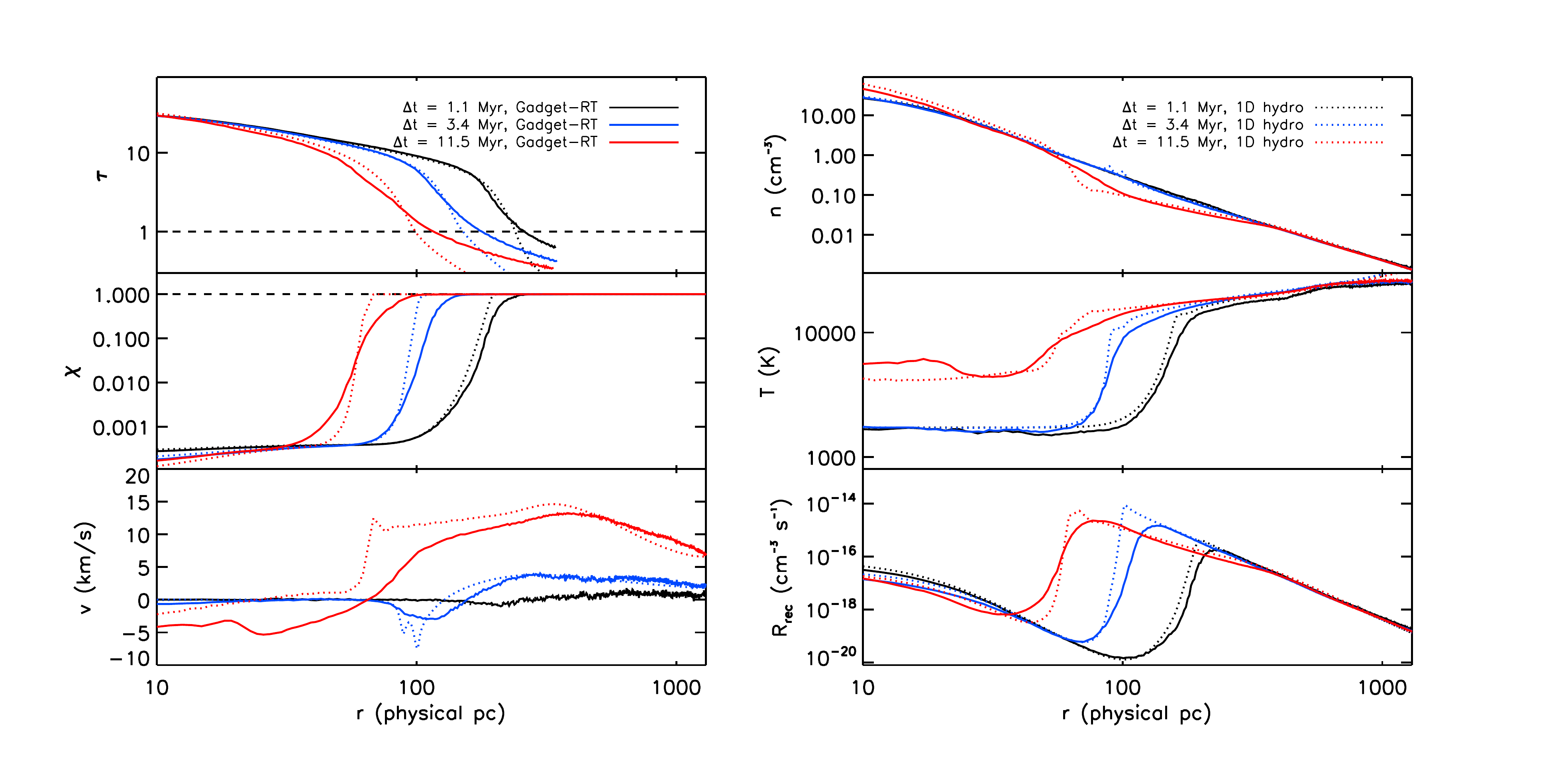}
  \caption{Radial profiles of the effective optical depth (top left panel), ionization fraction (middle left panel), radial velocity (bottom left panel), density (top right panel), gas temperature (middle left panel), and recombination rate (bottom right panel) from the GADGET-RT code of this work (solid) and 1D radiation-hydro code of Ahn and Shapiro 2007 (dotted). The results are compared for $\Delta t =$ 1.1 Myr (black), 3.4 Myr (blue), and 11.5 Myr (red). The radius on the $x$-axis is in the {\it physical} unit.}
  \label{KA_GADGET_compare}
  \end{center}
\end{figure*}

We test the accuracy of the GADGET-RT code for a spherically symmetric configuration that the one-dimensional radiation-hydrodynamics code of \citet{2007MNRAS.375..881A} can reproduce. In \citet{2007MNRAS.375..881A}, the 1D code was used to assess the effects of EIBR on a minihalo with the minimum-energy truncated isothermal sphere (TIS) profile \citep{1999MNRAS.307..203S,2001MNRAS.325..468I}. The 1D code accurately captures the evolution of I-fronts both in the supersonic R-type and the subsonic D-type phases. The 1D code has been tested for a number of problems with existing analytical solutions \citep[See Appendix C of][]{2007MNRAS.375..881A}.

We first create the initial conditions for the 1D code using the fitting formula in Appendix A of \citet{1999MNRAS.307..203S}. We adopt $M=10^6~M_{\odot}$ for the mass inside the truncation radius, $r_t=170~physical~pc$, and $z_{\rm col}=10$ for the redshift of collapse. We do {\it not} truncate the initial density profile at $r=r_t$, but instead extend it to $r=10 r_t$ extrapolating the fitting formula. This extended TIS profile decays toward large-$r$ direction in a reasonable way, allowing us to test the code for the outskirt of the minihalo. The density profile at $\Delta t = 1.1$ Myr shown in Figure~\ref{KA_GADGET_compare} is close to the initial conditions. We keep track of $N_{\rm sh}=10,000$ radial shells linearly spaced from $r=10^{-3}~r_t$ to $10~r_t$. This is the same spatial resolution adopted in \citet{2007MNRAS.375..881A}. We bound the outer-most shell with the pressure of that shell at the initial time-step. This pressure becomes practically negligible as soon as the ionization of outer shells photo-heats the gas above $10,000$ K from $\sim2,000$ K. 

We then create the corresponding initial conditions for the GADGET-RT code. We set the box size to be $20r_t$ and put the center of the halo at the center of the box. We randomly place particles using the extended TIS density profile as the probability function both for the dark matter and gas particles. The effective pressure for the dark matter is converted into the random velocity dispersion following the Boltzmann distribution. 

In the 1D code, the optical depth to the background radiation at the $i$th shell from the center at the frequency $\nu$ is given by the angular average over the lines of sight, $\vec{l}$:
\bea 
\tau_{\nu,i} = (4\pi)^{-1} \int d\Omega \int_{l=0}^{l_{\rm max}} dl  \sum_{X} n_{X,j}(r) \sigma_{X,\nu},
\eea
where $dr$ is the thickness of the shell, $n_{X}$ is the number density of a species $X$, $\sigma_{X,\nu}$ is the cross-section of the species $X$ for the frequency $\nu$, and the baryonic species $X$ include H, He, $\rm{He}^+$, $\rm{H}^-$, $\rm{H}_2$, and $\rm{H}_2^+$. Here $l_{\rm max}$ is the distance from the $i$th shell to the outer-most shell. $r$ is given by
\bea  \label{eq:r}
r = \sqrt{r_i^2+l^2+2lr_i\mu},
\eea
where $\mu = \hat{l} \cdot \hat{r} $. Solving Equation~(\ref{eq:r}) for $l$ setting $r=r_{N_{\rm sh}}$ gives $l_{\rm max}$. 

Since the angular integral in Equation~(\ref{eq:tau average}) is symmetric for the azimuthal direction, it can be simplified as the following.
\bea \label{eq:tau average}
\tau_{\nu,i} = \frac{1}{2} \int^{1}_{-1} d\mu \int_{l=0}^{l_{\rm max}} dl  \sum_{X} n_{X}(r) \sigma_{X,\nu}.
\eea
We use interpolation to define $n_{X}(r)$ for $r_1<r<r_{N_{\rm sh}}$. And, we use the Simpson's Rule to evaluate integrals. For EIBR, we adopt the same parameters used in the standard run (M\_I0\_z10) that the spectrum is given by $10^5$ K blackbody spectrum and $J_{21}$ = 1.

Figure~\ref{fig:ptl_map} shows the particle maps with the velocities and ionization statuses of the particles in the GADGET-RT simulation shown for four snapshots at $\Delta t = 0$ (upper left panel), 3.43 (upper right panel), 5.73 (lower left panel), and 11.5 Myr (lower right panel). At $\Delta t > 0$, the transition between the region populated with black arrows and that populated with red arrows marks an I-front propagating toward the minihalo center. A ring of black arrows pointing toward the center marks a shock that formed in reaction to the increased pressure at the outskirt of the halo. At $\Delta t = 11.5$ Myr, an out-flow of gas is also observed. These phenomena are all consistent with findings in \citet{2007MNRAS.375..881A}. 

For quantitative comparison, we compare the radial profiles of six physical quantities from the two simulations in Figure~\ref{KA_GADGET_compare}. The effective optical depth, $\tau_{\rm eff}$, in the top right panel is defined by $\tau_{\rm eff}=-\log(\mathcal{T})$ where $\mathcal{T} =(1/6)\Sigma_{X=\pm x,y,z}\exp(-N_X\sigma)$ is the average transmissivity from the six column densities for $\pm x, \pm y,$ and $\pm z $ directions calculated in the simulation. For the 1D code, the effective optical depth can be calculated precisely from the neutral hydrogen density profile. Along with $\tau_{\rm eff}$, we also compare the radial profiles of the ionized fraction, radial velocity, density, temperature, and recombination rate. 

$\tau_{\rm eff}$ is slightly overestimated in the outer part of the minihalo. This is because the cloud of neutral gas in the minihalo saturate at least one of the six sky pixels in the perspective of a shielded particle with H I column density, making it completely optically thick to the EIBR even when the minihalo is quite distant and should cover less of the sky than that pixel does. This however requires the location of the shielded particle to be not only outside of the cloud, but away from it by a few time the size of the cloud. That is well behind the I-front populated by highly ionized gas, where the overestimation of  $\tau_{\rm eff}$ does not make any significant error.

For this reason, we generally find a good agreement between the two codes for quantities other than $\tau_{\rm eff}$. Transition zones of the quantities at the I-front tend to be more spread in the GADGET-RT code because the resolution of the GADGET-RT code is unable to perfectly resolve the sharp I-front as in the 1D code. However, the outer fully ionized part of the halo shows an excellent agreement for all the quantities. For the purpose of looking into the fate of ionized gas behind I-fronts, this test result guarantees the reliability of the GADGET-RT code.

\bibliographystyle{apj}
\end{CJK}
\bibliography{reference}

\end{document}